\documentclass{article}
\pdfoutput=1

\usepackage{enumitem}
\usepackage{hyperref}
\setlist{nolistsep}
\usepackage{amssymb,amsmath,bm} 
\usepackage{subcaption,caption}
\usepackage{color}

\usepackage{mathtools}
\usepackage{tikz}
\makeatletter
\newcommand\mathcircled[1]{%
  \mathpalette\@mathcircled{#1}%
}

\newcommand{\Matlab}{{\sc Matlab}}
\newcommand{\dt}{\partial_{t}}

\title{A novel nonlocal partial differential equation model of endothelial progenitor cell cluster formation during the early stages of vasculogenesis}

\author{Chiara Villa \and
Alf Gerisch    \and
Mark A. J. Chaplain}

\author{Chiara Villa\thanks{School of Mathematics and Statistics, University of St Andrews, UK 
  (cv23@st-andrews.ac.uk)}
  \and
Alf Gerisch\thanks{Department of Mathematics, Technische Universität Darmstadt, Germany
 (gerisch@mathematik.tu-darmstadt.de) }
\and
Mark A. J. Chaplain\thanks{School of Mathematics and Statistics, University of St Andrews, UK 
  (majc@st-andrews.ac.uk)}
}

\begin{document}
\date{}
\maketitle
\begin{abstract}
The formation of new vascular networks is essential for tissue development and regeneration, in addition to playing a key role in pathological settings such as ischemia and tumour development. Experimental findings in the past two decades have led to the identification of a new mechanism of neovascularisation, known as cluster-based vasculogenesis, during which endothelial progenitor cells (EPCs) mobilised from the bone marrow are capable of bridging distant vascular beds in a variety of hypoxic settings \textit{in vivo}. 
This process is characterised by the formation of EPC clusters during its early stages and, while much progress has been made in identifying various mechanisms underlying cluster formation, we are still far from a comprehensive description of such spatio-temporal dynamics.
In order to achieve this, we propose a novel mathematical model of the early stages of cluster-based vasculogenesis, comprising of a system of nonlocal partial differential equations including key mechanisms such as endogenous chemotaxis, matrix degradation, cell proliferation and cell-to-cell adhesion. 
We conduct a linear stability analysis on the system and solve the equations numerically. We then conduct a parametric analysis of the numerical solutions of the one-dimensional problem to investigate the role of underlying dynamics on the speed of cluster formation and the size of clusters, measured via appropriate metrics for the {cluster} width and compactness. We verify the key results of the parametric analysis with simulations of the two-dimensional problem.
Our results, which qualitatively compare with data from \textit{in vitro} experiments, elucidate the complementary role played by endogenous chemotaxis and matrix degradation in the formation {of clusters, suggesting chemotaxis is responsible for the cluster topology while matrix degradation is responsible for the speed of cluster formation. 
{Our results also indicate that the nonlocal cell-to-cell adhesion term in our model, even though it initially causes cells to aggregate, is not sufficient to ensure clusters are stable over long time periods. Consequently, new modelling strategies for cell-to-cell adhesion are required to stabilise \textit{in silico} clusters.}
We end the paper with a thorough discussion of} promising, fruitful future modelling and experimental research perspectives.
\end{abstract}


\section{Introduction}
The formation of new vascular networks is essential for tissue development and regeneration, as a functional vasculature is critical for tissue homeostasis. It is responsible for the delivery of oxygen and nutrients as well as the disposal of waste products. 
In addition, neovascularisation of local tissue is critical in a variety of pathological processes, among which are {retinopathy}, tumour growth, wound healing and soft-tissue ischemia. The development of a vascular network in localised solid tumours is particularly well known to promote further tumour growth and metastases. 
Therefore a better understanding of the mechanisms governing neovascularisation can help improve current therapeutic strategies, as well as identify new ones. 


The two fundamental processes of neovascularisation are known as \emph{vasculogenesis}, the \textit{de novo} formation of a new vascular network, and \emph{angiogenesis}, the formation of new blood vessels from pre-existing ones.
While the distinction between these two processes has been clear since the early days of the field of study of vascular development~\cite{sabin1917origin}, for a long time the term vasculogenesis was only associated with endothelial progenitor cells (EPCs)-mediated embryonic vasculogenesis, while blood vessels formation in adult organisms was believed to be formed predominantly, if not exclusively, via angiogenesis~\cite{risau1995vasculogenesis}. 
In the past 20 years great progress was made in understanding neovascularisation processes, in particular with the discovery of bone marrow-derived EPCs and postnatal vasculogenesis~\cite{asahara2004endothelial,asahara1997isolation,asahara1999bone,kolte2016vasculogenesis,poole2001role,risau1995vasculogenesis,shi1998evidence}, for which a specific mechanism, here referred to as \emph{cluster-based vasculogenesis}, has recently been proposed{~\cite{blatchley2019hypoxia}}.
This newly identified mechanism of neovascularisation has been observed only in hypoxic settings \textit{in vivo}, from development~\cite{proulx2010cranial} to soft-tissue ischemia~\cite{tepper2005adult} to tumour vascularisation~\cite{vajkoczy2003multistep}, and has been studied in detail by Blatchley \textit{et al.}~\cite{blatchley2019hypoxia} in hypoxic gradients \textit{in vitro}.

\subsection{Cluster-based vasculogenesis}
{During this process, EPCs residing in the bone marrow are recruited in the circulation to the hypoxic site~\cite{asahara1999bone,shi1998evidence}, where they first organise into clusters. They then proceed to sprout from these clusters and, in combination with angiogenesis, form the new vascular network merging with pre-existing vasculature. }
The term ``cluster-based'' vasculogenesis was therefore proposed by Blatchley and coworkers~\cite{blatchley2019hypoxia} to highlight the key step of the process which well distinguishes it from classical ``single-cell'' vasculogenesis, during which sparse mature endothelial cells (ECs) directly reorganise into a vascular network~\cite{serini2003modeling}.

\begin{figure}[h!]
\centering
\includegraphics[width=0.8\linewidth]{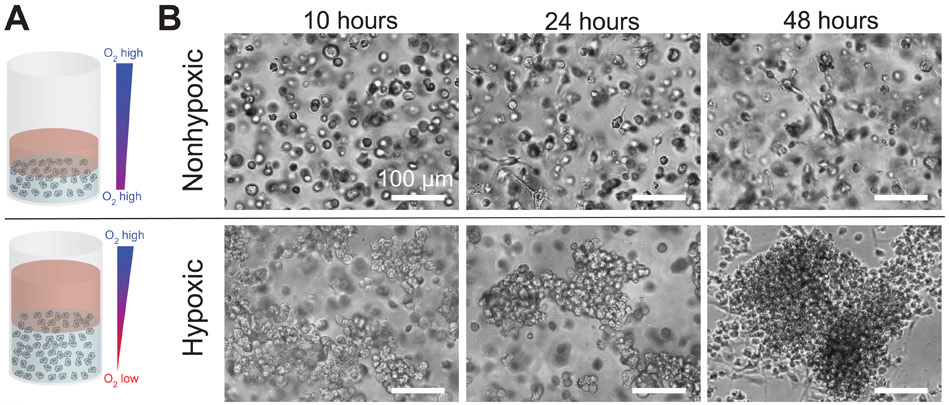}
\caption{\label{P1:figbio} {(A) Illustration summarising the 3D \textit{in vitro} experimental set ups considered by Blatchley \textit{et al.}~\cite{blatchley2019hypoxia} to study EPC cluster formation in hypoxic gradients (bottom) and in control nonhypoxic environments (top). (B) Corresponding experimental images displaying cluster formation after 48hours in hypoxic environments (bottom), while no clusters formed in nonhypoxic conditions (top).
Illustrations and images are taken from Figure 1 in~\cite{blatchley2019hypoxia} under Creative Commons licence \url{https://creativecommons.org/licenses/by-nc/4.0/}. }}
\end{figure}

\begin{figure}[htb!]
\centering
\includegraphics[width=0.44\linewidth]{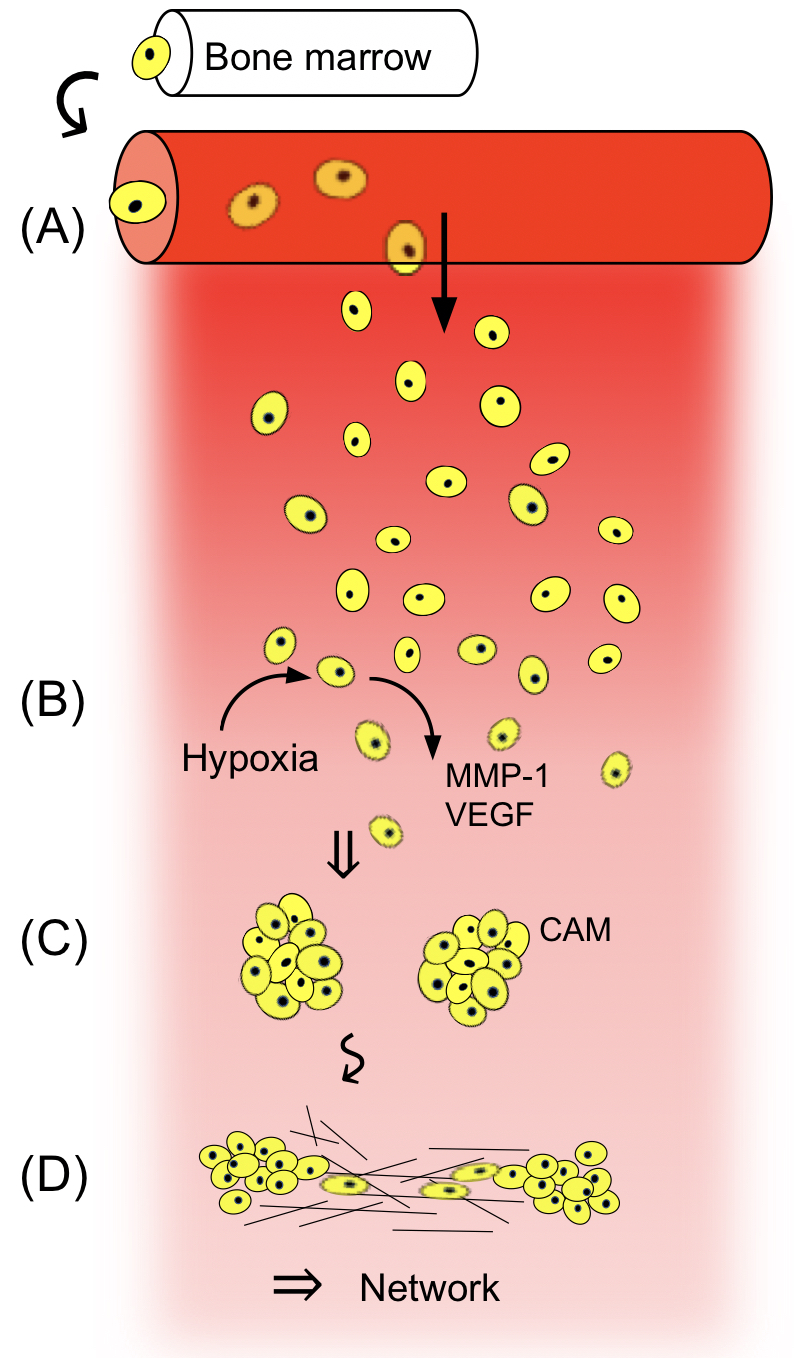}
\caption{\label{fig:bio} Summary of the steps of cluster-based vasculogenesis: (A) EPCs in the bone-marrow enter the circulation and reach the hypoxic site; (B) local hypoxia fosters EPC production of matrix degrading enzymes (MMP-1) and chemotactic agents (VEGF); (C) clusters form and are stabilised by cell-adhesion molecules (CAM); (D) cell-matrix interactions increase, sprouting from clusters occurs and a vascular network forms.}
\end{figure}

While EPCs recruitment to angiogenic sites has been well studied, we are still far from understanding how clusters form and what drives sprouting from these clusters. 
Akita \textit{et al.}~\cite{akita2003hypoxic} reported that EPC differentiation, secretion of angiogenic factors -- such as Vascular Endothelial Growth Factor (VEGF) -- and migration were enhanced by hypoxic conditioning, resulting in increased EPC cluster formation in hypoxia \textit{in vitro}. Nevertheless, it was not until recently that Blatchley and coworkers proposed, together with the name here used, a two-step mechanism of cluster formation and stabilisation in hypoxic environments~\cite{blatchley2019hypoxia}{, illustrated in Figure~\ref{fig:bio}. 
The proposed mechanism} relies on the observation that in highly hypoxic environments the production of matrix metalloproteases (MMPs) is upregulated by the presence of reactive oxygen species {(Figure~\ref{fig:bio}B). This results in }significant extracellular matrix (ECM) degradation responsible for EPCs aggregation into clusters, which are then stabilised by cell-to-cell adhesion {(Figure~\ref{fig:bio}C)}. EPC clusters were observed between 24 and 48 hours from the beginning of the experiments, presenting diameters in the range $100-400\,\mu$m{-- see Figure~\ref{P1:figbio}}.
Finally, after 48 hours, cell-to-matrix interactions lead to sprouting and network formation, particularly fostered by highly viscoelastic -- where not degraded -- ECM {(Figure~\ref{fig:bio}D)}. The final network, observed after 72 hours, was reported to be up to $500\,\mu$m in length, larger than previously observed ones formed via angiogenesis or single-cell vasculogenesis. 
Overall, cluster-based vasculogenesis has been shown to be of great importance in highly hypoxic tissues, as EPC clusters form in these regions far from pre-existing blood vessels and, after sprouting, function as a bridge for distant vascular beds, allowing for neovascularisation in physiological and pathological settings in which angiogenesis alone would not have sufficed. 

While this is a great step forward in the understanding of cluster-based vasculogenesis, further work is required to reach an exhaustive comprehension of the process, as well as to unlock its full potential in therapeutic interventions in a variety of pathological process. On this regard, mathematical modelling can help elucidate the mechanisms behind network formation, serving as a proof of concept mean for newly developed theories~\cite{servedio2014not}, as well as steer experimental investigations towards the most promising research perspectives. 

\subsection{Mathematical modelling of vasculogenesis}\label{sec:into:math}

While the mathematical model presented here is the first one formulated to study cluster-based vasculogenesis, many have been proposed to investigate {\textit{in vitro}} single-cell vasculogenesis.
{Most models in the literature refer back to the work of Serini \textit{et al.}~\cite{serini2003modeling} who investigate single-cell vasculogenesis in matrigel.
During the early stages of this process, ECs were observed to undergo rapid motion, during which they maintained a round shape, until collision with their closest neighbours (3-6 hours). They then proceeded to reorganise into a continuous multicellular network, which can be represented as a collection of nodes connected by capillary chords (of characteristic length $\ell\approx200 \mu$m), exhibiting a more elongated morphology, until network stabilisation (9-15 hours). Note that cell proliferation and death of mature ECs can be neglected on such a short timescale \textit{in vitro}.
We first review key single-cell vasculogenesis models in the literature, referring the interested reader to more detailed reviews~\cite{ambrosi2005review,boas2018cellular,czirok2013endothelial,scianna2013review}, and then discuss model requirements to study cluster-based vasculogenesis. }

\paragraph{{Partial differential equation models of single-cell vasculogenesis}}
Continuum deterministic models in the extant literature can be categorised {as Persistence and Endogenous Chemotaxis (PEC) models~\cite{ambrosi2004cell,coniglio2004percolation,gamba2003percolation,kowalczyk2004stability,liu2021asymptotic,serini2003modeling}, modelling the early stages of single-cell vasculogenesis, and mechanochemical models~\cite{manoussaki2003mechanochemical,murray1993mechanical,murray2003mechanochemical,namy2004critical}, better suited to describe the late stages of the process,} with the exception of the work by Tosin \textit{et al.}~\cite{tosin2006mechanics} who proposed a comprehensive model. 

{First introduced by Serini and coworkers~\cite{gamba2003percolation,serini2003modeling}, PEC models consist of a conservation equation for the EC density, a momentum equation describing changes in the EC velocity and a reaction-diffusion equation describing VEGF dynamics. {In particular ECs are assumed to undergo persistent motion {\it i.e.} movement characterised by inertia in the velocity field, and this movement is modulated by friction between cells and substrate, density-dependent pressure and chemotaxis (with the VEGF being secreted by the cells themselves). }
}
Such models have so far helped to extrapolate information underlying the origin and structure of newly formed vascular networks, for instance relating the characteristic chord length $\ell$ to the VEGF diffusion coefficient $D_c$ and decay rate $\lambda_c$ with the formula $\ell= \sqrt{D_c (\lambda_c)^{-1}}$~\cite{ambrosi2004cell,ambrosi2005review,gamba2003percolation,serini2003modeling},
or investigating the minimum -- and maximum -- initial cell density required for network assembly{~\cite{ambrosi2005review,coniglio2004percolation,gamba2003percolation,kowalczyk2004stability,serini2003modeling}. 
In particular, as the initial cell density increases we observe a transition from a phase in which several disconnected structures appear to one in which a single connected structure appears. 
Detailed analyses of this phenomenon, known as percolative transition, revealed the critical EC density for transition to be around $100$ cells/mm$^2$~\cite{coniglio2004percolation,gamba2003percolation}.
Further increasing the initial cell density, instead of a well-defined network with thin chords, one may observe a cell monolayer with void areas called lacunae, a phenomenon referred to as ``Swiss-cheese'' transition~\cite{ambrosi2005review,serini2003modeling}.
Linear stability analysis revealed chemotaxis to be the key destabilising force for lacunae fomation and pressure to be the main stabilising one~\cite{kowalczyk2004stability}. Thanks to the pressure term, the Swiss-cheese transition can be predicted by PEC models about the critical initial cell density of $300$ cells/mm$^2$~\cite{ambrosi2005review}.}

On the other hand, mechanochemical models of vasculogenesis {placed special emphasis on the role played by cell traction forces, thus being particularly suited for the later stages of vasculogenesis during which the mechanical interaction between the ECs and the ECM cannot be neglected. Mechanical models or vasculogenesis,  following the work of Murray and coworkers~\cite{murray1983mechanical}, consist of a conservation equation for the EC density and one for the ECM density, a force-balance equation to describe changes in the cell-matrix displacement as cells pull on the ECM and, in the case of mechanochemical models, an additional reaction-diffusion equation describing VEGF dynamics~\cite{ambrosi2005review}. Both ECs and ECM are assumed to be advected according to changes in the cell-matrix displacement, and ECs are also assumed to undergo strain-dependent movement~\cite{manoussaki2003mechanochemical,murray2003mechanochemical,namy2004critical}, haptotaxis~\cite{namy2004critical} and chemotaxis~\cite{manoussaki2003mechanochemical}. 
The ECM is modelled as a linear elastic or viscoelastic material, the substratum is introduced via external viscous drag or elastic forces, and cell traction either saturates or goes negative for large densities, the latter understood as modelling density-dependent pressure~\cite{namy2004critical}.
Linear stability analysis predicts instability of the homogeneous steady state, and in general lacunae formation, if the cell traction coefficient is high enough or the ECM Young modulus is low enough, and if the initial cell density is below a threshold value
~\cite{ambrosi2005review}. Haptotaxis was also indicated to have a destabilising effect~\cite{namy2004critical
}, while the restoring forces and long range effects have a stabilising effect~\cite{ambrosi2005review}.
Note that while these models may predict cluster formation in the case of very high cell traction or very low ECM stiffness, such modelling frameworks are not compatible with the application here considered as they do not include any of the key mechanisms of early-stages cluster-based vasculogenesis.}

{Finally, Tosin \textit{et al.}~\cite{tosin2006mechanics} proposed a PEC model with an external drag force in the momentum equation, slowing cells down as they move on the ECM, itself modelled as a linear elastic material. 
The model indicated cell adhesiveness to the substratuum to have a key role in network formation, with too high adhesiveness resulting in no network and too low adhesiveness affecting the stability of chords so that cells would eventually clusterise due to chemotaxis.}

\paragraph{{Cellular Potts models of single-cell vasculogenesis}}{
Despite their lower analytical tractability, cellular Potts (CP) models have been widely used to study single-cell vasculogenesis over the years~\cite{boas2018cellular,scianna2013review}. These are lattice-based models, with simulations obtained via a Monte-Carlo method following energy minimisation principles, which track cell and ECM dynamics at the mesoscale. 
They are therefore particularly suited to investigate mechanisms occurring at the cell level, such as cell shape and cell-to-cell adhesion, which are difficult to describe with continuum models of macroscale dynamics and which may be crucial for single-cell vasculogenesis as the process involves a low number of cells~\cite{merks2006cell}. 
These models predicted that elongated, adhesive cells can self-organised into vascular structures~\cite{palm2013vascular} and that these structures are faster achieved and stabilised thanks to endogenous chemotaxis~\cite{merks2006cell}. Furthermore, these models confirm a correlation between the characteristic chord length and the VEGF diffusion rate, as previously indicated by PEC models.
In the absence of cell elongation, the network formation is compromised and cells form islands. This observation is independent of whether elongation is imposed as a constraint~\cite{merks2006cell} or it is a consequence of strong cell-to-cell adhesion~\cite{merks2004cell}, of preferential attraction of cells to elongated structures~\cite{szabo2010role,szabo2007network,szabo2008multicellular}, or of cell traction, ECM stiffening and durotaxis~\cite{ramos2018capillary,van2014mechanical}. In the latter case, network formation was predicted for high enough cell traction compared to the ECM Young modulus, just like in mechanochemical models. 
Alternatively to cell elongation, cell polarisation resulting from cell adhesion-mediated saturation of chemotaxis may prevent clusters from forming~\cite{merks2005dynamic,merks2008contact,singh2015role}. Overall the work of Merks and coworkers pointed towards cell elongation, or polarisation and endogenous chemotaxis being essential for network formation.}

\paragraph{{Modelling the early stages cluster-based vasculogenesis}}{
While much work has been conducted in order to better understand the mechanisms at the basis of single-cell vasculogenesis, and many of which are relevant to cluster-based vasculogenesis, recent experimental observations~\cite{blatchley2019hypoxia} seem to suggest that also other processes need to be considered when studying EPC cluster formation during the early stages of cluster-based vasculogenesis. 
While CP models may be preferred when studying a system comprising of few cells, EPCs actively proliferate and we therefore choose to study this process through the lenses of a continuum deterministic model, which is also more ameanable to analytical investigation. 
The model should include endogenous chemotaxis, MMP-mediated ECM degradation, cell-to-cell adhesion, EPC diffusion, proliferation and death, together with annexed ECM, VEGF and MMP dynamics. 
Note that while cell proliferation and ECM degradation have been included in previous models of vascular network formation~\cite{boas2018cellular,daub2013cell,holmes2000mathematical,manoussaki2003mechanochemical,tranqui2000mechanical}, these have been formulated to study angiogenesis rather than vasculogenesis, which for instance require different initial conditions and exogenous rather than endogenous chemotaxis.} 

{PEC models included chemotaxis in the momentum equation, but this implies that cells accelerate in chemical gradients, an assumption that might be unrealistic given the highly viscous, non-inertial environment of the ECM~\cite{merks2006cell,merks2008contact}. We thus choose to follow standard modelling of chemotaxis in the flux term of the EPC density mass-balance equation, as done in mechanochemical models~\cite{ambrosi2005review,manoussaki2003mechanochemical}, following the classic Patlak-Keller-Segel (PKS) model of diffusion and chemotaxis~\cite{keller1970initiation,painter2019mathematical}. As the PKS model is well-known for having solutions that blow up in finite time, a number of variations have been proposed in the literature~\cite{bubba2019chemotaxis,kowalczyk2005preventing,painter2019mathematical}, and we will hereby consider a modified version of the forms proposed by Hillen and Painter~\cite{hillen2001global,painter2002volume} whereby chemotaxis is saturated in a tightly packed environment. While this results in a local description of chemotaxis, various nonlocal ones have also been proposed over the years, reviewed for instance in~\cite{chen2020mathematical,hillen2009user,painter2015nonlocal}.}

{While CP models allowed for a mesoscopic description of cell-to-cell adhesion, during the development of PEC and mechanochemical models this biological process had not yet been explicitly modelled at the macroscopic scale. In 2006 Armstrong and coworkers published a seminal paper~\cite{armstrong2006continuum} proposing a continuum nonlocal description of intrapopulation and interpopulation cell adhesion. This was later adapted by Gerisch and Chaplain~\cite{gerisch2008mathematical} to account for cell-to-cell and cell-to-matrix adhesion. The model relies on the presence of an advective flux of the cell density at position $\bf x$, calculated nonlocally by scouting for availability of adhesion sites in a sensing region centered in $\bf x$. 
We refer the interested reader to~\cite{hillen2020nonlocal,hillen2018global,sherratt2009boundedness} for results on the existence and uniqueness of solutions to the model equation proposed by Armstrong \textit{et al.}~\cite{armstrong2006continuum}.
This has become a popular continuum description of cell adhesion, with a number of variations proposed to study tumour invasion and cell movement across the ECM~\cite{bitsouni2017mathematical,buttenschon2018space,chaplain2011mathematical,domschke2014mathematical,gerisch2010mathematical,painter2010impact,sherratt2009boundedness}, development and cell-sorting dynamics~\cite{armstrong2009adding,carrillo2019population,gerisch2010mathematical,painter2015nonlocal}. Attention has also been given over the years to variations of the nonlocal modelling of adhesion by considering different ways to enforce limits on the cell density and furthermore to include nonlinear cross diffusion, that is substituting linear diffusion modelling random movement with nonlinear diffusion modelling movement of cells down a density-dependent pressure gradient, into the overall model, see~\cite{burger2020segregation,carrillo2019population,MadzvamuseGerischBarreira2017,murakawa2015continuous}.
The principal effect of these modifications is that they lead to sharper interfaces between the densities of different cell types or the ECM or can even lead to strict segregation of them. Since such effects are not part of the primary dynamics which we will focus on in our model of the early stages of cluster-based vasculogenesis, we will not include these variations here, see however the research perspectives in Section~\ref{sec:con}.}

\subsection{Study aims and paper structure}\label{sec:aims}
{The main aim of this work is to develop a continuum deterministic model of vasculogenesis that is suitable for \textit{in silico} studies of the EPC cluster formation. 
We thus} extend mathematical models of vascular network assembly to investigate the early stages of cluster-based vasculogenesis, during which mechanical interactions between EPCs and the ECM can be neglected.  We present a continuum deterministic model of EPC cluster formation, comprising of a system of partial differential equation modelling dynamics such as endogenous chemotaxis, MMP-mediated ECM degradation, nonlocal cell-to-cell and cell-to-matrix adhesion, EPC diffusion, proliferation and death, together with annexed ECM, VEGF and MMP dynamics. Our model provides a theoretical basis for a comprehensive description of the mechanisms underlying cluster formation. 
{In such perspective, another objective of this study is to clarify the role played by different dynamics and elucidate the determinants of cluster size with the introduction of appropriate metrics. 
We investigate this \textit{in primis} by means of a linear stability analysis and numerical simulations relying on a baseline parameter set drawn from the literature. 
We then proceed to provide a first overview of potential model behaviour and an indication of the importance of various parameters or processes on the pattern formation potential. We do this by simulating the model using the baseline parameter set but with different parameters, one at a time, taking values from a suitable range of values as mostly identified from existing literature.
Finally, in view of our findings, we seek to identify crucial modelling and experimental research perspectives, together with mathematical investigations that could yield significant results and complement experimental studies of cluster-based vasculogenesis.
}

In Section~\ref{sec:met} we describe the mathematical model in detail, together with the analytical method used to investigate the system and the numerical method used to obtain solutions, and introduce the metrics defined to study cluster size. In Section~\ref{sec:res} we report the analytical results, followed by detailed numerical investigations of the cluster formation process and the size of clusters {for the one-dimensional (1D) and two-dimensional (2D) problems}, which are qualitatively compared with the experimental findings of Blatchley \textit{et al.}~\cite{blatchley2019hypoxia}. In Section~\ref{sec:con} we summarise the main findings and discuss future research perspectives.

\section{Methods}\label{sec:met}
\subsection{Mathematical model}\label{sec:met:mod}
Let $t \in \mathbb{R}_{\geq 0}$ indicate time and ${\bf x} \in \mathbb{R}^2$ position in space. 
We will consider the dynamics in a 2D spatial domain $\Omega\subset \mathbb{R}^2$, as well as the corresponding 1D problem, for which we make use of the notation $x\in \mathbb{R}$ to indicate space. Unless indicated otherwise, all definitions introduced in this section for the 2D problem hold for the corresponding 1D one. The density of EPCs at time $t$ and position ${\bf x}$ is given by $n(t,{\bf x})$, in units of cell~cm$^{-3}$, and the ECM density by $\rho(t,{\bf x})$, in units of nM. Similarly we indicate the concentration of a matrix-degrading enzyme, such as MMP-1, by $m(t,{\bf x})$, in units of $\mu$g~cm$^{-3}$, and that of a chemotactic agent, such as VEGF-A, by $c(t,{\bf x})$, in units of ng~cm$^{-3}$. We also introduce the vector of dependent variables ${\bf v}(t,{\bf x}):= \big(n(t,{\bf x}),\rho(t,{\bf x}),m(t,{\bf x}),c(t,{\bf x})\big)^\intercal$. Our model then consists of a system of mass-balance equations, one for each dependent variable introduced. 

\subsubsection{EPC density dynamics}\label{sec:sec:n}
The mass-balance equation for the density of EPCs is of the form:
\begin{equation}\label{eq:n}
\dt n +\nabla \cdot [ {\bf J}_d(n) + {\bf J}_c(n,c) + {\bf J}_a({\bf v}) ] = pn(1-\vartheta_1n-\vartheta_2\rho) \,,
\end{equation}

\noindent where ${\bf J}_d(n)$ models spatial diffusion to account for random movement of cells, ${\bf J}_c(n,c)$ indicates the chemotactic flux in response to VEGF gradients and ${\bf J}_a({\bf v})$ the advective flux due to cell-to-cell and cell-to-matrix adhesion, while the term on the right-hand side of the equation models cell proliferation and death.
In particular, we consider a modified version of the standard logistic growth, as proposed by Gerisch and Chaplain~\cite{gerisch2008mathematical}, in which EPCs proliferate at rate $p\in \mathbb{R}_{\geq 0}$ and die due to competition for space -- occupied by both cells and ECM -- and resources. Parameters $\vartheta_1\in\mathbb{R}_{>0}$ and $\vartheta_2\in \mathbb{R}_{>0}$ indicate the fraction of one unit volume of physical space occupied by EPCs at unit density and by ECM at unit density respectively, such that  $(\vartheta_1n+\vartheta_2\rho)$ indicates the total fraction of locally occupied space. 

\paragraph*{Spatial diffusion and chemotaxis}
{Following the motivations introduced in Section~\ref{sec:into:math}, we make}  use of the following definitions {for the diffusive and chemotactic flux terms}:
\begin{equation}\label{eq:flux:dc:def}
{\bf J}_d(n) = -D_n \nabla n \,, \qquad {\bf J}_c(n,c) = \chi \, f(n,\rho)\, n \nabla c \,.
\end{equation}
\noindent  In definitions~\eqref{eq:flux:dc:def} spatial diffusion follows Fick's law, with diffusivity $D_n\in \mathbb{R}_{\geq0}$, while chemotaxis is modelled as an advective flux of cells up the gradient of the chemotactic agent concentration $c$, modulated by the chemotactic sensitivity of cells. This is proportional to the chemotactic sensitivity coefficient $\chi\in \mathbb{R}_{\geq0}$ {and the function $f(n,\rho)$, which accounts for volume exclusion. This is given by}
\begin{equation}\label{def:f}
f(n,\rho)=(1-\vartheta_1n-\vartheta_2\rho)_+
\end{equation}
\noindent {where we have used $(\cdot)_+ := max(0,\cdot)$ -- see also ~\cite{gerisch2008mathematical,hillen2001global,painter2002volume} . According to definition~\eqref{def:f}, the chemotactic sensitivity of the cells is proportional to the locally available space. If the local space is overcrowded, cells will struggle to sense the chemotactic gradient,} to the extent that if the space is locally full no chemotaxis can occur.

\paragraph*{Cell-to-cell and cell-to-matrix adhesion}
{Following the motivations introduced in Section~\ref{sec:into:math}, we} model cell-to-cell and cell-to-matrix adhesion in the continuous form proposed by {Gerisch and Chaplain~\cite{gerisch2008mathematical}}. The advective flux of {the EPC density} $n$ due to cell-to-cell and cell-to-matrix adhesion in~\eqref{eq:n} is therefore given by

\begin{equation}\label{eq:flux:a}
{\bf J}_a({\bf v}) = n\, \mathcal{A}[{\bf v}(t,\cdot)] \,
\end{equation}

\noindent where $\mathcal{A}[{\bf v}(t,\cdot)] $ is the adhesion velocity at some point $\bf x$,{ that is the velocity of cells at $\bf x$ due to adhesive interactions with their environment,} which is an operator acting on ${\bf v}(t,\cdot)$ defined as a function of $\bf x$. For problems in 1D 
 this is defined as

\begin{equation}\label{eq:a:def:1d}
\mathcal{A}[{\bf v}(t,\cdot)](x) := \frac{1}{R}\int_0^R \sum_{j=0}^1  \eta(j)\, \Gamma(r)g\big({\bf v}(t,x+r\eta(j))\big)\, dr \;,
\end{equation}

\noindent while for problems in 2D 
 it is defined as

\begin{equation}\label{eq:a:def:2d}
\mathcal{A}[{\bf v}(t,\cdot)]({\bf x}) := \frac{1}{R}\int_0^R r \int_0^{2\pi}  {\bm{\eta}}(\theta) \, \Gamma(r)g\big({\bf v}(t,{\bf x}+r{\bm{\eta}}(\theta))\big)\, d\theta \,dr \;.
\end{equation}

\noindent In equation \eqref{eq:a:def:1d} $\eta(j)=(-1)^j$ with $j=0,1$ indicates the 1D right and left unit outer normal vector, while in equation \eqref{eq:a:def:2d} the 2D unit outer normal vector corresponding to angle $\theta$ is given by $\bm{\eta}(\theta) = (\cos\,\theta,\sin\,\theta)^\intercal$.
In definitions \eqref{eq:a:def:1d} and \eqref{eq:a:def:2d}, we have that the sensing region of cells at position ${\bf x}\in\mathbb{R}^d$ ($d=1,2$) is the $d$-dimensional ball centred in $\bf x$ with radius $R>0$, called the sensing radius. Then $\Gamma(r)$ is the radial dependency function, indicating how strong the adhesion velocity at a point $\bf x$ is influenced by points at a distance $r\leq R$ from the centre $\bf x$ of the sensing region. 
{Since this should not alter the magnitude itself of the adhesion velocity, $\Gamma(r)$ must satisfy}
\begin{equation}\label{eq:omega:cond}
\int_0^R 2\, \Gamma(r)\,dr = 1 \qquad \text{and} \qquad \int_0^R 2\pi r\, \Gamma(r)\,dr = 1 
\end{equation}
\noindent {in the 1D and 2D problems respectively, \textit{i.e.} in equations \eqref{eq:a:def:1d} and \eqref{eq:a:def:2d} respectively.}
Assuming $\Gamma(r)$ decays linearly with $r$ to be zero at the boundary of the sensing region, we here make use of the forms proposed by Gerisch and Chaplain~\cite{gerisch2008mathematical}

\begin{equation}\label{eq:omega:def}
\Gamma(r) = \frac{1}{R} \Big( 1-\frac{r}{R}\Big) \qquad \text{and} \qquad \Gamma(r) = \frac{3}{\pi R^2} \Big( 1-\frac{r}{R}\Big)
\end{equation}

\noindent in the 1D and 2D problems respectively{, chosen so that~\eqref{eq:omega:cond} is satisfied}. Finally, in both \eqref{eq:a:def:1d} and \eqref{eq:a:def:2d} the term $g({\bf v}(t,\cdot))$ represents the {nonlocal impact of the system's state at some point within the sensing region at ${\bf x}$ on the} velocity of the cells at $\bf x$ due to adhesions to other cells or the ECM in the sensing region {at} $\bf x$. This is given by

\begin{equation}\label{eq:g}
g({\bf v}) := g(n,\rho) = ( S_{nn}n+S_{n\rho}\rho)\,{f(n,\rho)} \;,
\end{equation}

\noindent in which $S_{nn} \in \mathbb{R}_{\geq 0}$ and $S_{n\rho} \in \mathbb{R}_{\geq 0}$ are the cell-to-cell and cell-to-matrix adhesion coefficients respectively, {while the function $f(n,\rho)$ is defined in~\eqref{def:f}.
Under definition~\eqref{eq:g}, the velocity of cells at position $\bf x$ in the direction of a point -- say -- $\bf y$ in the sensing region of $\bf x$ due to cell-to-cell adhesion is directly proportional to the cell density $n$ at $\bf y$, and that due to cell-to-matrix adhesion is directly proportional to the ECM density $\rho$ at $\bf y$. This is because a higher cell or ECM density correlates with a higher number of adhesion sites. 
Meanwhile, under definitions~\eqref{eq:g} and~\eqref{def:f}, the velocity is also proportional to the available space at position $\bf y$. 
{This accounts for volume exclusion, as cells will be unable to sense
adhesive sites at spatial locations with high densities of EPCs and/or ECM and, hence, will not migrate in those directions.}

\subsubsection{ECM density dynamics}\label{sec:sec:rho}

We let the ECM be degraded by matrix-degrading enzymes at a rate $\gamma\in\mathbb{R}_{\geq0}$ and account for ECM remodelling at a rate $\mu\in \mathbb{R}_{\geq0}$, resulting in the following mass-balance equation for the ECM density:

\begin{equation}\label{eq:rho}
\dt \rho = -\gamma\rho m +\mu (1-\vartheta_1n-\vartheta_2\rho)_+ \,,
\end{equation}

\noindent where the parameters $\vartheta_1$ and $\vartheta_2$ have already been introduced in {Section~\ref{sec:sec:n}}, indicating ECM remodelling is here understood as a restructuring phenomenon that only occurs if space is available~\cite{domschke2014mathematical}. {This is assumed to be independent of the cell density $n$ as ECM remodelling \textit{in vivo} is generally mediated by other cell types present in the tissue, such as mesenchymal stem cells, fibrocytes and fibroblast~\cite{bellini2007role,bianchetti2012extracellular,diaz2014cd34+,mcanulty2007fibroblasts,stenmark2002hypoxic}, which we do not include in our modelling framework.}

\subsubsection{MMP concentration dynamics}\label{sec:sec:m}

We let the matrix-degrading enzyme (MMP) be produced by the EPCs at a rate $\alpha_m\in \mathbb{R}_{\geq0}$, undergo Fickian diffusion with diffusivity $D_m\in \mathbb{R}_{\geq0}$, and decay at rate $\lambda_m \in \mathbb{R}_{\geq0}$. Then the MMP concentration $m(t,{\bf x})$ satisfies:

\begin{equation}\label{eq:m}
\dt m - D_m \Delta m = \alpha_m n - \lambda_m m \,.
\end{equation}


\subsubsection{VEGF concentration dynamics}\label{sec:sec:c}

Similarly to the matrix-degrading enzyme, we let the chemotactic agent (VEGF) be produced by the EPCs at a rate $\alpha_c\in \mathbb{R}_{\geq0}$, undergo Fickian diffusion with diffusivity $D_c\in \mathbb{R}_{\geq0}$, and decay at rate $\lambda_c \in \mathbb{R}_{\geq0}$. This results in the following mass-balance equation for the VEGF concentration $c(t,{\bf x})$:

\begin{equation}\label{eq:c}
\dt c - D_c \Delta c = \alpha_c n - \lambda_c c \,.
\end{equation}


\subsubsection{Boundary conditions}\label{sec:bc}
While equation \eqref{eq:rho} describes the dynamics of the ECM in the closed spatial domain $\bar{\Omega}=\Omega\cup \partial\Omega$, equations \eqref{eq:n}, \eqref{eq:m} and \eqref{eq:c} are posed on the open set $\Omega$, and are complemented with zero-flux boundary conditions. 
{These boundary conditions imply that no mass is exchanged with the outside of the spatial domain, \textit{i.e.} we have a closed system. 
For the nonlocal terms \eqref{eq:a:def:1d} and \eqref{eq:a:def:2d} this means that they cannot sense the system's state outside of the spatial domain $\Omega$.
We thus impose that in these terms the function $g({\bf v}(t,\cdot))$ equals zero if it is to be evaluated for a point outside of $\Omega$, see also \cite{domschke2014mathematical}, and thus obtain well-defined nonlocal terms \eqref{eq:a:def:1d} and \eqref{eq:a:def:2d} throughout the spatial domain.
The above definition of the nonlocal terms near the spatial domain boundary resembles what is termed repulsive (or naive) boundary condition in \cite{hillen2020nonlocal}. 
This reference also gives further possibilities for the definition of nonlocal terms on bounded domains and discusses their mathematical and biological implications.}

\subsubsection{Initial conditions}\label{sec:ic}
As proposed by Serini \textit{et al.}~\cite{serini2003modeling}, we construct the initial conditions to mimic sparsely distributed cells on the ECM. In particular, the initial cell density is given by the sum of $K\in\mathbb{N}$ randomly distributed bell-shaped bumps. In particular, we construct these bumps as Gaussian-like functions with maximum height and full width at half maximum (FWHM) both equal to the size of an average cell diameter $a\in \mathbb{R}_{>0}$. We let the initial ECM density be constant and the initial MMP and VEGF concentrations be null. Thus we have

\begin{equation}\label{def:ic}
n(t,{\bf x})= \sum_{i=1}^K G_i({\bf x};a)  \,, \qquad \rho(0,{\bf x})=\rho_0>0 \,, \qquad m(0,{\bf x})=c(0,{\bf x})=0 \,,
\end{equation}

\noindent where $G_i({\bf x};a)$ indicates the Gaussian-like function centered at the (randomly selected) ${\bf x}^{(i)} \in\Omega$ and is given by

\begin{equation}\label{def:G}
G_i({\bf x};a) := a \exp\bigg[ -\frac{4  \ln{2}}{a^2}\, \lvert \, {\bf x} - {\bf x}^{(i)} \rvert ^2 \,\bigg] \,.  
\end{equation}
 
\noindent {Initial conditions~\eqref{def:ic} and~\eqref{def:G} are defined for the 2D problem and we make use of analogous definitions for the 1D problem, obtained by simply replacing ${\bf x}\in\Omega\subset\mathbb{R}^2$ with ${x}\in\Omega\subset\mathbb{R}$.} In equation \eqref{def:G} we have used the formula FWHM$=2\sqrt{2 \ln{2}} \sigma$ where $\sigma$ is the standard deviation of the Gaussian.

\subsection{Nondimensional model and baseline parameter set}\label{sec:met:nondim}

We nondimensionalise the system of equations \eqref{eq:n} and \eqref{eq:rho}-\eqref{eq:c}, together with definitions~\eqref{eq:flux:dc:def}-\eqref{eq:g}, \eqref{def:ic} and~\eqref{def:G}, by letting

\begin{equation*}
\hat{t} = \frac{t}{\tau}\,, \quad \hat{{\bf x}} = \frac{{\bf x}}{L}\,, \quad  \hat{n} = \frac{n}{N}\,, \quad  \hat{\rho} = \frac{\rho}{P}\,, \quad  \hat{m} = \frac{m}{M}\,, \quad  \hat{c} = \frac{c}{C}\,.
\end{equation*}

\noindent We use $L=0.1$ cm as characteristic length scale, in accordance with previous vasculogenesis works~\cite{manoussaki2003mechanochemical,serini2003modeling} and for easy visual comparison with the experimental results reported by Blatchley and coworkers~\cite{blatchley2019hypoxia}. We then take reference time scale $\tau:=L^2/D$, where $D$ is a characteristic diffusion coefficient $D\sim 10^{-6}$ cm$^2$s$^{-1}$~\cite{bray2000cell}, resulting in a reference time scale $\tau=10^4$s. 
The reference cell density is chosen to be $N:=n_M=\vartheta_1^{-1}$ and we take $\vartheta_1=10^{-9}$ cm$^3$/cell, the average volume occupied by an endothelial cell~\cite{rubin1989endothelial}. We use a reference ECM density of $P=10^{-1}$ nM~\cite{anderson2005hybrid,anderson2000mathematical,terranova1985human} and define the parameter $\vartheta_2:=P^{-1}$. We take the reference VEGF density to be $C=20$ ng cm$^{-3}$, in the range of values generally considered in \textit{in vitro} set ups~\cite{hanjaya2009vascular,lee2007autocrine,serini2003modeling}.
Finally, Blatchley \textit{et al.}~\cite{blatchley2019hypoxia} reported concentrations of MMP-1 in the range $1-100\,\mu$g~ml$^{-1}$, so we take the intermediate concentration as reference MMP density, \textit{i.e.} $M=10\,\mu$g~cm$^{-3}$. 
Let us introduce the following nondimensional parameters: 

\begin{align*}
&\hat{D}_n = \frac{D_n}{D}\,, \quad \hat{\chi} = \frac{\chi C}{D}\,, \quad \hat{R} = \frac{R}{L}\,, \quad  \hat{S}_{nn} = \frac{S_{nn}}{D\vartheta_1}\,, \quad  \hat{S}_{n\rho} = \frac{S_{n\rho}}{D\vartheta_2}\,, \quad  \hat{p} = p\tau\,, \quad  \hat{\gamma} = \gamma M\tau\,, \quad  \hat{\mu} = \mu\tau\vartheta_2\,, \\ 
&\hat{D}_m = \frac{D_m}{D}\,, \quad  \hat{\alpha}_m = \frac{\alpha_m\tau}{M\vartheta_1}\,, \quad  \hat{\lambda}_m = \lambda_m\tau\,, \quad \hat{D}_c = \frac{D_c}{D}\,, \quad  \hat{\alpha}_c = \frac{\alpha_c\tau}{C\vartheta_1}\,, \quad  \hat{\lambda}_c = \lambda_c\tau \,, \quad  \hat{a} = \frac{a}{L}\,, \quad \hat{\rho_0}=\frac{\rho_0}{P} \,.
\end{align*}

\noindent Then the overall nondimensionalised system becomes, dropping hats for convenience,

\begin{equation}\label{eq:system}
\hspace{-10pt}\begin{cases}
&\dt n = D_n \Delta n - \chi \nabla \cdot \big( n \,f(n,\rho) \, \nabla c\big) - \nabla \cdot \big( n \, \mathcal{A}[ {\bf v}(t,\cdot)]\big) +pn(1-n-\rho)  \\[10pt]
&\dt \rho = -\gamma\rho m + \mu (1-n-\rho)_+  \hspace{7.95cm} \\[10pt]
& \dt m = D_m \Delta m +\alpha_m n -\lambda_m m \hspace{7.85cm}  \\[10pt]
& \dt c = D_c \Delta c +\alpha_c n -\lambda_c c \hspace{8.7cm}  
\end{cases} 
\end{equation}

\noindent where \eqref{eq:system}$_1$, \eqref{eq:system}$_3$ and \eqref{eq:system}$_4$ are posed on $(t,{\bf x})\in(0,\infty)\times \Omega$ and are complemented by zero{-flux boundary conditions} on $\partial\Omega$, while \eqref{eq:system}$_3$ is posed on $(t,{\bf x})\in(0,\infty)\times  \bar{\Omega}$.
In equation \eqref{eq:system} the operator $\mathcal{A}[ {\bf v}(t,\cdot)]$ takes the form \eqref{eq:a:def:1d} in 1D and \eqref{eq:a:def:2d} in 2D, with $\Gamma(r)$ still defined as in \eqref{eq:omega:def}, and $g({\bf v})$ {is given as in \eqref{eq:g} with $f(n,\rho)$ now given by}
\begin{equation}\label{eq:f2}
f(n,\rho) = (1-n-\rho)_+ \;.
\end{equation}
\noindent The parameters $R$, $S_{nn}$ and $S_{n\rho}$ are the nondimensional ones introduced above, and the system~\eqref{eq:system} is complemented with initial conditions \eqref{def:ic}-\eqref{def:G}, in which the parameters $a$ and $\rho_0$ now corresponds to the nondimensional ones introduced above. 
The baseline parameter set, with the corresponding nondimensional (ND) parameter values, is reported in Table~\ref{Tab:par} -- see~\ref{sec:parameters} for details.

\bigskip
\begin{table}[h!]
\centering
\caption{Baseline parameter set (ND = Nondimensional value)}\label{Tab:par}
\begin{tabular}{|c|c|c|l|c|}
\hline
\textbf{Parameter} & \textbf{Dimensional value} & \textbf{ND} & \textbf{Description} & \textbf{Ref.} \\
\hline
$\vartheta_1$ & $10^{-9}$ cm$^3$/cell & & Average cell volume &~\cite{rubin1989endothelial} \\
$\vartheta_2^{-1}$ & $10^{-1}$ nM & & Reference ECM density &~\cite{anderson2005hybrid,anderson2000mathematical,terranova1985human} \\
$D$ & $10^{-6}$ cm$^2$s$^{-1}$ & & Reference diffusion rate &~\cite{bray2000cell} \\
$D_n$ & $10^{-9}$ cm$^2$s$^{-1}$ & $10^{-3}$ & EPC diffusion coefficient &~\cite{ambrosi2005review} \\
$\chi$  & $1.4\times 10^{-7}$ cm$^5$ng$^{-1}$s$^{-1}$ & 2.8 & Chemotactic coefficient &~\cite{jain2013hybrid} \\
$R$  & $5 \times 10^{-3}$ cm & 0.05 & Cell sensing radius &~\cite{sen2009matrix} \\
$S_{nn}$ & $10^{-16}$ cm$^5$ s$^{-1}$ & 0.1 & Cell-to-cell adhesion coefficient &~\cite{gerisch2008mathematical} \\
$S_{n\rho}$ & $10^{-6}$ cm$^2$ nM$^{-1}$ s$^{-1}$ & 0.1  & Cell-to-matrix adhesion coefficient  &~\cite{gerisch2008mathematical} \\
$p$ & $10^{-5}$  s$^{-1}$ & 1 & Cell proliferation rate &~\cite{kinev2013endothelial} \\ 
$a$ & $10^{-3}$ cm & $10^{-2}$ & Average cell diameter ($\sqrt[3]{\vartheta_1}$)  &~\cite{rubin1989endothelial} \\
$\gamma$ &  $9\times 10^5$ cm$^3$g$^{-1}$s$^{-1}$ & 0.2 & ECM degradation rate &~\cite{kim2010interaction} \\
$\mu$ & $0.2 \times 10^{-5}$ nM s$^{-1}$ & 0.2 & Rate of ECM remodelling &~\cite{deakin2013mathematical,domschke2014mathematical,gerisch2008mathematical} \\
$\rho_0$ & $0.5 \times 10^{-1}$ nM & $0.5$ & Initial ECM density  & Section~\ref{sec:res:lsa} \\
$D_m$ & $8\times 10^{-9}$ cm$^2$s$^{-1}$ & $8\times10^{-3}$ &  MMP diffusion coefficient &~\cite{saffarian2004interstitial}\\
$\alpha_m$ & $0.5 \times 10^{-12} \, \mu$g s$^{-1}$ & $0.5$ & MMP production rate &~\cite{anderson2000mathematical,deakin2013mathematical,domschke2014mathematical,gerisch2008mathematical} \\
$\lambda_m$ & $5\times 10^{-5}$ s$^{-1}$ & $0.5$ & MMP decay rate &~\cite{kim2010interaction} \\
$D_c$ & $10^{-7}$ cm$^2$s$^{-1}$ & $0.1$ & VEGF diffusion coefficient &~\cite{ambrosi2005review,gamba2003percolation,miura2009vitro,serini2003modeling} \\
$\alpha_c$ & $5\times10^{-12}$ ng s$^{-1}$ & 2.5 & VEGF production rate & ~\cite{yen2011two} \\
$\lambda_c$ & $\lambda_c = 2.7 \times 10^{-4}$ s$^{-1}$  & $2.7$ & VEGF decay rate&~\cite{serini2003modeling,singh2015role} \\
\hline
\end{tabular}
\end{table}
\bigskip

\subsection{Linear stability analysis}\label{sec:met:lsa}
We perform a linear stability analysis (LSA) on the spatially homogeneous steady states, say, $\bf \bar{v}$ of the nondimensional model~\eqref{eq:system}-\eqref{eq:f2}, to gain insights into possible mechanisms responsible for aggregation dynamics. During the LSA we first introduce a small spatially homogeneous perturbation ${\bf v} = {\bf \bar{v}} + {\bf \tilde{v}}(t)$, with $|{\bf \tilde{v}}|\ll 1$, in \eqref{eq:system} and linearise. By assuming the small perturbation is proportional to $\exp{(\sigma t)}$, we derive a dispersion relation for $\sigma$ and impose Re$(\sigma)<0$, to ensure spatially homogeneous steady states are stable to spatially homogeneous perturbations. We then repeat these steps under a {small} spatially inhomogeneous perturbation ${\bf v} = {\bf \bar{v}} + {\bf \tilde{v}}(t,x)$, with $|{\bf \tilde{v}}|\ll 1$, assuming it is proportional to $\exp{(\sigma t +i k x)}$ -- or ${\bf \tilde{v}}(t,{\bf x})\propto \exp{(\sigma t +i\,{\bf k}\cdot {\bf x})}$ in 2D. Once we obtain a dispersion relation $\sigma(k^2)$ -- where $k^2 = | {\bf k} |^2 $ in 2D -- we study the conditions under which Re$(\sigma(k^2))>0$ for some $k^2$, as in such regimes we expect spatially inhomogeneous perturbations to grow in time and patterns to arise. We conduct this analysis both in 1D and 2D for the model~\eqref{eq:system}-\eqref{eq:f2}, as well as for the corresponding problem in absence of {volume exclusion}, \textit{i.e.} substituting definition{~\eqref{eq:f2} for $f(n,\rho)$ with}
\begin{equation}\label{eq:f0}
f(n,\rho) = 1 \,.
\end{equation}

\subsection{Numerical method}\label{sec:met:num}
We solve the nondimensional system~\eqref{eq:system}-\eqref{eq:f2} in $\Omega = (0,1)$ in 1D and $\Omega = (0,1)\times (0,1)$ in 2D, with zero-flux boundary conditions and initial conditions~\eqref{def:ic}-\eqref{def:G}. All numerical simulations have been performed in \Matlab{}. The numerical scheme follows the method of lines by first discretising the nonlocal model in space (with $1000$ grid cells in 1D and $100\times 100$ grid cells in 2D), yielding an initial value problem for a large system of {stiff} ordinary differential equations. This system is then solved using the {linearly-implicit} time integration scheme ROWMAP~\cite{weiner1997rowmap}, implemented in a Fortran subroutine and called from \Matlab{}. For the discretisation in space we use a second-order finite volume approach which makes use of flux-limiting {in order to ensure an accurate and at the same time non-negative approximation of the taxis and adhesion terms; note that} here in particular we employ the Koren flux limiter ($k=1/3$, $\delta=0.25$). {This finite-volume discretization in space, except for the approximation of the nonlocal term, is fully described in \cite{GerischChaplain2006} and references therein. A detailed account of the approximation of the nonlocal term in a periodic boundary condition setting is given in \cite{gerisch2010approximation}. The essential observation there is that, for the simultaneous evaluation of the nonlocal term on all grid cell interfaces, a matrix-vector product with a circulant matrix (spatially 1D case) or a block-circulant matrix (spatially 2D case) must be evaluated and that this can be done efficiently using Fast Fourier Transform (FFT) techniques. In fact, this efficient approximation of the nonlocal term, simultaneously on the full computational grid, is key to an overall efficient numerical scheme for the full PDE system. 
In our case here, however, we have that the nonlocal term comes with zero-flux boundary conditions and the matrix-vector product in this case involves matrices with Toeplitz structure -- see also \cite{domschke2014mathematical}. 
In this case, FFT techniques cannot be applied directly. However, these (block) Toeplitz matrices have a banded structure with a band width which is small thanks to the fact that the size of the sensing region of the nonlocal term is small compared to the size of the domain $\Omega$. Thus these (block) Toeplitz matrices can be embedded in slightly larger (block) circulant matrices and the desired result of the matrix-vector product can again be obtained efficiently using FFT techniques. We elaborate on this embedding in the Supplementary Material. Recent work by Colombi et al. \cite{ColombiFallettaSciannaEtAl2021} considers a nonlocal term similar to ours and investigates numerical approaches to its approximation. Note, however, that in their case the nonlocal term is evaluated in a single spatial location (changing with time) as opposed to the computationally much more involved task of approximating the nonlocal term on all grid cell interfaces simultaneously as in our case.}

\subsection{Cluster size metrics}\label{sec:met:cw}
In order to gain insight into the role played by different biological, chemical and mechanical factors in dictating the cluster size, we define two different measures of cluster size with complementary information, as similarly done in~\cite{palmer2003nanostructured}. We define these for the nondimensional 1D problem, but analogous definitions can be considered for the 2D problem.
Assume that $Q\in\mathbb{N}$ clusters have formed at time $t=T$ and let $\omega\subset\Omega$ be the subdomain supporting these clusters, \textit{i.e.}

\begin{equation}\label{def:omega}
\omega :=\text{supp}\, n(T,{x}) \,.
\end{equation}

\noindent Then $\omega$ can be partitioned into $Q$ subdomains $\omega_1$, ..., $\omega_Q$, \textit{i.e.} we have

\begin{equation}\label{def:omegai}
\bigcup\limits_{i=1}^{Q} \omega_i = \omega \qquad \text{and} \qquad \omega_i \bigcap \omega_j = \emptyset  \;\;\; \text{for} \;\;\; i,j=1,...,Q\,, \;\; i\neq j \,,
\end{equation}

\noindent where each $\omega_i$ ($i=1,...,Q$) corresponds to the support of a cluster. We let the average cluster \textit{width} $W$ and average cluster \textit{compactness} $C$ be defined by

\begin{align}
&W := \frac{1}{Q}\sum_{i=1}^{Q} W_i \,, \quad \text{where} \quad W_i := |\omega_i| \quad i=1,...,Q \,,\label{def:W} \\
&C := \frac{1}{Q}\sum_{i=1}^{Q} C_i \,, \quad \text{where} \quad C_i =  \frac{\int_{\omega_i}n(T,{x})\,  \text{d}x}{W_i} \quad i=1,...,Q \,. \label{def:C}
\end{align}

\noindent Under definitions~\eqref{def:omega}-\eqref{def:W}, the width $W_i$ ($i=1,...,Q$) of each cluster is a measure of the length of its support, which may be understood as an indicator of the diameter of the cluster assuming the 1D case is reflective of the corresponding 2D problem. 
Note that the analogous 2D definition to \eqref{def:W} would inform us on the area covered by each cluster, from which the average cluster diameter could be calculated. However, this would need to be complemented with an additional metric for cluster elongation (\textit{e.g.} the ratio between the diameter of the circle circumscribing $\omega_i$ and that of the one inscribed in $\omega_i$), in order to obtain an exhaustive description of the cluster's topology. 
Under definition~\eqref{def:C}, in which $n(T,x)$ is the nondimensional cell density, the compactness $C_i$ ($i=1,...,Q$) of each cluster is a measure of the {average cell density within cluster $i$.} 
Cluster compactness allows us to distinguish between simple cell aggregates and well-defined clusters, identified as such only if $C$ is higher than $0.5$, corresponding to at least half the local volume being occupied by cells.
Under the choices of nondimensionalisation and of spatial domain for the numerical simulations, we expect $0\leq W_i \leq 1$ and -- under cell incompressibility assumptions -- $0\leq C_i \leq 1$ for all $i=1,..,Q$. Figure~\ref{fig:CWbio} summarises the biological interpretation of possible combinations of $W$ and $C$. {In practice, the cluster domains $\omega_i$ used to calculate $W$ and $C$ are identified via image segmentation by thresholding, with a threshold value for the cell density set to be $10^{-4}$ (two orders of magnitude smaller than the nondimensional average cell diameter $a$), below which $n(t,x)$ is approximated to zero. }

\begin{figure}[htb!]
\begin{minipage}[c][7cm][t]{.5\textwidth}
\centering
\includegraphics[width=1\linewidth]{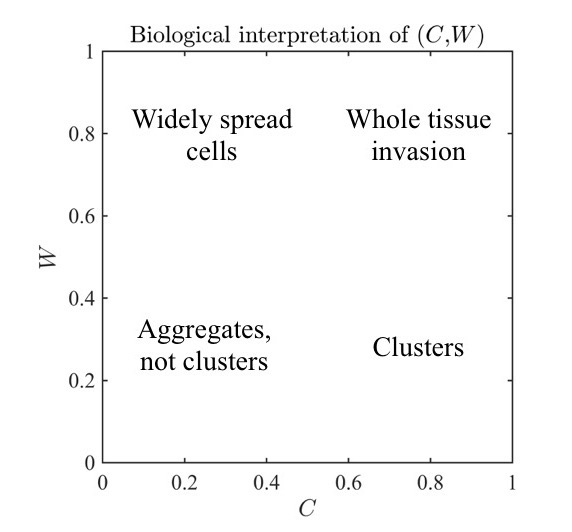}
\end{minipage}
\begin{minipage}[c][6.5cm][t]{.5\textwidth}
\centering
\includegraphics[width=0.8\linewidth]{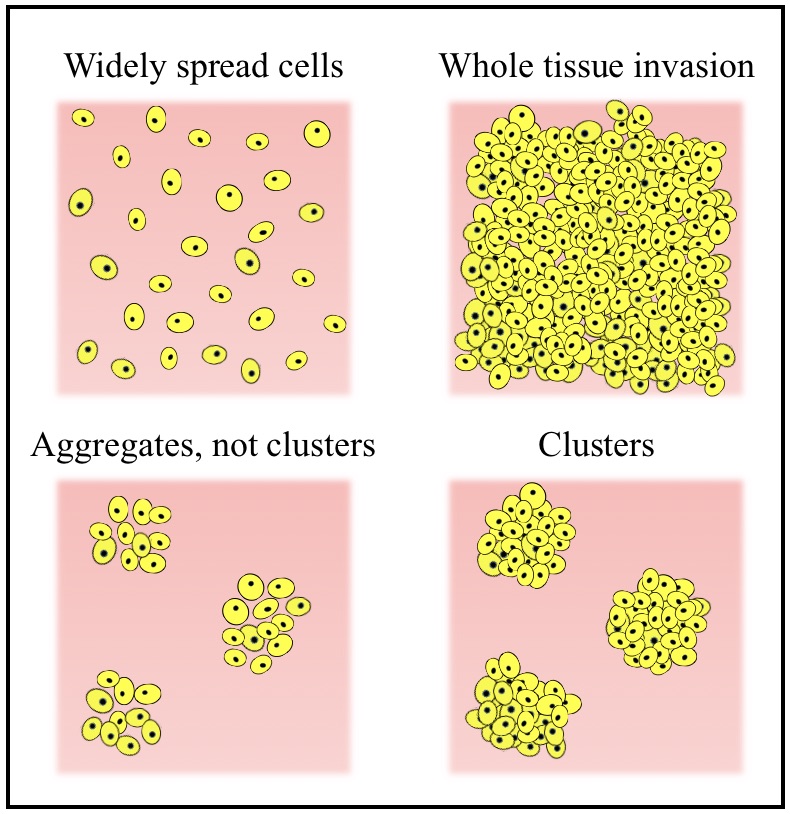}
\end{minipage}
\vspace{10pt}
\caption{\label{fig:CWbio} Biological interpretation of possible combinations of cluster width $W$~\eqref{def:W} and cluster compactness $C$~\eqref{def:C}: high $W$ and low $C$ capture the scenario in which few cells are widely 
spread across the domain (top left region of the $C$-$W$ plane); low $W$ and low $C$ capture the presence of loose cell aggregates, (bottom left region); low $W$ and high $C$ capture  the presence of clusters (bottom right region); high $W$ and high $C$ capture the scenario in which many cells are widely spread across the domain, corresponding to the case of tissue invasion in which no clusters can be identified (top right region).}
\end{figure}

\section{Results}\label{sec:res}
{As outlined in Section~\ref{sec:aims}, we here investigate the determinants of cluster formation and cluster size, by means of a linear stability analysis and numerical simulations, thank to which we provide an overview potential model behaviours under one-at-a-time perturbations from the baseline parameter set.} 
We report in Section~\ref{sec:res:lsa} the key results of the LSA, as described in Section~\ref{sec:met:lsa}, and in Sections~\ref{sec:res:bps}-\ref{sec:res:2d} the results of the numerical simulations obtained using the method described in Section~\ref{sec:met:num}. In Section~\ref{sec:res:bps} we report the results of the 1D model under the baseline parameter set in Table~\ref{Tab:par} and qualitatively investigate the determinants of cluster formation in Section~\ref{sec:res:main}. In Section~\ref{sec:res:size} we report the results of the {parametric} analysis conducted to elucidate the relation between different parameters and the average cluster width and compactness, as defined in Section~\ref{sec:met:cw}. Finally, we report the results of the 2D model in Section~\ref{sec:res:2d}.

\subsection{Linear stability analysis results}\label{sec:res:lsa}

{We first conducted a LSA on the biologically significant spatially homogeneous steady states of the system~\eqref{eq:system} under definitions  \eqref{eq:a:def:1d}-\eqref{eq:g} and~\eqref{eq:f2}, both in 1D and 2D. Then, we investigated any changes in the LSA results in the absence of volume exclusion in the chemotactic and cell adhesion terms of equation~\eqref{eq:system}$_1$, that is under definition~\eqref{eq:f0} for $f(n,\rho)$ instead of~\eqref{eq:f2}.} The full LSA can be found in {Section~S1 of} the supplementary document `SuppInfo' with the key results being as follows. {We find that} the two spatially homogeneous steady states are such that either the whole domain is solely occupied by cells ($\bar{n}=1$ and $\bar{\rho}=0$) or solely occupied by ECM ($\bar{n}=0$ and $\bar{\rho}=1$){, the latter of which we do not consider further as it is not biologically relevant for our current study}. In absence of ECM degradation ($\gamma=0$), in addition to the two above, there are infinitely many ``intermediate" spatially homogeneous steady states satisfying $0<\bar{n},\bar{\rho}<1$, with $\bar{n}+\bar{\rho}=1$. These ``intermediate'' steady states are {linearly} stable to {small} spatially inhomogeneous perturbations. When we neglect the influence of volume exclusion in the chemotactic and cell adhesion terms of equation~\eqref{eq:system}$_1$, we find that chemotaxis and cell-to-cell adhesion can combine to make the biologically significant steady states ($n > 0$), which are the same as in the presence of volume exclusion, unstable to {small} spatially inhomogeneous perturbations if the cell density is sufficiently large (see the `SuppInfo' document for more details). Thus we expect patterns to arise when cell-to-cell adhesions and chemotactic mechanisms dominate the dynamics, which may occur, for instance, when the domain is not densely packed with cells and ECM. We therefore take the initial ECM density in~\eqref{def:ic} to be $\rho_0 = 0.5$ in our nondimensional baseline parameter set, in order to increase the chances that clusters form in the model. 

\subsection{Cluster formation under the baseline parameter set}\label{sec:res:bps}
We report in Figure~\ref{P1:fig0} the cell density $n(t,x)$ obtained from numerical simulations of the 1D model under the baseline parameter set and in Figure~\ref{fig:CW} the corresponding cluster width $W$ and compactness $C$. The plots displayed in these figures indicate that our model predicts three stages of cluster formation in 1D. First EPCs form aggregates which reach minimum cluster width of about $W=0.2$ around $t=17$ (\textit{cf.} left panel in Figure~\ref{P1:fig0} and Figure~\ref{fig:CW}). Then the cells in these aggregates continue to proliferate increasing their compactness while keeping the cluster width unchanged up to about $t=50$ (\textit{cf.} central panel in Figure~\ref{P1:fig0} and Figure~\ref{fig:CW}). Finally, the cells continue to proliferate until the whole domain is occupied by cells (\textit{cf.} right panel in Figure~\ref{P1:fig0} and Figure~\ref{fig:CW}). 

\begin{figure}[h!]
\includegraphics[width=1\linewidth]{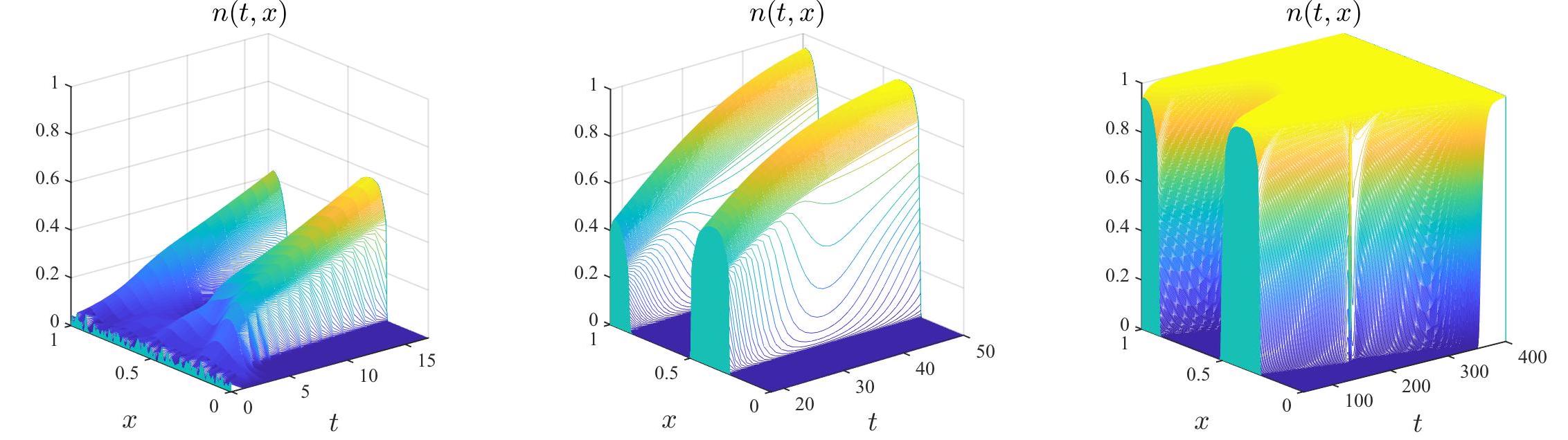}
\caption{\label{P1:fig0} Plots of the cell density $n(t,x)$ obtained solving the system~\eqref{eq:system}, together with definitions~\eqref{eq:a:def:1d}, \eqref{eq:omega:def} and~\eqref{eq:f2}, initial conditions~\eqref{def:ic} and~\eqref{def:G}$_1$, complemented with zero{-flux} boundary conditions, under the baseline parameter set in Table~\ref{Tab:par}. The solution is plotted in the time intervals $t=[0,17]$ (left panel), $t=[17,50]$ (central panel) and $t=[50,400]$ (right panel).}
\end{figure}

The simulated dynamics of cluster formation nicely match the experimentally observed ones by Blatchley \textit{et al.}~\cite{blatchley2019hypoxia}, {see Figure~\ref{P1:figbio}}, although in our simulations they occur on a slightly slower timescale. In fact in~\cite{blatchley2019hypoxia} the minimum cluster size was reached around 24 hours (about nondimensional time $t=8.7$) and kept unchanged while clusters increased compactness up to 48 hours ($t=17.3$) before late stage dynamics kick in.
The numerically obtained cluster width corresponds to about $200\,\mu$m, which {is within the range of cluster diameters} observed in~\cite{blatchley2019hypoxia}. 
Finally, after $t=50$ we observe the equivalent of whole tissue invasion, suggesting our modelling framework cannot properly capture cluster stabilisation in the long run, at least in the 1D case.

\begin{figure}[h!]
\centering
\includegraphics[width=0.45\linewidth]{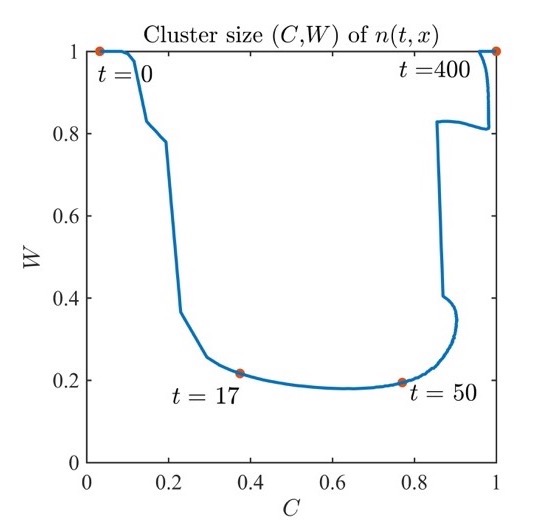}
\caption{\label{fig:CW} Average cluster width $W$~\eqref{def:W} and cluster compactness $C$~\eqref{def:C} (left panel) measured over time on the cell density $n(t,x)$ (right panel) obtained solving the system~\eqref{eq:system}, together with definitions~\eqref{eq:a:def:1d}, \eqref{eq:omega:def} and~\eqref{eq:f2}, initial conditions~\eqref{def:ic} and~\eqref{def:G}$_1$, complemented with zero{-flux} boundary conditions, under the baseline parameter set in Table~\ref{Tab:par}, reported in Figure~\ref{P1:fig0}.}
\end{figure}

\subsection{Qualitative investigation of the determinants of cluster formation}\label{sec:res:main}
In view of the results reported in Section~\ref{sec:res:bps}, we now investigate the role played by chemotaxis, ECM degradation and cell-to-cell adhesion in cluster formation by varying the relevant parameter values and observing changes in the solution up until $t=50$, starting from the same initial conditions considered in the previous section. {The long-time dynamics (up to $t=400$) of the solutions plot in Figure~\ref{P1:fig1} are reported in the supplementary Figure~S2.}

\paragraph*{The {primary} role of chemotaxis and ECM degradation} The plots reported in the second row of Figure~\ref{P1:fig1} reveal the role of chemotaxis and ECM degradation in cluster formation, according to the dynamics described in Section~\ref{sec:met:mod}. 
Under the baseline parameter set we observe cluster formation (\textit{cf.} first plot in second row of Figure~\ref{P1:fig1}), as discussed in the previous paragraph and summarised in Figure~\ref{P1:fig0}.
{In absence of ECM degradation, even though cell aggregates of the same width form, the maximum cell density remains below $0.3$ (\textit{cf.} second plot in second row of Figure~\ref{P1:fig1}) and actually decays over longer periods of time (see supplementary Figure~S2).} 
 On the other hand, in absence of chemotaxis no cell aggregation occurs and we either observe total invasion of the domain by the cells (\textit{cf.} third plot  in second row of Figure~\ref{P1:fig1}) or, in absence of ECM degradation, a simple spatial redistribution of the cells (\textit{cf.} fourth plot  in second row of Figure~\ref{P1:fig1}). These results indicate that chemotaxis and ECM degradation are both crucial to cluster formation, with chemotaxis playing a key role in cell aggregation and ECM degradation being responsible for these aggregates to grow into well-defined and compact clusters. 

\paragraph*{The {secondary} role of cell-to-cell adhesion} 
{Let us now compare the plots in the second row of Figure~\ref{P1:fig1} with those in the rest of the figure, which have been obtained by varying the cell-to-cell adhesion coefficient $S_{nn}$, the value of which was chosen \textit{a priori} due to lack of proper estimates in the current literature.}
We immediately observe that{ for small values $S_{nn}\le 0.1$}, under the baseline parameter set, cell-to-cell adhesion does not play any particular role in cluster formation, as demonstrated by the fact that the simulations in absence of cell-to-cell adhesion {closely resemble} those with $S_{nn}=0.1$ (\textit{cf.} first and second row of Figure~\ref{P1:fig1} {and Figure~S3 in `SuppInfo' document}). 
On the other hand, increasing the order of magnitude of the cell-to-cell adhesion coefficient results in the {initial} formation of small-scale aggregates. {Moreover, the maximum density reached by these aggregates increases as $S_{nn}$ increases (\textit{cf.} third and fourth row of Figure~\ref{P1:fig1}). This however does not seem to affect the long-time dynamics of the solution, which remain analogous to those described in the previous paragraph. For instance, under the baseline parameter set except for $S_{nn}=1$, the initial aggregates merge -- likely due to chemotaxis -- into wider clusters and then proceed to invade the whole tissue (\textit{cf.} first plot in third row of Figure~\ref{P1:fig1} and of Figure~S2). For $S_{nn}=10$ we do not yet observe small aggregates merging into larger clusters at $t=50$ -- likely due to cell-to-cell adhesion overpowering chemotaxis -- and tissue invasion is simply delayed (\textit{cf.} first plot in fourth row of Figure~\ref{P1:fig1} and of Figure~S2).
Overall these results suggest that the continuum nonlocal description of cell-to-cell adhesion considered in our model, while it may capture the aggregating effect of cell-to-cell adhesion for $S_{nn}$ high enough, does not capture the stabilising effect that we are seeking for in this modelling framework.}

\begin{figure}[h!]
\includegraphics[width=1\linewidth]{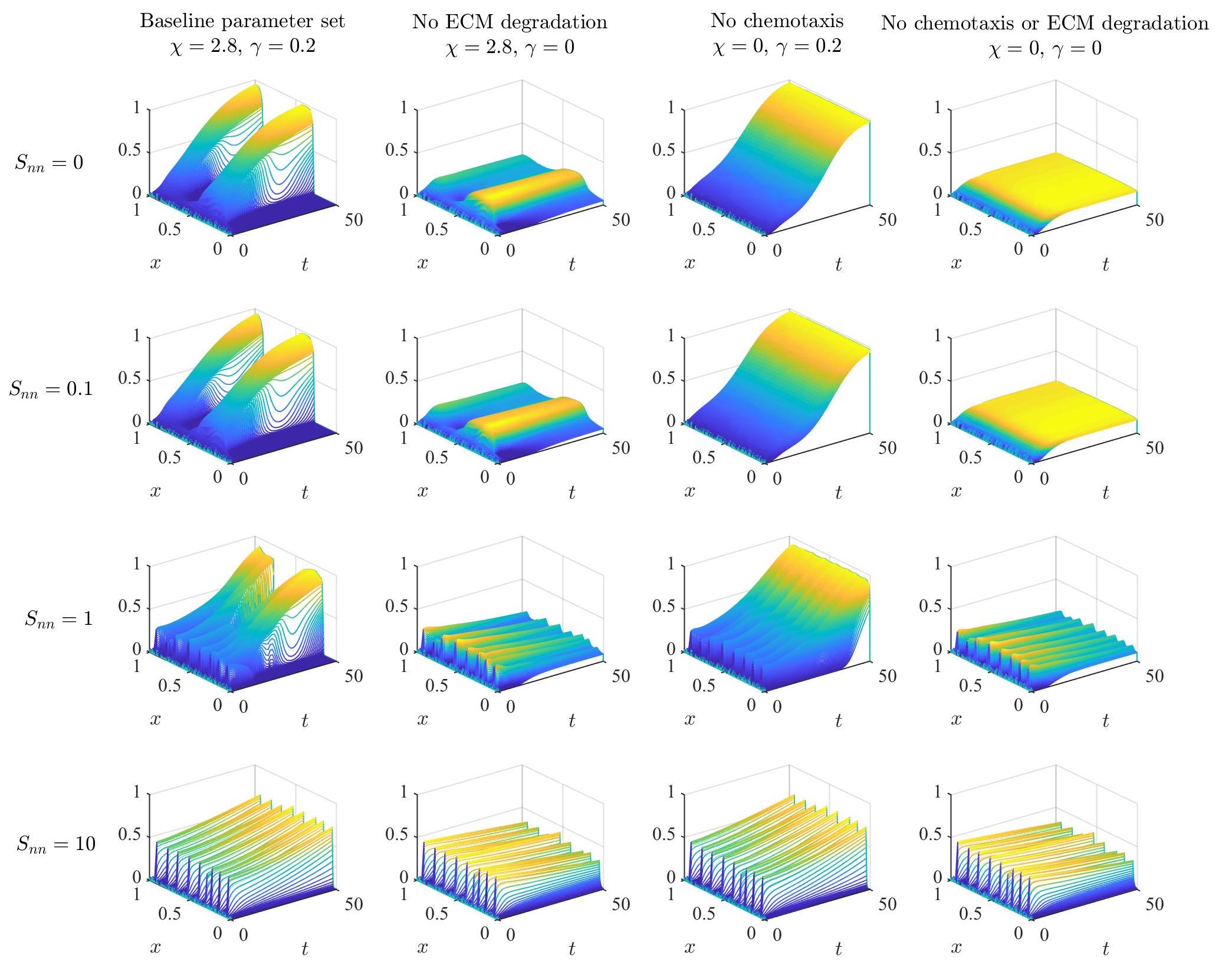}
\caption{\label{P1:fig1}\textbf{First row:} Plots of the cell density $n(t,x)$ up to $t=50$ obtained solving the system~\eqref{eq:system}, together with definitions~\eqref{eq:a:def:1d}, \eqref{eq:omega:def} and~\eqref{eq:f2}, initial conditions~\eqref{def:ic} and~\eqref{def:G}$_1$, complemented with zero{-flux} boundary conditions, in absence of cell-to-cell adhesion, \textit{i.e} for $S_{nn}=0$: under the baseline parameter set (first column), in absence of ECM degradation, \textit{i.e.} for $\gamma=0$ (second column), in absence of chemotaxis, \textit{i.e.} $\chi=0$ (third column), and in absence of both chemotaxis and ECM degradation, \textit{i.e.} $\chi=\gamma=0$ (fourth column). \textbf{Second, third and {fourth} rows:} Same as first row but in the presence of cell-to-cell adhesion, with $S_{nn}=0.1$ (second row), $S_{nn}=1$ (third row) and $S_{nn}=10$ ({fourth} row) respectively. }
\end{figure}

\subsection{Quantitative investigation of the determinants of cluster size}\label{sec:res:size}
We study changes in the measures $W$ and $C$, defined according to \eqref{def:W} and \eqref{def:C}, at $t=50$ under variations of each parameter in the baseline parameter set. In particular, we consider the effect of halving and doubling the magnitude of each parameter in equations \eqref{eq:system}$_1$ and \eqref{eq:system}$_2$ in Figure~\ref{Fig:Size:nrho}, and those in equations \eqref{eq:system}$_3$ and \eqref{eq:system}$_4$ in Figure~\ref{Fig:Size:mc}. Each boxplot had been obtained using data from 100 simulations under the same parameter set, staring from randomised initial conditions as in \eqref{def:ic} and \eqref{def:G}$_1$. 

The cluster width $W$ measured over 100 simulations under the baseline parameter width ranges between $0.15$ and $0.41$ with median (and mean) around $0.23$, as portrayed in Figures~\ref{Fig:Size:nrho} and \ref{Fig:Size:mc} (central boxplot in each $W$ plot). {The nondimensional width corresponds to a diameter in the range $150-410 \mu$m, which is consistent with the experimentally observed cluster sizes} in~\cite{blatchley2019hypoxia} ({\textit{cf.} Figure~\ref{P1:figbio}}).

\paragraph*{The role of chemotaxis}
Boxplots of the width $W$ and compactness $C$ of clusters for different values of the chemotactic sensitivity $\chi$ are displayed in Figure~\ref{Fig:Size:nrho}b. While higher values of $\chi$ correlate with slightly smaller clusters
, yet maintaining a mean width around 0.2, lowering the magnitude of $\chi$ seems to result in a wider range of values of $W$ with much higher median, as well as higher compactness $C$. This supports the notion that lowering the chemotactic sensitivity of the cells hinders cluster formation and -- assuming all other dynamics are present -- fosters tissue invasion, which is in line with the results presented in Section~\ref{sec:res:main}.

\paragraph*{The role of ECM degradation}
Boxplots of the width $W$ and compactness $C$ of clusters for different values of the ECM degradation rate $\gamma$ are displayed in Figure~\ref{Fig:Size:nrho}f. Notice that, while the median width $W$ is maintained around 0.2, increasing the magnitude of $\gamma$ results in a smaller range of values measured for both $W$ and $C$, as well as higher values of cluster compactness. In addition, this trend suggests that further decreasing $\gamma$ will lead to higher values of $W$ and lower values of $C$ measured (see supplementary Figure S4c in `SuppInfo' document), which is in line with the observed dynamics in absence of ECM degradation in Section~\ref{sec:res:main} (\textit{cf.} Figure~\ref{P1:fig1}, second and fourth columns). Overall, this highlights the key role ECM degradation has in  cluster formation, establishing a relation between the rate $\gamma$ and the cluster compactness $C$.

\paragraph*{The role of matrix remodelling} 
We see in Figure~\ref{Fig:Size:nrho}g that increasing the magnitude of the matrix remodelling rate $\mu$ {results in a larger range of values measured for both $W$ and $C$, as well as lower values of cluster compactness, while the median width $W$ is maintained around 0.2. These are }opposite effects to those obtained increasing its degradation rate $\gamma$ (\textit{vid.} Figure~\ref{Fig:Size:nrho}f), which is coherent with the opposite nature of these dynamics.

\paragraph*{The role of cell-to-cell adhesion}
The plots in Figure~\ref{Fig:Size:nrho}c confirm that small variations in the cell-to-cell adhesion coefficient $S_{nn}$ do not influence the size of clusters, as expected from the results in Section~\ref{sec:res:main}. Additional numerical tests considering different orders of magnitude of $S_{nn}$ (see supplementary Figure~S1b in `SuppInfo' document) revealed a slight increase in the median $W$, probably due to the initial presence of smaller clusters (\textit{cf.} Figure~\ref{P1:fig1}, third row) which at $t=50$ may still be in the process of merging depending on their spatial distribution. {These results do not suggest that cell-to-cell adhesion is a key mechanism in determining cluster -- and therefore network -- topology.}

\paragraph*{The role of cell diffusion and cell-matrix interactions}
The boxplots of $W$ and $C$ of clusters for different values of the cell diffusion coefficient $D_n$, displayed in Figure~\ref{Fig:Size:nrho}a, show that changing the magnitude of $D_n$ has very little effect on the cluster compactness $C$ and median width $W$. There is however an increasing range of values of $W$ measured as $D_n$ increases, indicating that low diffusivity correlates with more precisely defined clusters, while high diffusivity results in more variability in cluster size. Moreover, this variability allows for larger values of $W$ to be measured, suggesting that much higher diffusivity may result in tissue invasion (see supplementary Figure S4a in `SuppInfo' document).
The same observations can be conducted on the boxplots in Figure~\ref{Fig:Size:nrho}d, obtained by varying the magnitude of the cell-to-matrix adhesion coefficient. This is in line with the notion that lower cell-matrix interactions facilitate cluster formation, while much higher cell-matrix interactions promote tissue invasion.

\paragraph*{The role of cell proliferation} 
In Figure~\ref{Fig:Size:nrho}e we see that slower proliferation -- \textit{i.e.} lower $p$ -- correlates with a wide range of lower values of cluster compactness $C$ measured, while the median width is maintained around 0.2. On the other hand, faster proliferation -- \textit{i.e.} higher $p$ -- results in a wide range of higher values of cluster width $W$ and a small range of high values of compactness $C$. Note that these data portray different stages of the cluster formation process, as demonstrated in Figure~\ref{fig:CW} (central panel): initially aggregates form without being very condensed (low-to-medium $W$ and $C$), then they increase their compactness while keeping steady width (low $W$ and high $C$), and eventually grow further invading the surrounding space (medium-to-high $W$ and high $C$). This suggests that the rate of proliferation of EPCs might play a key role in determining the speed of the cluster formation process.

\paragraph*{The role of initial ECM density} 
The plots in Figure~\ref{Fig:Size:nrho}h indicate that changes in the initial ECM density $\rho_0$ does not affect the {long-time} spatio-temporal dynamics. In fact, while different values of $\rho_0$ results in slightly different values of the median width and compactness of clusters observed at $t=50$, these values still capture the same biological scenario (\textit{i.e.} cluster are recognisable at $t=50$).

\begin{figure}
\begin{minipage}[c][18cm][t]{.5\textwidth}
  \vspace{-20pt}
  \centering
  \includegraphics[width=\linewidth]{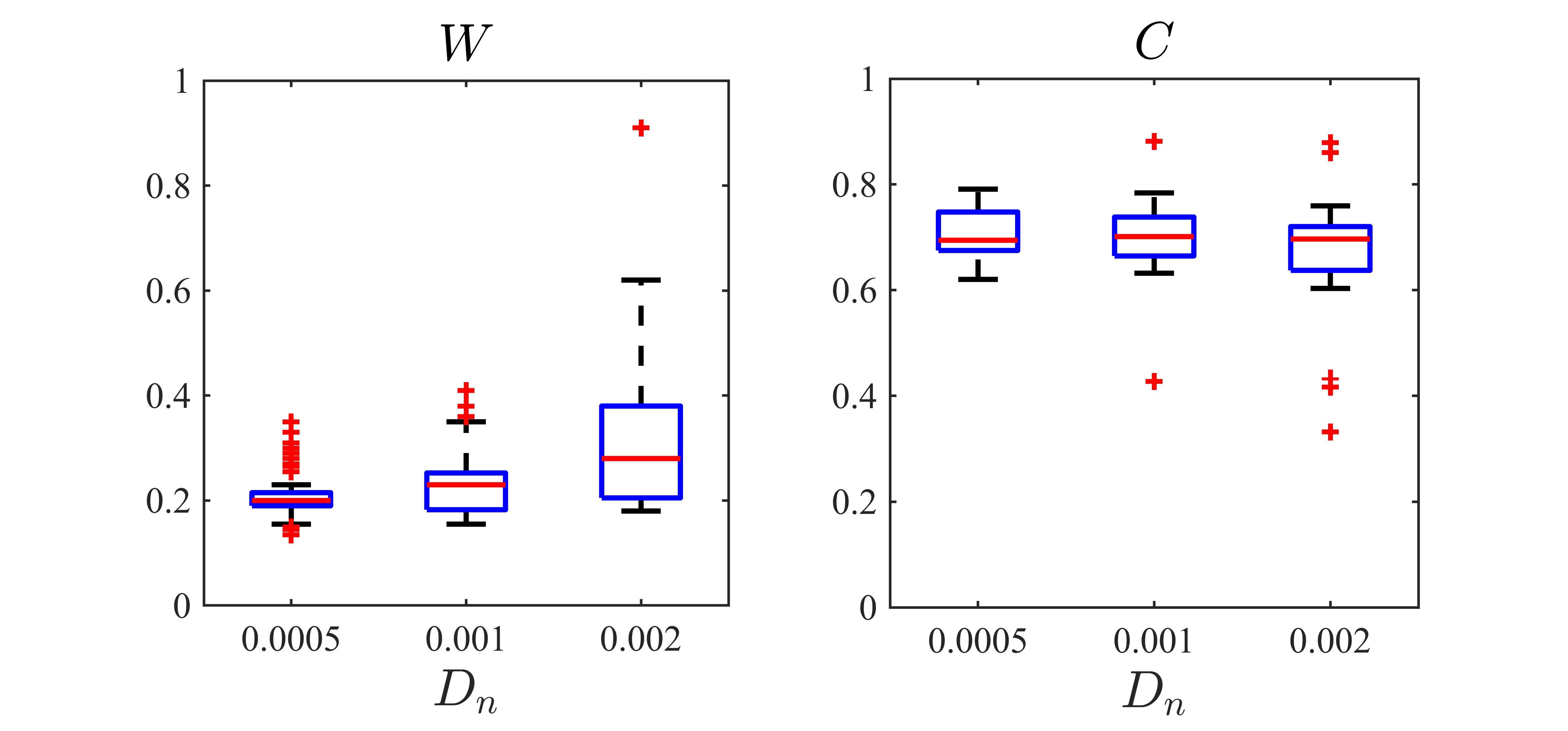}
  \subcaption{Varying cell diffusion $D_n$}
  \label{fig:size:Dn}\par\vfill
    \vspace{1\baselineskip}
  \includegraphics[width=\linewidth]{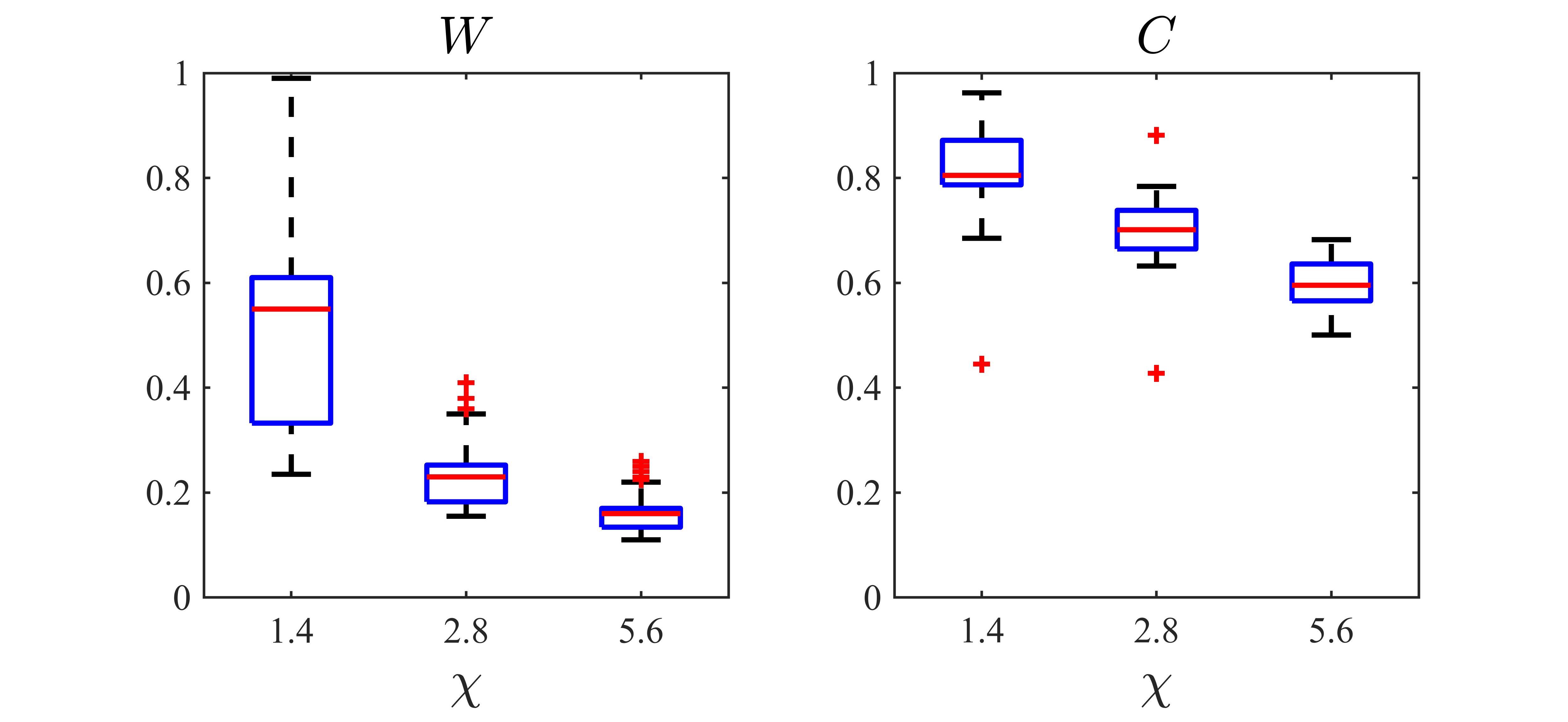}
  \subcaption{Varying chemotactic sensitivity $\chi$}
  \label{fig:size:chi}\par\vfill
    \vspace{1\baselineskip}
  \includegraphics[width=\linewidth]{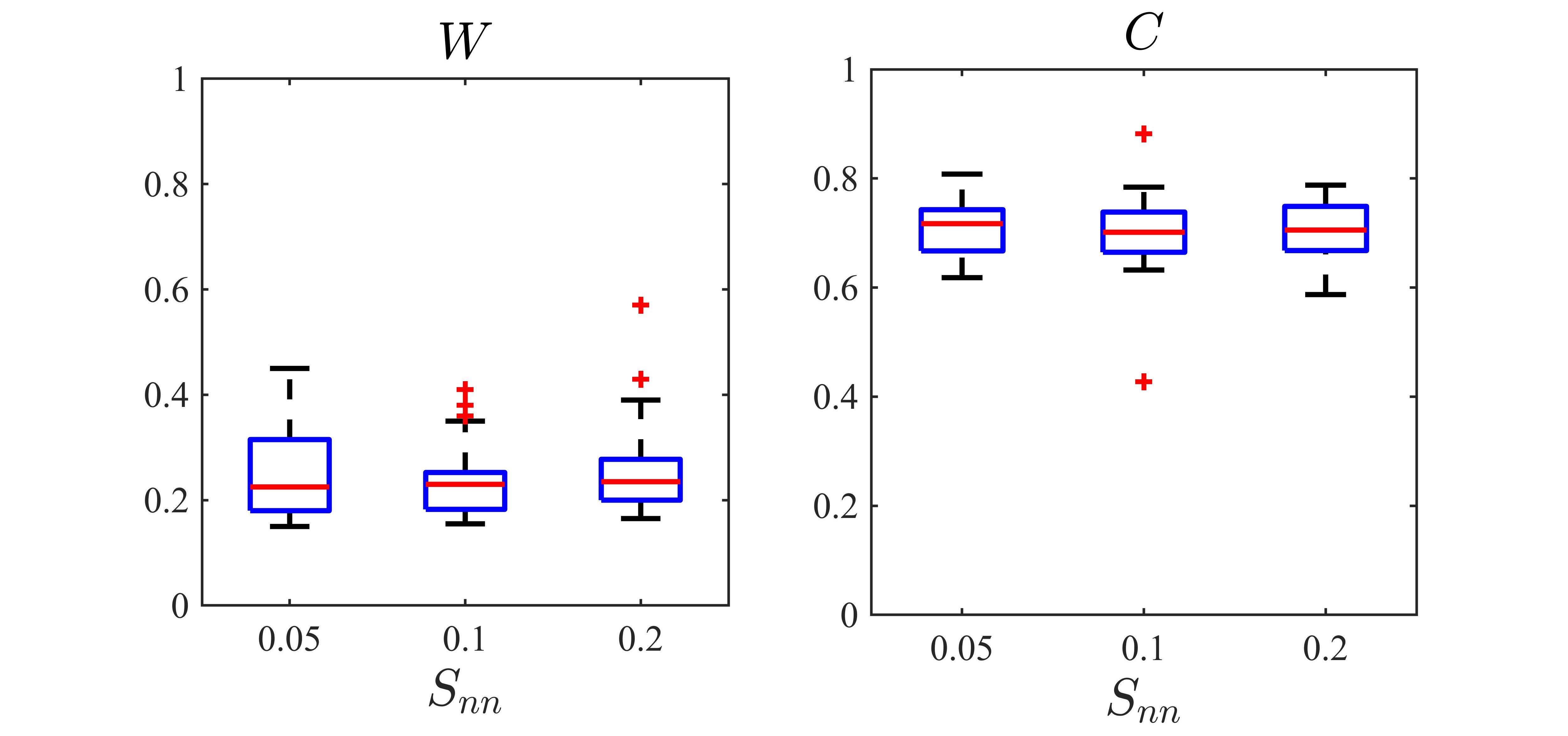}
  \subcaption{Varying cell-to-cell adhesion coefficient $S_{nn}$}
  \label{fig:size:Snn}\par\vfill
    \vspace{1\baselineskip}
  \includegraphics[width=\linewidth]{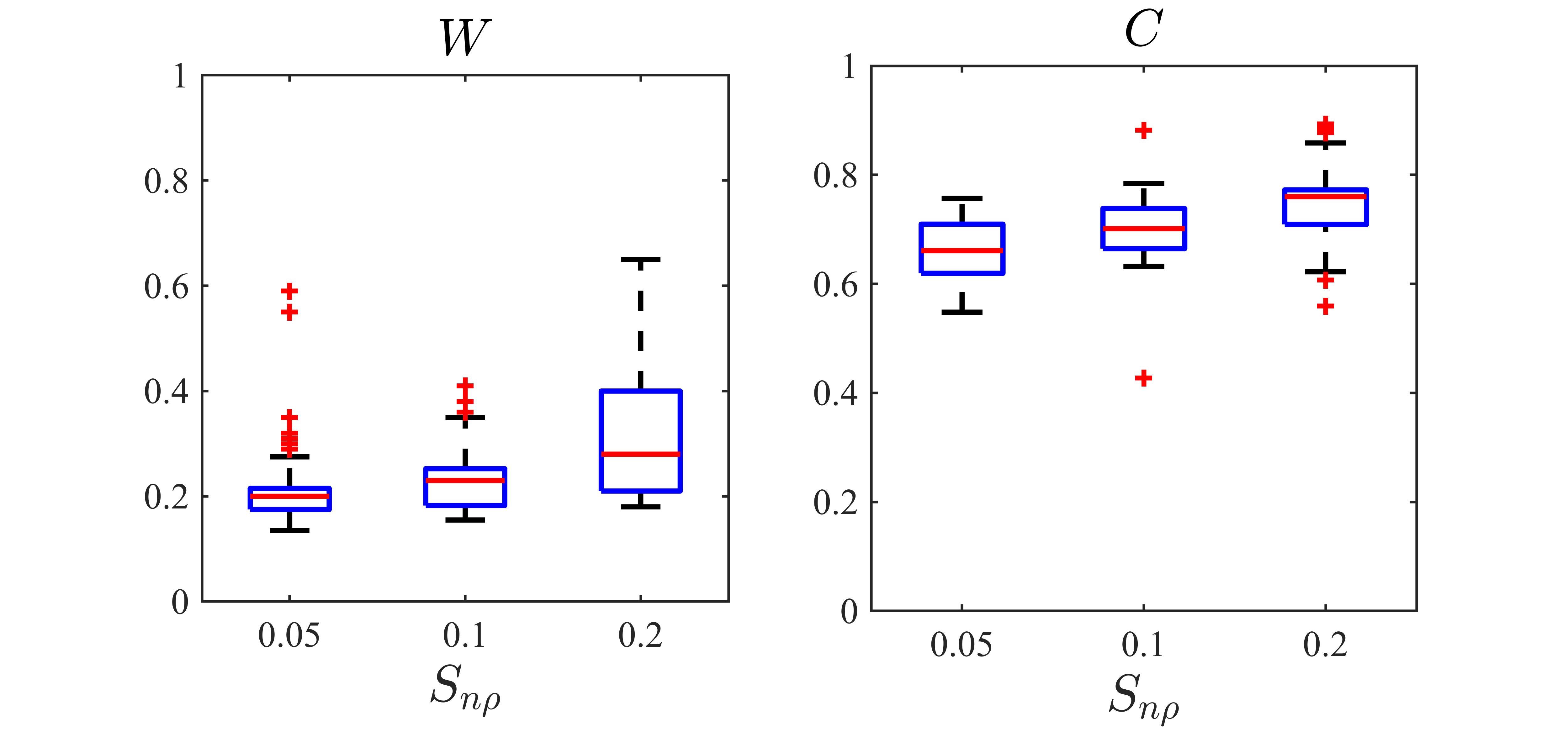}
  \subcaption{Varying cell-to-matrix adhesion coefficient $S_{n\rho}$}
  \label{fig:size:Snp}
\end{minipage}%
\begin{minipage}[c][18cm][t]{.5\textwidth}
  \vspace{-20pt}
  \centering
  \includegraphics[width=\linewidth]{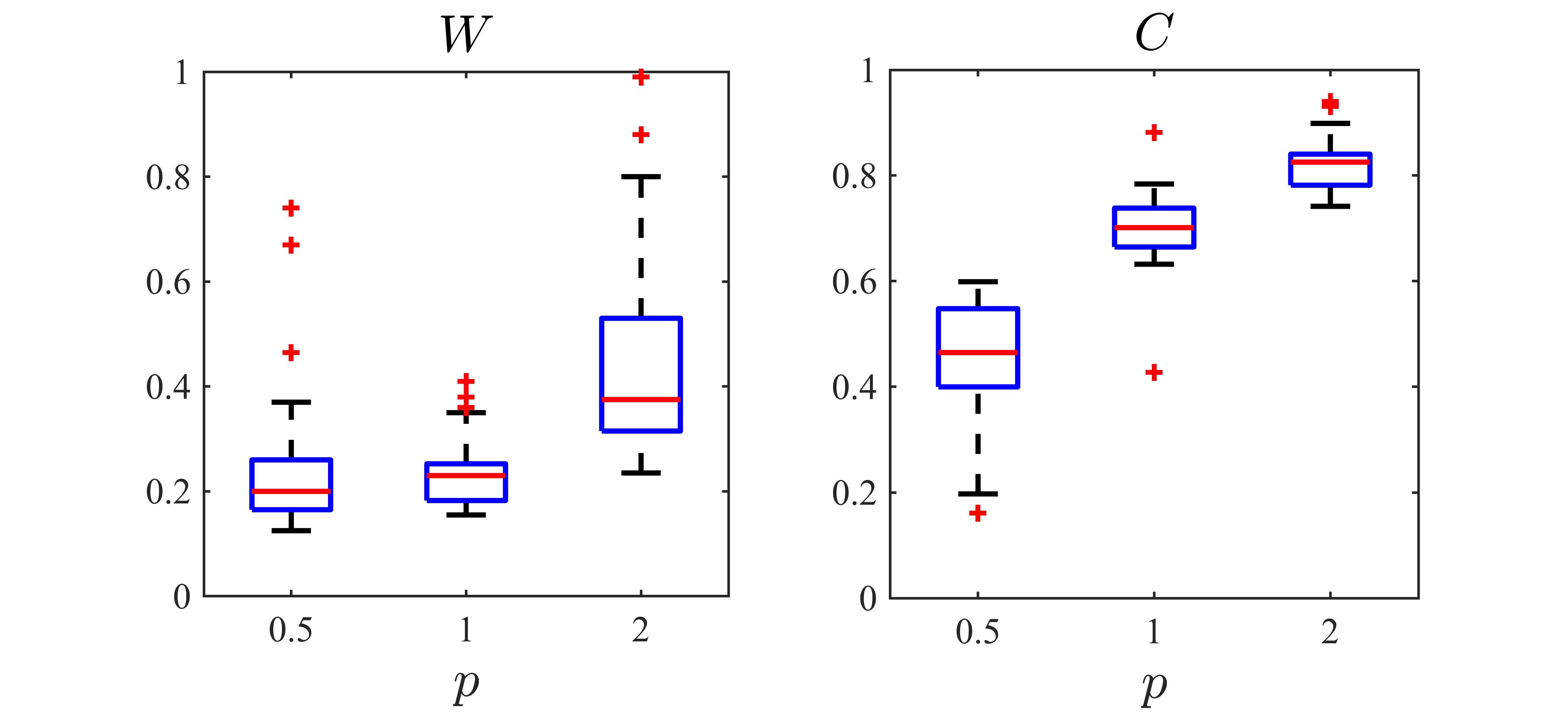}
  \subcaption{Varying cell proliferation rate $p$}
  \label{fig:size:p}\par\vfill
    \vspace{1\baselineskip}
  \includegraphics[width=\linewidth]{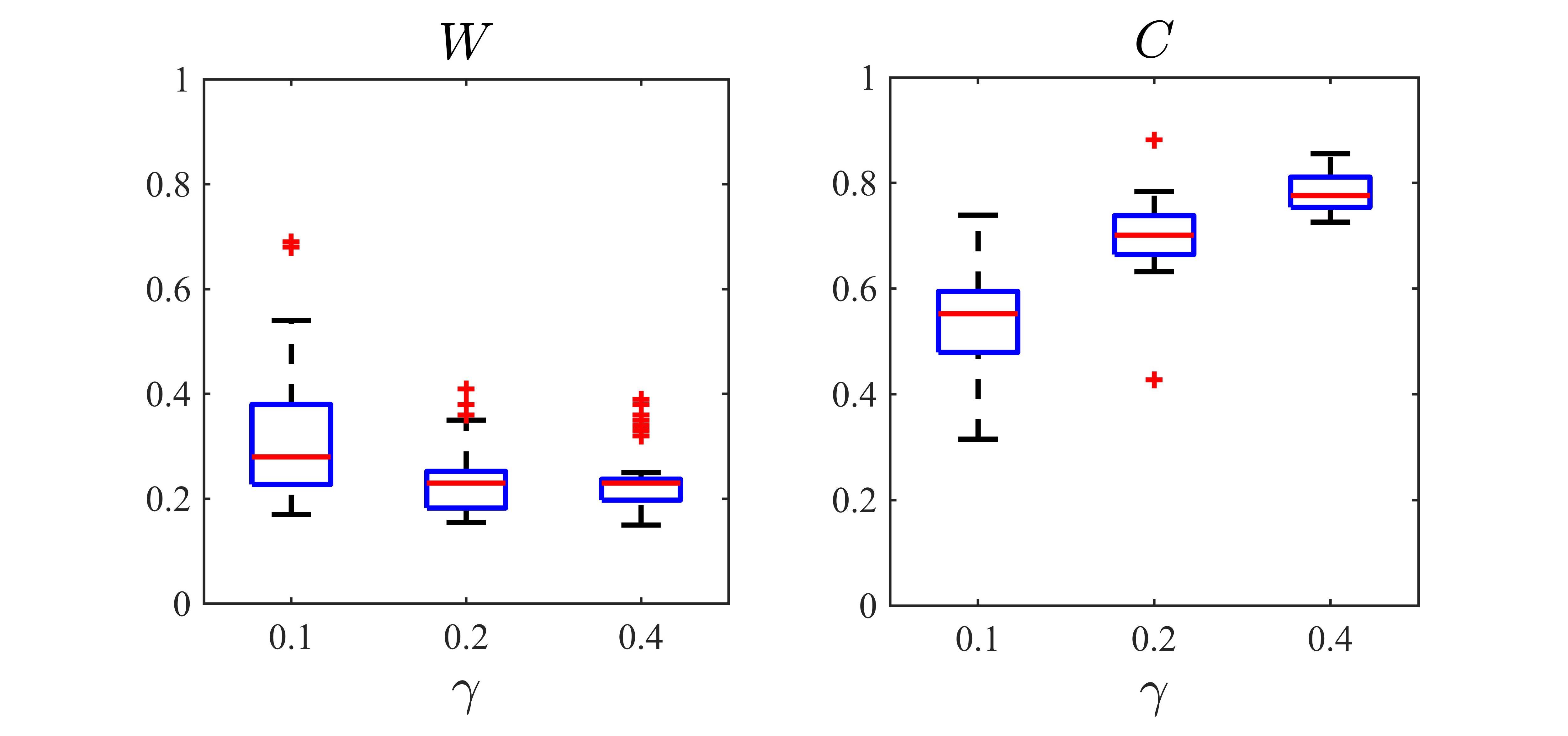}
  \subcaption{Varying ECM degradation rate $\gamma$}
  \label{fig:size:gamma}\par\vfill
    \vspace{1\baselineskip}
  \includegraphics[width=\linewidth]{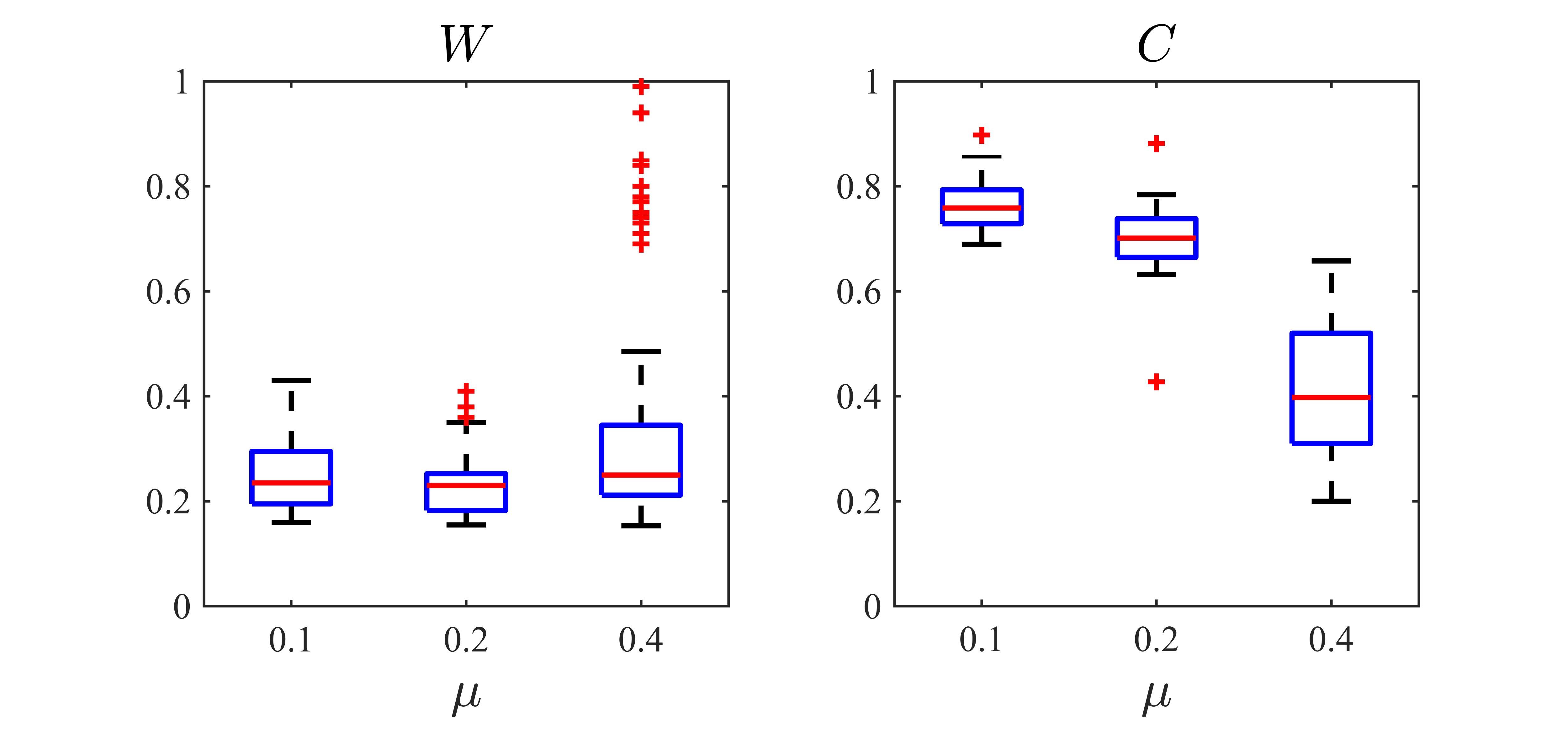}
  \subcaption{Varying ECM remodelling rate $\mu$}
  \label{fig:size:mu}\par\vfill
    \vspace{1\baselineskip}
  \includegraphics[width=\linewidth]{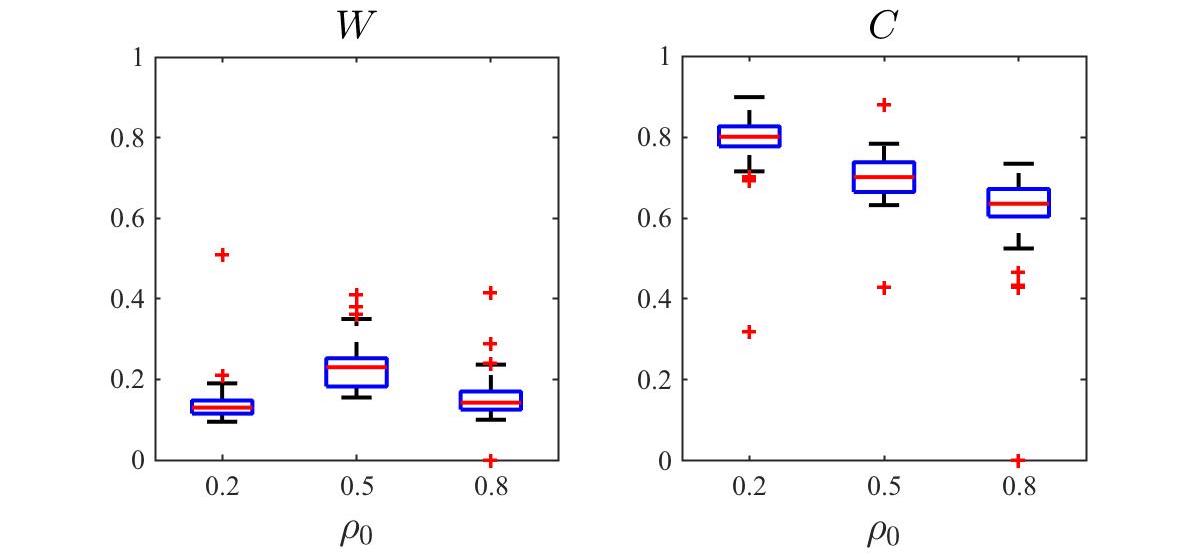}
  \subcaption{Varying initial ECM density $\rho_0$}
  \label{fig:size:rho0}
\end{minipage}
\vspace{2\baselineskip}
\caption{\label{Fig:Size:nrho} Cluster width $W$ and compactness $C$, defined in \eqref{def:W} and \eqref{def:C}, under variations from the baseline parameter set (BPS), in Table~\ref{Tab:par}, of each parameter in equations \eqref{eq:system}$_1$ and \eqref{eq:system}$_2$. 
In (a) we have boxplots of $W$ and $C$ measured on the numerical solution of the system~\eqref{eq:system} at $t=50$, for $D_n$ taking its value in the BPS (center), half (left) and double (right) its value in the BPS. Each boxplot collects data from 100 simulations under randomised initial conditions \eqref{def:ic}-\eqref{def:G}$_1$.
In (b)-(g) we have the same as in (a) but varying parameters $\chi$ (b), $S_{nn}$ (c), $S_{n\rho}$ (d), $p$ (e), $\gamma$ (f), $\mu$ (g) and $\rho_0$ (h).}
\end{figure}

\paragraph*{{The role of VEGFs dynamics}}
Figure~\ref{Fig:Size:mc}e displays changes in cluster width $W$ and compactness $C$ as the rate production of VEGFs $\alpha_c$ varies. Note that these results well mirror those obtained by varying chemotactic sensitivity (\textit{cf.} Figures \ref{Fig:Size:nrho}b and \ref{Fig:Size:mc}e), which is coherent with the notion that higher VEGF production rates correlate with stronger chemotactic dynamics, already established to play a key role in cluster formation.
In addition, the size of clusters seems to be proportional to the VEGF diffusion coefficient $D_c$, as demonstrated by the plots in Figure~\ref{Fig:Size:mc}d in which we see that increasing the magnitude of $D_c$ results in higher $W$ and $C$. This trend, however, suggests that much higher values of $D_c$ may result in tissue invasion, rather than cluster formation (see supplementary Figure S4e in `SuppInfo' document).
Finally, as suggested by Figure~\ref{Fig:Size:mc}f, changes in the VEGF decay rate $\lambda_c$ do not seem to particularly effect cluster size {over the range of parameter values considered}.

\paragraph*{{The role of MMPs dynamics}}
Figure~\ref{Fig:Size:mc}b displays changes in cluster width $W$ and compactness $C$ as the rate production of MMPs $\alpha_m$ varies. Similarly to what was observed for $\alpha_c$ in relation to $\chi$, we see that these boxplots closely resemble those obtained varying $\gamma$ (\textit{cf.} Figures~\ref{Fig:Size:nrho}f and~\ref{Fig:Size:mc}b), which is coherent with the notion that higher MMP production rates correlate with stronger degrading dynamics, already established to be responsible for turning aggregates into clusters -- that is, increasing their compactness. 
In addition, we see in Figure~\ref{Fig:Size:mc}c that increasing the magnitude of the MMPs decay rate $\lambda_m$ yields opposite effects to those obtained increasing their production $\alpha_m$, further confirming the role MMP-mediated ECM degradation has in cluster formation.
On the other hand, changes in the MMPs diffusivity $D_m$ do not seem to affect cluster size, as demonstrated in Figure~\ref{Fig:Size:mc}a -- verified under different orders of magnitude of $D_m$ (see supplementary Figure S4d in `SuppInfo' document). This suggests not much will be gained by distinguishing between membrane-bound and diffusive MMPs, at least in this modelling framework {and within the range of parameter values considered}.

\begin{figure}[h!]
\begin{minipage}[c][13.5cm][t]{.5\textwidth}
  \vspace*{\fill}
  \centering
  \includegraphics[width=\linewidth]{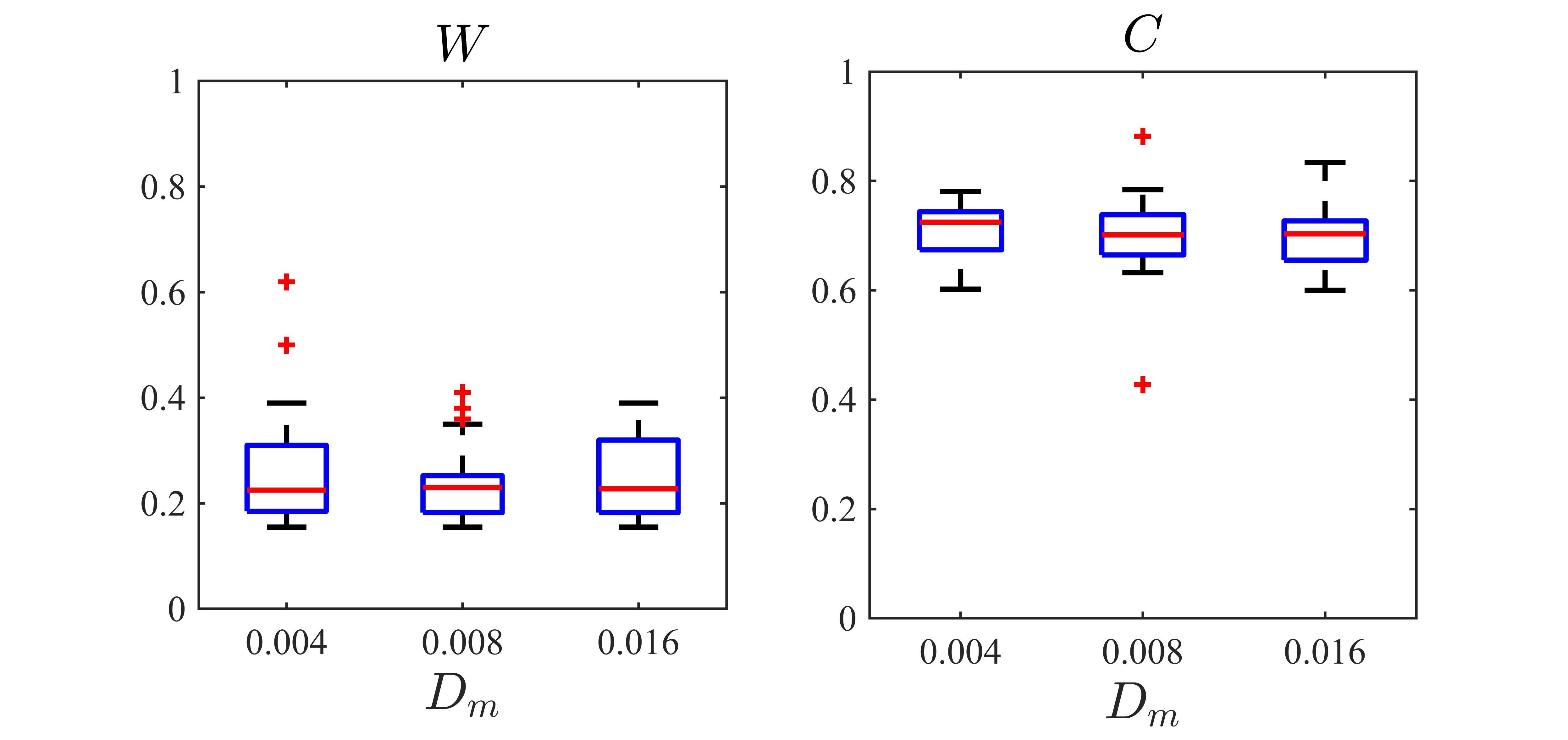}
  \subcaption{Varying MMP diffusion $D_m$}
  \label{fig:size:Dm}\par\vfill
   \vspace{1\baselineskip}
  \includegraphics[width=\linewidth]{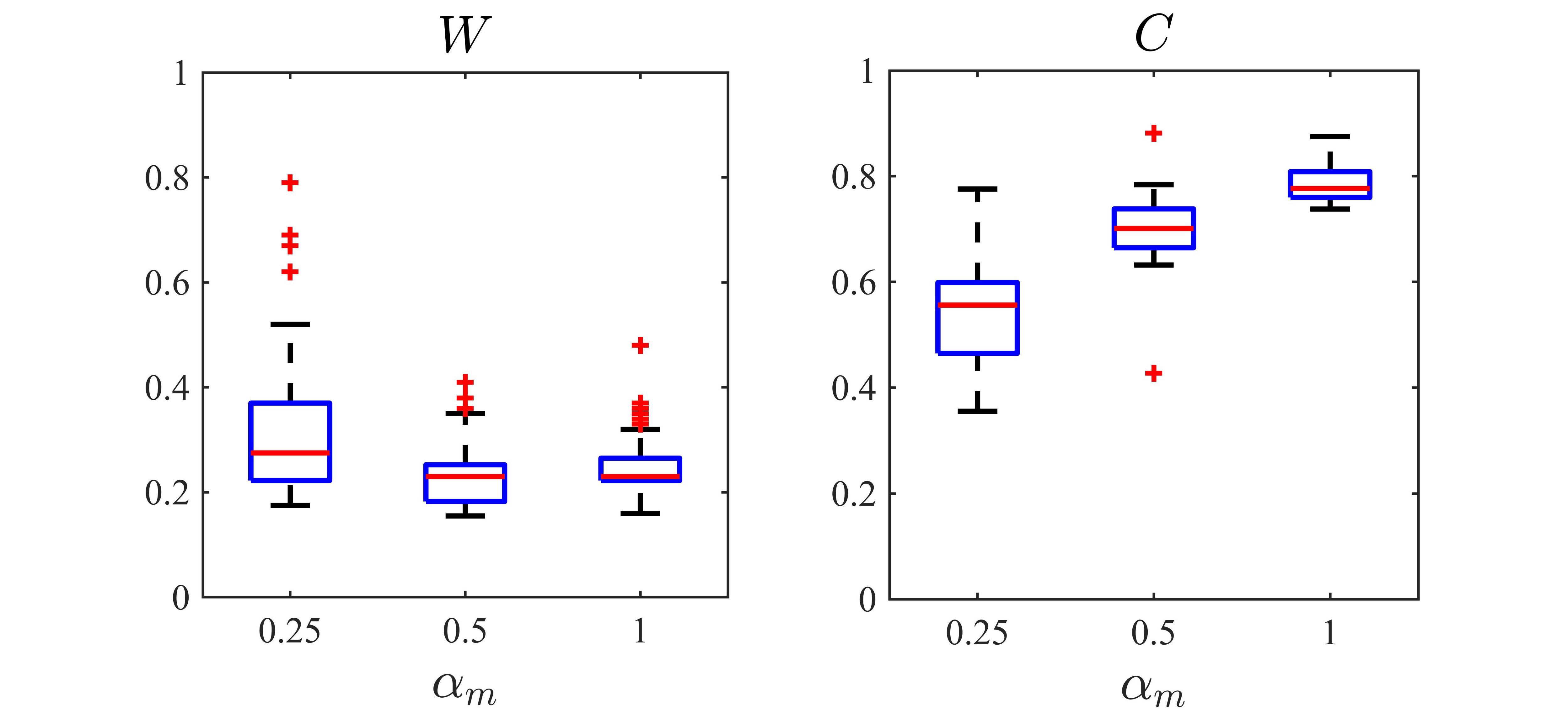}
  \subcaption{Varying MMP production rate $\alpha_m$}
  \label{fig:size:alpham}\par\vfill
   \vspace{1\baselineskip}
  \includegraphics[width=\linewidth]{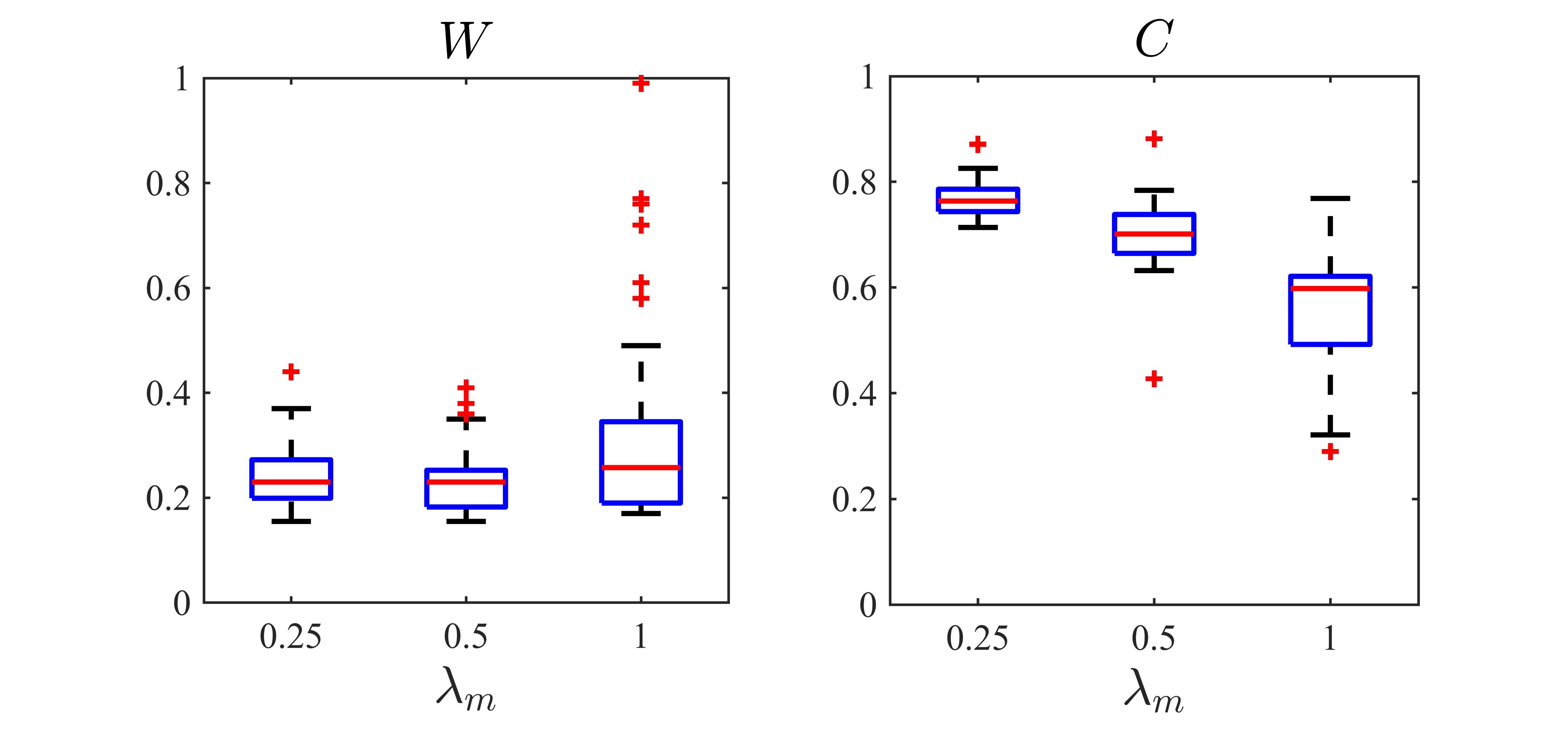}
  \subcaption{Varying MMP decay rate $\lambda_m$}
  \label{fig:size:lambdam}
\end{minipage}%
\begin{minipage}[c][13.5cm][t]{.5\textwidth}
  \vspace*{\fill}
  \centering
  \includegraphics[width=\linewidth]{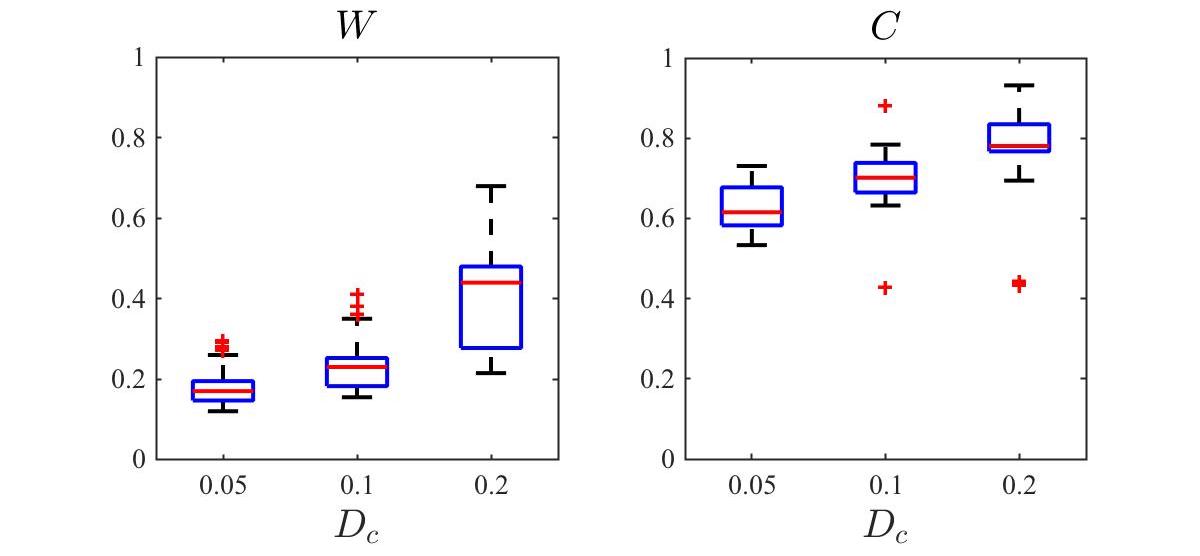}
  \subcaption{Varying VEGF diffusion $D_c$}
  \label{fig:size:Dc}\par\vfill
   \vspace{1\baselineskip}
  \includegraphics[width=\linewidth]{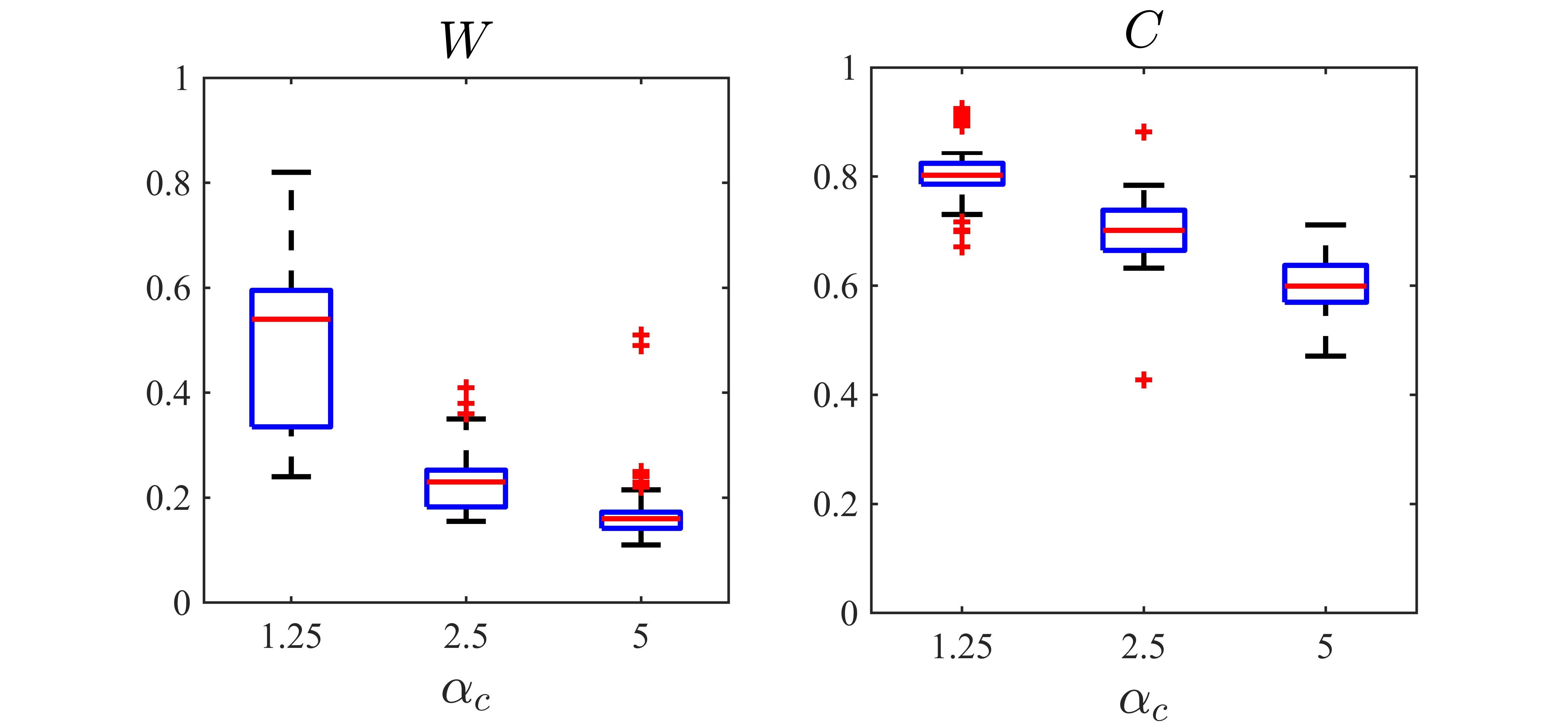}
  \subcaption{Varying VEGF production rate $\alpha_c$}
  \label{fig:size:alphac}\par\vfill
   \vspace{1\baselineskip}
  \includegraphics[width=\linewidth]{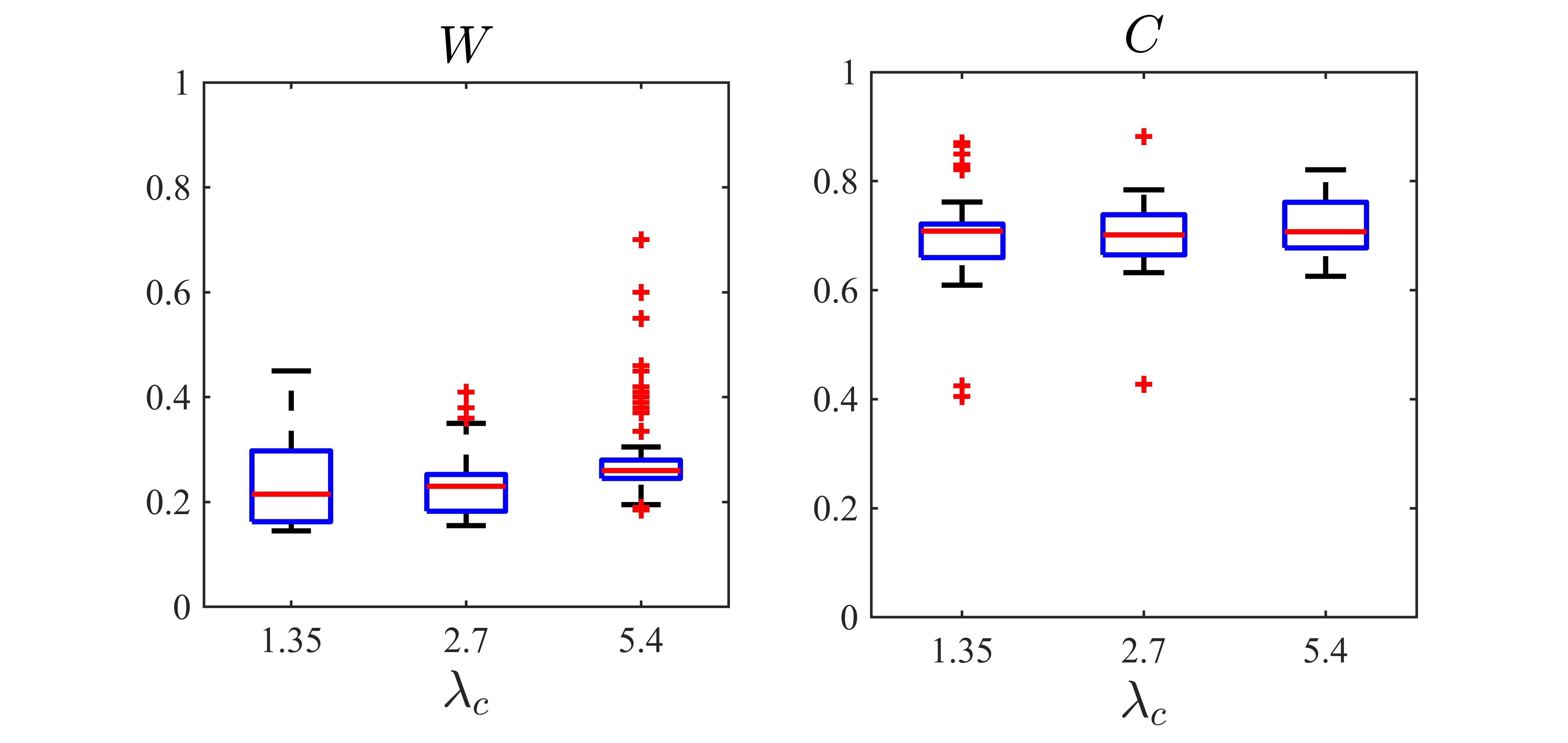}
  \subcaption{Varying VEGF decay rate $\lambda_c$}
  \label{fig:size:lambdac}
\end{minipage}
\vspace{2\baselineskip}
\caption{\label{Fig:Size:mc} Cluster width $W$ and compactness $C$, defined in \eqref{def:W} and \eqref{def:C}, under variations from the baseline parameter set in Table~\ref{Tab:par} of each parameter in equations \eqref{eq:system}$_3$ and \eqref{eq:system}$_4$. 
In (a) we have boxplots of $W$ and $C$ measured on the numerical solution of the system~\eqref{eq:system} at $t=50$, for $D_m$ taking its value in the BPS (center), half (left) and double (right) its value in the BPS. Each boxplot collects data from 100 simulations under randomised initial conditions \eqref{def:ic}-\eqref{def:G}$_1$.
In (b)-(f) we have the same as in (a) but varying parameters $\alpha_m$ (b), $\lambda_m$ (c), $D_c$ (d), $\alpha_c$ (e) and $\lambda_c$ (f).}
\end{figure}

\subsection{Two-dimensional clusters}\label{sec:res:2d}
Let us now consider the 2D problem. In this section we focus on the most interesting results obtained in Section~\ref{sec:res:size} and investigate the role played by chemotaxis, ECM degradation and cell proliferation in the formation of 2D clusters. 
{We also consider the results in absence of ECM remodelling, discussing its biological interpretation in relation to \textit{in vitro} and \textit{in vivo} assays.}

\paragraph{2D clusters under the baseline parameter set}
Under the baseline parameter set reported in Table~\ref{Tab:par}, 2D cluster formation follows slightly different spatio-temporal dynamics to those observed in the 1D case, as demonstrated by the plots in the second row of Figures~\ref{P3:gamma}, \ref{P3:chi} and~\ref{P3:p}. 
At $t=20$ the cell density has already reached maximum local compactness in some regions, while aggregation dynamics are still at their early stages (\textit{cf.} first panel in second row of Figures~\ref{P3:gamma}, \ref{P3:chi} or~\ref{P3:p}), and the minimum cluster size is observed no earlier than $t=140$ (\textit{cf.} {fourth} panel in second row of Figures~\ref{P3:gamma}, \ref{P3:chi} or~\ref{P3:p}).
The cluster observed at this stage has a nondimensional diameter of about 0.2, which agrees with the results in 1D and is therefore {also consistent} with experimental observations reported by Blatchley \textit{et al.}~\cite{blatchley2019hypoxia}. After the cluster has formed, no tissue invasion is observed -- see supplementary Figure S5 in `SuppInfo' document.

\paragraph{The role of ECM degradation {in 2D}}
The plots reported in Figure~\ref{P3:gamma} demonstrate that ECM degradation promotes the formation of 2D clusters, as predicted by the results of the 1D model presented in Sections~\ref{sec:res:main} and~\ref{sec:res:size}.
Lower ECM degradation rates correlate with slower aggregation dynamics and lower compactness of such aggregates (\textit{cf.} first row of Figure~\ref{P3:gamma}) -- to the extent that in absence of ECM degradation no clusters form (see supplementary Figure S6 in `SuppInfo' document) -- and higher ECM degradation rates correlate with faster cluster formation with well-defined and compact clusters observed at much earlier times (\textit{cf.} third row of Figure~\ref{P3:gamma}). 
{As for the 1D model, }we obtain the same results under analogous variations of the MMP secretion rate $\alpha_m$ (see supplementary Figure S7 in `SuppInfo' document).

\begin{figure}[h!]
\includegraphics[width=1\linewidth]{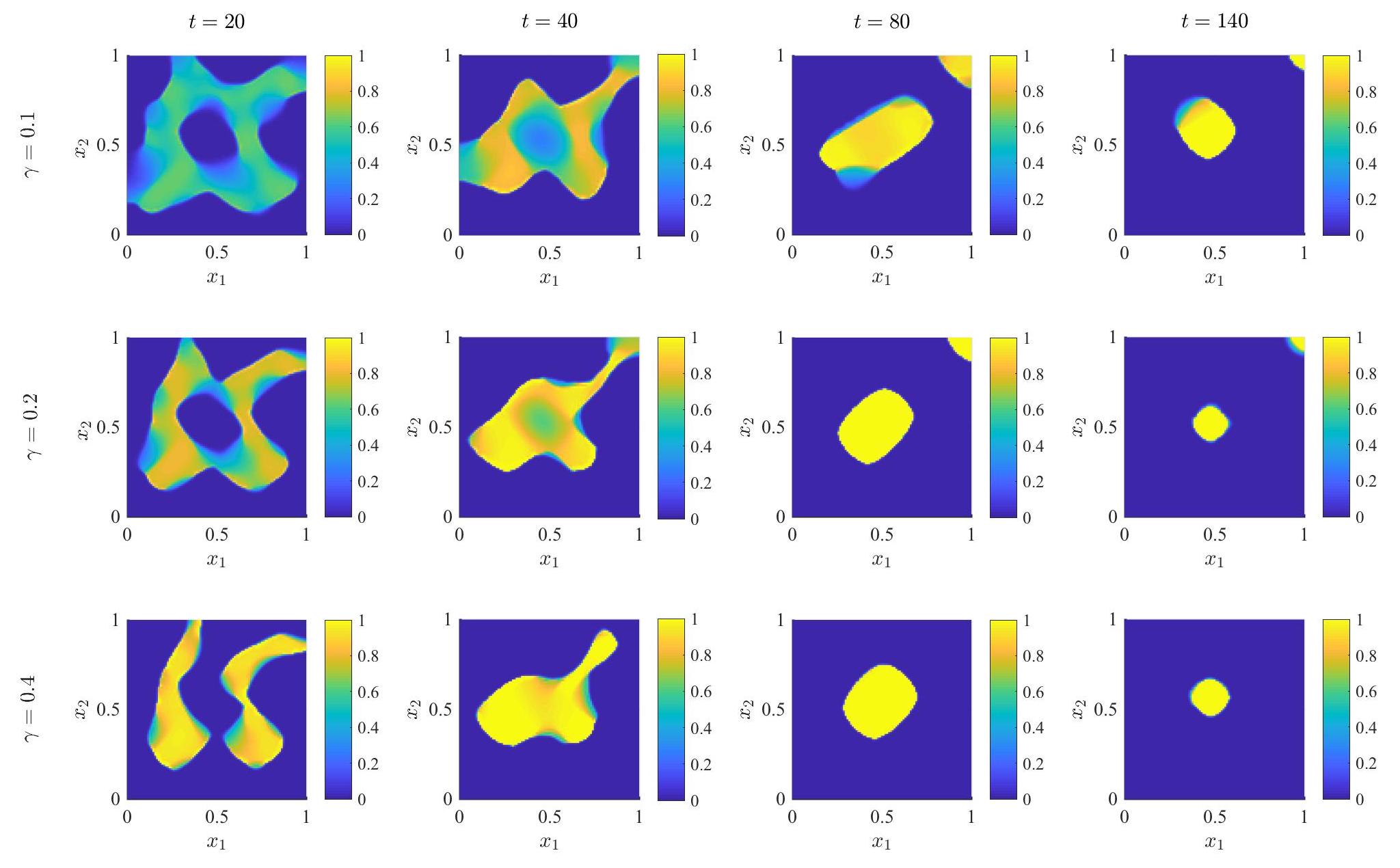}
\caption{\label{P3:gamma} \textbf{First row:} Plots of the cell density $n(t,{\bf x})$ obtained solving the system~\eqref{eq:system}, together with definitions~\eqref{eq:a:def:2d}, \eqref{eq:omega:def} and~\eqref{eq:f2}, initial conditions~\eqref{def:ic} and~\eqref{def:G}$_2$, complemented with zero{-flux} boundary conditions, under the parameter choices reported in Table~\ref{Tab:par}, except for $\gamma=0.02$. The solution is plotted at time $t=20$ (first panel), $t=40$ (second panel), $t=80$ (third panel) and $t=140$ (fourth panel). \textbf{Second and third row:} Same as first row, except for $\gamma=0.2$ (second row) and $\gamma=2$ (third row).}
\end{figure}

\paragraph{The role of chemotaxis {in 2D}}
The plots reported in Figure~\ref{P3:chi} demonstrate that endogenous chemotaxis promotes aggregation dynamics, as predicted by the results of the 1D model presented in Sections~\ref{sec:res:main} and~\ref{sec:res:size}, and reveal that  chemotaxis is an important determinant of 2D cluster topology. 
Lower values of the chemotactic sensitivity $\chi$ correlate with slower and weaker aggregation dynamics, so that no well-defined clusters can be observed (\textit{cf.} first row of Figure~\ref{P3:chi}). Higher values of $\chi$ correlate with faster and stronger aggregation dynamics, with well-defined clusters observed at earlier times (\textit{cf.} third row of Figure~\ref{P3:chi}) and cluster diameter -- once the clusters have reached minimum size -- smaller than that observed with lower values of $\chi$ (\textit{cf.} last panel in the second row and last panel in the third row of Figure~\ref{P3:chi}).
{As for the 1D model, }we obtain the same results under analogous variations of the VEGF secretion rate $\alpha_c$ (see supplementary Figure S8 in `SuppInfo' document).

\begin{figure}[h!]
\includegraphics[width=1\linewidth]{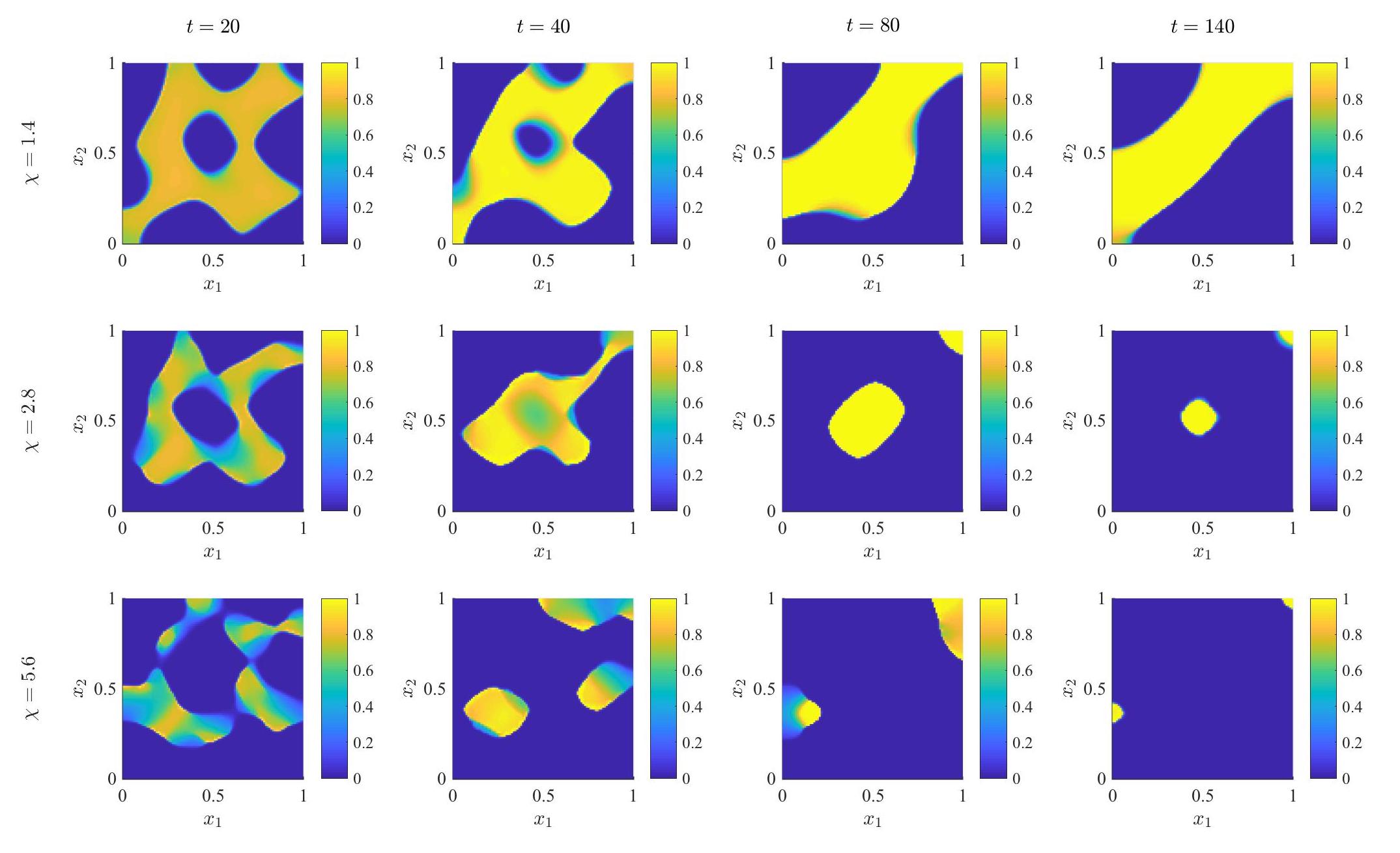}
\caption{\label{P3:chi} \textbf{First row:} Plots of the cell density $n(t,{\bf x})$ obtained solving the system~\eqref{eq:system}, together with definitions~\eqref{eq:a:def:2d}, \eqref{eq:omega:def} and~\eqref{eq:f2}, initial conditions~\eqref{def:ic} and~\eqref{def:G}$_2$, complemented with zero{-flux} boundary conditions, under the parameter choices reported in Table~\ref{Tab:par}, except for $\chi=1.4$. The solution is plotted at time $t=20$ (first panel), $t=40$ (second panel), $t=80$ (third panel) and $t=140$ (fourth panel). \textbf{Second and third row:} Same as first row, except for $\chi=2.8$ (second row) and $\chi=5.6$ (third row).}
\end{figure}

\paragraph{The role of cell proliferation {in 2D}}
{While the results of the 1D model presented in Section~\ref{sec:res:size} seemed to suggest that  the speed of the cluster formation process is proportional to the cell proliferation rate $p$, the plots reported in Figure~\ref{P3:p} in a 2D framework indicate otherwise. }
While the rate of cell proliferation may influence how the cells respond to spatial gradients at the beginning of the cluster formation process (\textit{cf.} plots in the second column of Figure~\ref{P3:p}), it does not seem to affect the overall spatio-temporal dynamics of 2D cluster formation. 
Upon these considerations, the results displayed in Figure~\ref{fig:size:p} may simply be a 1D projection of the 2D dynamics occurring around the same time -- consider for instance a 1D cross section (\textit{e.g.} $x_1=0.5$) of the plots in the second column of Figure~\ref{P3:p}.

\begin{figure}[h!]
\includegraphics[width=1\linewidth]{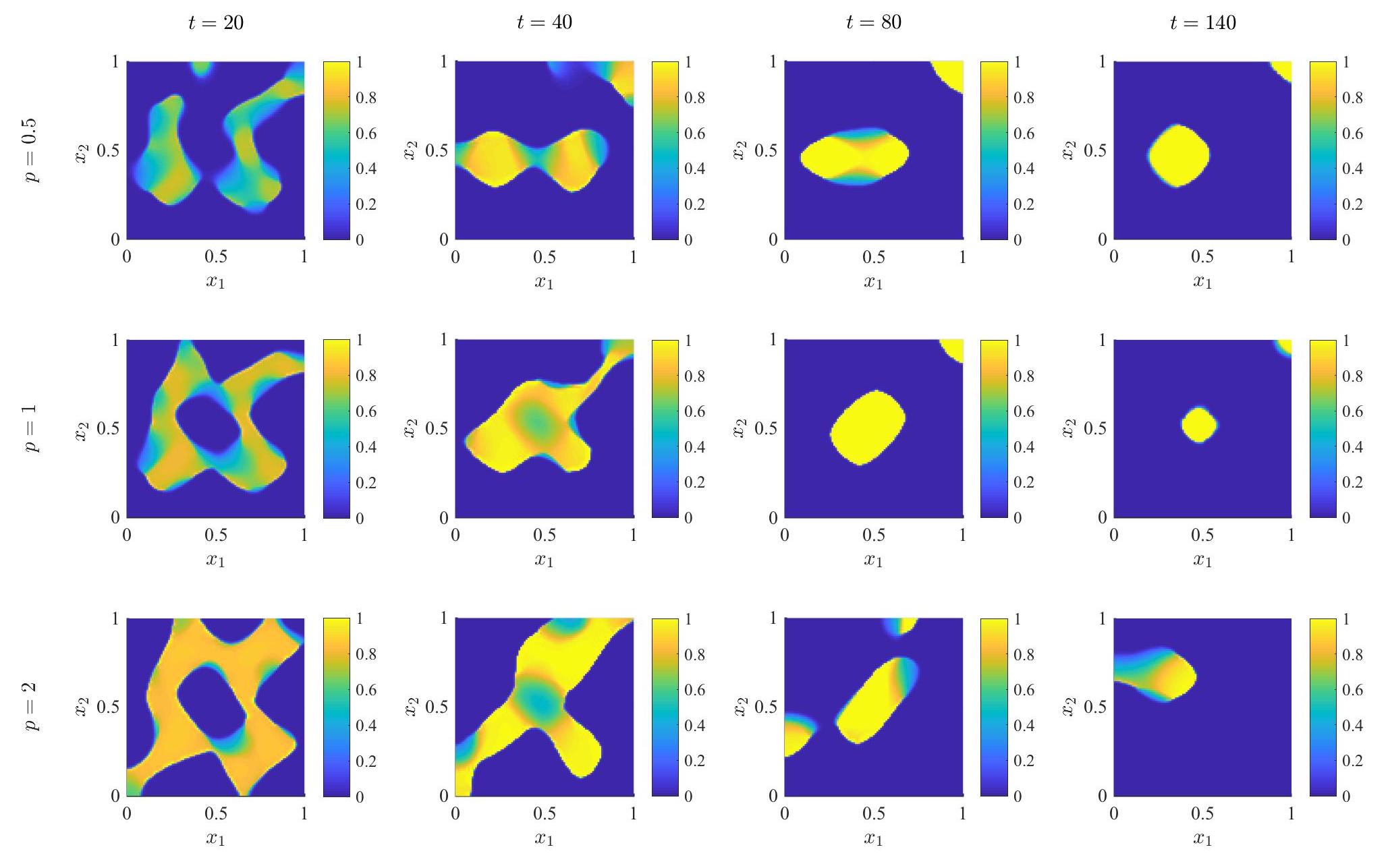}
\caption{\label{P3:p} \textbf{First row:} Plots of the cell density $n(t,{\bf x})$ obtained solving the system~\eqref{eq:system}, together with definitions~\eqref{eq:a:def:2d}, \eqref{eq:omega:def} and~\eqref{eq:f2}, initial conditions~\eqref{def:ic} and~\eqref{def:G}$_2$, complemented with zero{-flux} boundary conditions, under the parameter choices reported in Table~\ref{Tab:par}, except for $p=0.5$. The solution is plotted at time $t=20$ (first panel), $t=40$ (second panel), $t=80$ (third panel) and $t=140$ (fourth panel). \textbf{Second and third row:} Same as first row, except for $p=1$ (second row) and $p=2$ (third row).}
\end{figure}

\paragraph*{{The role of matrix remodelling and its relation to \textit{in vitro} studies}} {While matrix remodelling naturally occurs \textit{in vivo} thanks to the presence of other cells in biological tissue, this is not generally observed in \textit{in vitro} assays, which occur in isolated environments. 
Thus, in order to compare our results with those of \textit{in vitro} studies, we chose to set the ECM remodelling rate $\mu=0$, {the results are presented} in Figure~\ref{P3:mu}. 
{We observe that clusters still form in the absence of ECM remodelling, however they are smaller than in the baseline parameter set} (\textit{cf.} last panel in the second row of Figures~\ref{P3:gamma}-\ref{P3:p} and that of Figure~\ref{P3:mu}). 
{The smaller cluster sizes may result from the more dominant role} played by cell-to-cell adhesion following ECM degradation: {in absence of ECM remodelling, volume exclusion is less likely to affect the cell dynamics and, at the cell boundaries, cell-matrix interactions become negligible compared to cell-cell interactions.} 
We would then expect to observe clusters of a smaller diameter but higher maximum density, but because of death due to competition for space we instead observe a loss of cell mass. While these results are mathematically {consistent} with our model set up, the observed cluster size under the baseline parameter set was more coherent with \textit{in vitro} experimental observations in~\cite{blatchley2019hypoxia} than that observed for $\mu=0$. This could be explained by considering that our 2D set up is meant to reflect dynamics occurring in a 2D horizontal cross section of the 3D experimental domain reported in~\cite{blatchley2019hypoxia}, illustrated in Figure~\ref{P1:figbio}A. As the cells degraded the matrix and reorganised into clusters, they also fell towards the bottom of the hydrogel, as illustrated by the experimental data reported in Figure~\ref{P3:figbio}, in regions where it was not yet degraded. Thus, following the horizontal plane intersecting the cluster's center, we would likely observe  section crossing the center of a cluster, we would indeed observe cell-independent matrix remodelling.
}

\begin{figure}[htb!]
\includegraphics[width=1\linewidth]{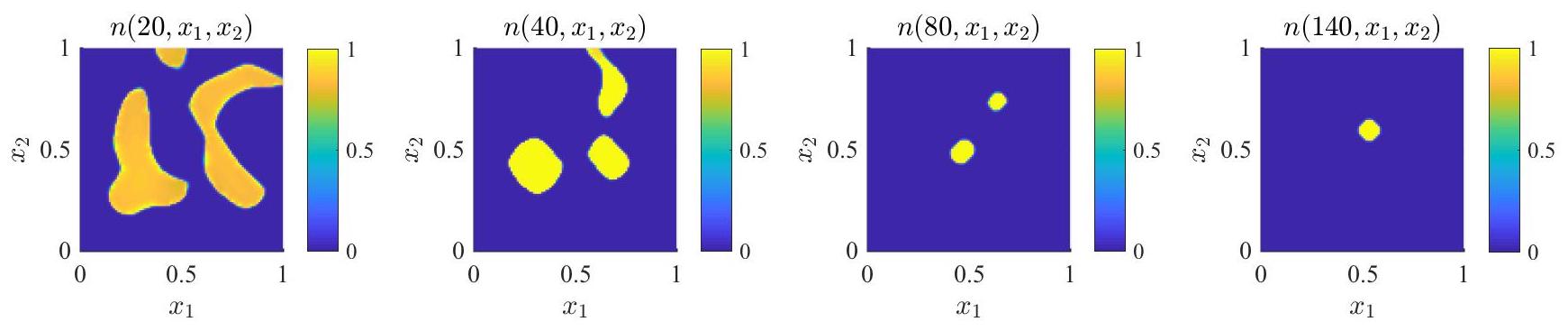}
\caption{\label{P3:mu} { Plots of the cell density $n(t,{\bf x})$ obtained solving the system~\eqref{eq:system}, together with definitions~\eqref{eq:a:def:2d}, \eqref{eq:omega:def} and~\eqref{eq:f2}, initial conditions~\eqref{def:ic} and~\eqref{def:G}$_2$, complemented with zero-flux boundary conditions, under the parameter choices reported in Table~\ref{Tab:par}, except for $\mu=0$. The solution is plotted at time $t=20$ (first panel), $t=40$ (second panel), $t=80$ (third panel) and $t=140$ (fourth panel). }}
\end{figure}

\begin{figure}[h!]
\centering
\includegraphics[width=0.45\linewidth]{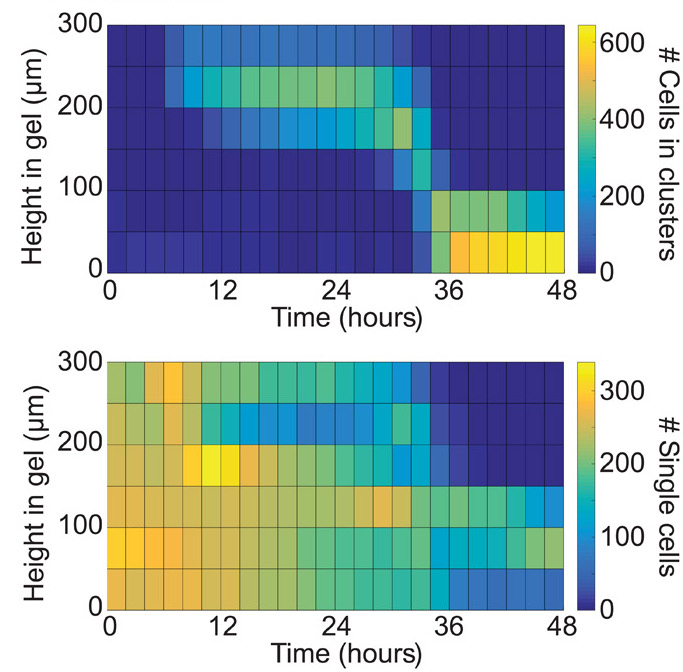}
\caption{\label{P3:figbio} {Visual summary of the number of single cells (bottom) and the number of cells in clusters (top) observed over time at different heights in the 3D matrigel in hypoxic gradients by Blatchley \textit{et al.}~\cite{blatchley2019hypoxia} -- experimental set up illustrated in Figure~\ref{P1:figbio}A (bottom). 
Figure taken from Figure 1E in~\cite{blatchley2019hypoxia} under Creative Commons licence \url{https://creativecommons.org/licenses/by-nc/4.0/}. }}
\end{figure}

\section{Discussion and research perspectives}\label{sec:con}
Despite the great progress made in the past 20 years in understanding the mechanisms behind EPC cluster-based vasculogenesis, much more needs to be achieved in order to unlock its full therapeutic potential.
Mathematical modelling provides theoretical means to shed light on the otherwise hidden role played by underlying dynamics in the origin and structure of the emergent vascular network, as previously achieved in the study of mature EC vascular network formation (single-cell vasculogenesis).
We therefore formulated a {nonlocal PDE} model of EPC cluster formation during the early stages of vasculogenesis, including mechanisms such as ECM degradation, cell proliferation and cell-to-cell adhesion, which were {recently} found to distinguish cluster-based vasculogenesis from single-cell vasculogenesis{~\cite{blatchley2019hypoxia}}.
Thanks to the introduction of appropriate metrics of cluster width and compactness, we investigated the role played by underlying dynamics in facilitating cluster formation, regulating the speed of the cluster-formation process and the size of clusters{-- see Table~\ref{Tab:res1d} for a summary of the results of the parametric analysis of the 1D problem. Furthermore, we verified that} the key {observations from the parametric analysis for the} 1D model still hold in a 2D framework, {with the exception of the role played by EPC proliferation}.

\begin{table}[h!]
\centering
\caption{{Summary of the results of the parametric analysis of the 1D problem conducted in Section~\ref{sec:res:size}. The biological meaning of each parameter is summarised in Table~\ref{Tab:par}. Cluster width $W$ and compactness $C$ are defined in Section~\ref{sec:met:cw}.}} \label{Tab:res1d}
\begin{tabular}{|c|l|c|}
\hline
\textbf{Parameter} & \textbf{Main effect of increasing the parameter value on cluster size} & \textbf{Ref. Figure} \\
\hline
$D_n$ &  Little effect favouring tissue invasion & \ref{fig:size:Dn}, S4a \\
$\chi$  & Better defined clusters (smaller range of width $W$ measured) & \ref{fig:size:chi} \\
$S_{nn}$ &  Little to no effect & \ref{fig:size:Snn}, S4b \\
$S_{n\rho}$ & Little effect favouring tissue invasion  & \ref{fig:size:Snp} \\
$p$ & Increased cluster formation speed (refuted in 2D)
 & \ref{fig:size:p} \\ 
$\gamma$ & Better defined clusters (higher median compactness $C$ measured) & \ref{fig:size:gamma}, S4c \\
$\mu$ & Less well defined clusters (lower median compactness $C$ measured) & \ref{fig:size:mu} \\
$\rho_0$ &  Little to no effect  &  \ref{fig:size:rho0} \\
$D_m$ &   Little to no effect & \ref{fig:size:Dm}, S4d \\
$\alpha_m$ &  Better defined clusters (higher median compactness $C$ measured) & \ref{fig:size:alpham} \\
$\lambda_m$ &  Less well defined clusters (lower median compactness $C$ measured) & \ref{fig:size:lambdam} \\
$D_c$ &  Bigger clusters (higher median width $W$ and compactness $C$ measured) & \ref{fig:size:Dc}, S4e \\
$\alpha_c$ &  Better defined clusters (smaller range of width $W$ measured) &  \ref{fig:size:alphac} \\
$\lambda_c$ &   Little to no effect & \ref{fig:size:lambdac} \\
\hline
\end{tabular}
\end{table}

\subsection{Chemotaxis, degradation and their link to hypoxia\label{P3:Ch6:hypoxia}}
Our results confirmed the role played by ECM degradation in the formation of EPC clusters both with the 1D and 2D models, providing additional theoretical support to the mechanism of EPC cluster formation proposed by Blatchley \textit{et al.}~\cite{blatchley2019hypoxia}.  {For example, the \textit{in vitro} experiments indicated that no clusters form in the absence of MMPs, which is precisely what we discover from the numerical solutions of our model.}
In addition, the investigation conducted in Sections~\ref{sec:res:size} and~\ref{sec:res:2d} indicates that the speed of cluster formation is proportional to the rate of ECM degradation -- or the rate of MMP secretion by the cells -- which nicely agrees their experimental observations.
Our numerical results further highlighted that ECM degradation alone may not suffice to explain the formation of clusters, as endogenous chemotaxis was shown to be responsible for aggregation dynamics, without which the cells would simply invade the whole tissue. In addition, the investigation conducted in Sections~\ref{sec:res:size} and~\ref{sec:res:2d} suggests that the size of clusters is (inversely) related to the chemotactic sensitivity of the cells -- {and the rate at which they secrete VEGF} -- indicating chemotaxis may be a key determinant of cluster topology.
{Note that, in view of the experimental evidence presented in~\cite{blatchley2019hypoxia} and~\cite{akita2003hypoxic}, the MMP production rate $\alpha_m$ and the VEGF production rate $\alpha_c$ may be correlated with the local level of hypoxia. It would therefore be interesting to let these production rates be given as functions of the local oxygen concentration. The results here obtained suggest that our model would then predict cluster formation to be fostered by higher levels of hypoxia, indeed agreeing with the referenced experimental observations~\cite{akita2003hypoxic,blatchley2019hypoxia}.} 

\subsection{Cell-to-cell adhesion and its aggregative vs stabilising effect}\label{P3:Ch6:adhesion}
{Cell-to-cell adhesion, on the other hand, does not seem to play an important role in cluster formation or cluster topology. 
While small values of the cell-to-cell adhesion coefficient have an negligible effect on the simulation results overall, larger values show at least a transient impact on cluster formation and the cluster topology arising.}
While this {weak effect} may seem counter-intuitive considering previous single-cell vasculogenesis works~\cite{boas2018cellular,merks2004cell,ramos2018capillary,scianna2013review}, these models do not include both chemotaxis and degradation and thus do not predict cluster formation as an intermediate step in network formation. In view of these works, however, we cannot exclude that {aspects of the } final network topology, \textit{e.g.} chord thickness, depend on {the amount of } cell-to-cell adhesion. 
{Furthermore, a}ccording to Blatchley \textit{et al.}~\cite{blatchley2019hypoxia}, cell-to-cell adhesion is not related to cluster topology, which is coherent with our model predictions, but is responsible for cluster stability.
The mathematical study of cluster-based vasculogenesis would {therefore} benefit from a more suited description of cell-to-cell adhesion, as we have seen in Section~\ref{sec:res:main} that our modelling choice allows us to capture the aggregative effect of cell-to-cell adhesion, but not its role in cluster stabilisation. 
{This could perhaps be achieved by, in addition to modelling cell-to-cell and cell-to-matrix adhesion nonlocally, modelling diffusion nonlinearly~\cite{carrillo2019population,murakawa2015continuous}. 
Alternatively, one could modify the volume exclusion term (\textit{i.e.} definition~\eqref{def:f}) to explicitly model cell-adhesion molecules, which would allow us to study the effect of cell-adhesion-mediated saturation of chemotaxis~\cite{merks2005dynamic,merks2008contact,singh2015role}. 
Otherwise, the stabilising effect of cell-to-cell adhesion} could be more easily achieved by adopting an individual-based or hybrid modelling approach~\cite{anderson2005hybrid,singh2015role,turner2002intercellular,turner2004discrete}, as done in previous {CP models of} single-cell vasculogenesis~\cite{boas2018cellular,merks2004cell,merks2008contact,ramos2018capillary,scianna2011multiscale,scianna2013review,szabo2008multicellular}.

\subsection{Experimental research perspectives}
{The results reported in this paper suggest that in order to gain a comprehensive understanding of EPC cluster formation during the early-stages of cluster-based vasculogenesis, it is necessary to consider both ECM degradation and chemotaxis. {So far, these processes have only been investigated separately in experimental studies~\cite{akita2003hypoxic,blatchley2019hypoxia} but it would be interesting to examine them in both normoxic and hypoxic conditions, in view of our \textit{in silico} results (\textit{cf.} Section~\ref{P3:Ch6:hypoxia}). Model validation would also allow for a more accurate baseline parameter set. 
In fact,} while the role played by ECM degradation and the size of clusters in our numerical simulations well matches the experimental observations reported by Blatchley and coworkers~\cite{blatchley2019hypoxia}, this -- and future -- mathematical models would benefit from the estimation of parameter values by fitting the model to experimental data in order to obtain a better match of the timescale of 2D cluster formation. 
{Note, for instance,} that the ten-fold increase in $\alpha_m$ which results in the observation of well-defined clusters at $t=20$ (see supplementary Figure S7) might be justified by the experimental conditions reported in~\cite{blatchley2019hypoxia} {(see~\ref{sec:parameters})}.
{In view of the results reported in Section~\ref{sec:res:2d} in absence of matrix remodelling, it would also be interesting to experimentally investigate the role ECM remodelling may plays in cluster formation and cluster size. This could lead to interesting observations regarding the empirical differences of cluster-based vasculogenesis during \textit{in vitro} and \textit{in vivo} studies, and suggest interesting focus points for therapeutic intervention. In order to achieve this, new \textit{in vitro} experimental set ups might be required, in order to avoid the remodelling-resembling effects of cells falling towards the bottom of the 3D hydrogel during the cluster-formation process.}

\subsection{Mathematical research perspectives}{
In this work we have conducted a parametric analysis of the model in order to elucidate potential model behaviour. Whereas this already gave meaningful information to discuss the importance of the various processes involved in cluster-based vasculogenesis there are more advanced mathematical tools available for the analysis of the dependence of the model outcome on the model parameters.
In particular, once accurate ranges of parameter values, possibly together with a distribution of the values over the range, have been estimated for various parameters from data spanning different levels of hypoxia, it would be of significant interest to conduct a global (or partially global) sensitivity analysis~\cite{marino2008methodology,qian2020sensitivity,renardy2019global} of the cluster width and compactness to parameter variability. 
In fact, while the  conducted parametric analysis highlighted the role played by each single parameter as it deviates from its baseline value, the global sensitivity analysis would provide a tool to investigate the effect of combined variations of multiple or all parameters over ranges indicative of different levels of hypoxia. This global approach is also significant given the nonlinear nature of the model~\cite{saltelli2010avoid,saltelli2019so}. Upon revision of the modelling strategy adopted to describe cell-to-cell adhesion, a global sensitivity analysis could be conducted in 2D in relation to the cluster compactness ($C$), area ($W$) and elongation (see Section~\ref{sec:met:cw}). Moreover, it would be particularly interesting to conduct a weakly nonlinear analysis~\cite{boonkorkuea2010nonlinear,cross2009pattern,han2017pattern,hoyle2006pattern} of the 2D model for a quantitative description of how chemotaxis is responsible for cluster topology. For instance, considering the results reported in Figure~\ref{P3:chi}, there might be a threshold value of the chemotactic sensitivity below which we do not observe 2D clusters.}

\subsection{Modelling research perspectives}{
In addition to the modelling research perspectives discussed in Sections~\ref{P3:Ch6:hypoxia} and~\ref{P3:Ch6:adhesion}, interesting investigations could be conducted by considering the effect of persistence of motion or cell-matrix mechanical interactions.}
\subsubsection{The possible interplay between degradation and persistence of motion}
{Previous CP models of single-cell vasculogenesis predicted cluster formation in absence of cell elongation~\cite{merks2006cell}, due to chemotaxis. Despite their exclusion of ECM degradation and proliferation, we expect such result to capture the role of elongation during cluster-based vasculogenesis, given that EPC motion during the early stages of the process is mostly amoeboid-like, and we observe cell polarisation only in the later stages~\cite{blatchley2019hypoxia}. 
On the other hand, persistence of motion, indicated to prevent cell clusterisation by PEC models of single-cell vasculogenesis~\cite{tosin2006mechanics}, was observed to be enhanced during amoeboid-like migration~\cite{serini2003modeling}.
 It would therefore} be interesting to investigate whether clusters would still form in this modelling framework with the inclusion of persistence of motion.
{In particular note that Tosin \textit{et al.}~\cite{tosin2006mechanics} included in their momentum equation a drag force generated by cells moving on the ECM and observed that lower cell-matrix adhesion resulted in cell clusterisation. We might thus expect ECM degradation to play a key role in lowering cell adhesiveness to the ECM in such modelling framework, thus leading to clusters.}

\subsubsection{Modelling the late-stage dynamics of cluster-based vasculogenesis}
We plan to extend the theoretical investigation of cluster-based vasculogenesis to later stages of this process, during which EPCs increase their interaction with the ECM and bridge clusters, forming the vascular network. At such stage the mechanical interaction between the cells and ECM becomes non-trivial and we are therefore going to consider a mechanochemical model similar to those previously proposed to study the late stages of single-cell vasculogenesis~\cite{manoussaki2003mechanochemical,murray2003mechanochemical,tosin2006mechanics}. 
{Various works in the current literature already address the formation of sprouts from existing clusters~\cite{boas2018cellular,merks2008contact,szabo2010role,szabo2008multicellular} and prior single-cell vasculogenesis models {investigating} the formation of a vascular network, rather than clusters, may still be relevant to these late stages.
{For instance, while cell-to-cell adhesion may saturate chemotaxis~\cite{merks2008contact}, VEGF gradients may still be strongly perceived at the cluster boundaries. This is especially the case if ECM-bound VEGF was the main signalling molecule, and in this case we would indeed expect steep VEGF gradients at the cluster boundaries~\cite{merks2006cell}.} Perhaps such a scenario can explain the change in cell morphology. The role cell elongation plays in the formation of a final well-defined network has already been extensively addressed by CP models of single-cell vasculogenesis~\cite{boas2018cellular,merks2004cell,merks2006cell,palm2013vascular,ramos2018capillary,scianna2013review,van2014mechanical} 
Blatchley \textit{et al.}~\cite{blatchley2019hypoxia} indeed reported an elongated cell morphology during sprouting, as well as an increased mechanical interaction with the ECM. The combination of cell traction on the ECM and strain-dependent movement of cells would also suffice in explaining bridging between clusters~\cite{ambrosi2005review,manoussaki2003mechanochemical,murray2003mechanochemical,namy2004critical}. Nevertheless mechanochemical models, as well as CP models including cell traction~~\cite{ramos2018capillary,van2014mechanical}, {generally} predict that higher ECM stiffness inhibits network formation, which is in contradiction with the experimental observations presented in~\cite{blatchley2019hypoxia}. 
{Existing single-cell vasculogenesis models of late-stage dynamics, however, do not include ECM degradation, a key element for the study of cluster-based vasculogenesis. 
On the other hand, ECM degradation} has been shown to have an important role during sprouting angiogenesis~\cite{boas2018cellular,daub2013cell,holmes2000mathematical,scianna2013review,tranqui2000mechanical}, a result which could very well translate into an active role in cluster-based vasculogenesis -- although the interplay between ECM degradation and cell-matrix interactions may be particularly complex.}
Upon formulation of a mathematical model which accurately predicts {cluster-based} vascular network assembly, this could be used to investigate the determinants of network size and topology{. For instance, one might explore }whether the size of clusters or cell-to-cell adhesion~\cite{merks2008contact,ramos2018capillary} will affect tube diameters, or whether VEGF diffusion and decay rates determine chord length~\cite{ambrosi2004cell,ambrosi2005review,gamba2003percolation,serini2003modeling}, and how these are affected by matrix stiffness.

\appendix
\section{Parameter values}\label{sec:parameters}
As already indicated in Section \ref{sec:met:nondim}, we use $L=0.1$ cm as characteristic length scale, in accordance with previous vasculogenesis works~\cite{manoussaki2003mechanochemical,serini2003modeling} and for easy visual comparison with the experimental results reported by Blatchley and coworkers~\cite{blatchley2019hypoxia}. We then take reference time scale $\tau:=L^2/D$, where $D$ is a characteristic diffusion coefficient $D\sim 10^{-6}$ cm$^2$s$^{-1}$~\cite{bray2000cell}, resulting in a reference time scale $\tau=10^4$s. 

\paragraph*{Endothelial progenitor cells (or endothelial cells)}

The reference cell density is chosen to be $N:=n_M=\vartheta_1^{-1}$ and we take $\vartheta_1$ to be the average volume occupied by an EC. 
Rubin and coworkers~\cite{rubin1989endothelial} measured the average ECs volume during different phases of the cell cycle, registering values in the range of $800-1800\,\mu$m$^3$, with a predominance of measurements around $1000\,\mu$m$^3$. Hence we take $\vartheta_1=10^{-9}$~cm$^3$/cell, corresponding to an average cell diameter of approximately $a= 10^{-3}$ cm. 
Measures and estimates of the diffusion coefficient of ECs fall in the range $10^{-6}$-$10^{-12}$ cm$^2$s$^{-1}$~\cite{ambrosi2005review}, so we take $D_n = 10^{-9}$  cm$^2$s$^{-1}$.
We consider the chemotactic sensitivity coefficient estimated by Jain and Jackson~\cite{jain2013hybrid}, corresponding to $\chi = 1.4\times 10^{-7}$ cm$^5$ng$^{-1}$s$^{-1}$.
Sen \textit{et al.}~\cite{sen2009matrix} estimated a maximum cell surface radius, upon morphological changes to better adhere to the underlying gel, of about $50\,\mu$m. Given cell-to-cell and cell-to-matrix adhesion occurs via adhesion molecules on the cell surface, we take the sensing radius $R=0.5\times10^{-2}$ cm. 
While the nonlocal term introduced in equations \eqref{eq:a:def:1d} and \eqref{eq:a:def:2d} allows us to well consider cell-to-cell and cell-to-matrix adhesion dynamics at tissue level, these are the result of smaller scale dynamics between adhesion molecules and receptors on cell surfaces -- \textit{vid.} for instance~\cite{albelda1990integrins,berrier2007cell,garrod1993cell} -- of which our modelling choice is a simplification. As a result, good estimates for cell-to-cell and cell-to-matrix coefficients $S_{nn}$ and $S_{n\rho}$ are currently lacking. We therefore consider nondimensional values chosen in~\cite{gerisch2008mathematical} for our baseline parameter set which correspond to $S_{nn}=10^{-16}$  cm$^5$ s$^{-1}$ and  $S_{n\rho}=10^{-6}$ cm$^2$ nM$^{-1}$ s$^{-1}$  respectively, acknowledging that model fitting to experimental data is required. Kinev and coworkers~\cite{kinev2013endothelial} reported dubling times of non-irradiated ECFCs -- the same class of EPCs employed by Blatchley \textit{et al.}~\cite{blatchley2019hypoxia} -- of about 19.5 hours, estimated from measured growth rates assuming exponential growth. Following their calculations, this corresponds to proliferation rates of about $p= 10^{-5}$ s$^{-1}$.

\paragraph*{Extracellular matrix}
We use a reference matrix density of $P=10^{-1}$ nM~\cite{anderson2005hybrid,anderson2000mathematical,terranova1985human} and define the parameter $\vartheta_2:=P^{-1}$. 
We take the ECM degradation rate by MMPs proposed by Kim and Friedman~\cite{kim2010interaction}, \textit{i.e.} $\gamma=9\times 10^5$~cm$^3$g$^{-1}$s$^{-1}$.
ECM remodelling is a complex process involving a variety of cells and molecules~\cite{chang2016restructuring,daley2008extracellular,lefebvre1999developmental,streuli1999extracellular}, so the remodelling term introduced in equation \eqref{eq:rho} is an oversimplification of the underlying dynamics. Therefore the lack of experimental values or estimates for the remodelling rate $\mu$ is not surprising, and we take the nondimensional value $0.2$ as similarly considered in~\cite{deakin2013mathematical,domschke2014mathematical,gerisch2008mathematical} for our baseline parameter set. Under the defined nondimensional parameters this corresponds to the dimensional value $\mu = 0.2 \times 10^{-5}$ nM s$^{-1}$.  

\paragraph*{Matrix degrading enzyme (MMP)}
Blatchley \textit{et al.}~\cite{blatchley2019hypoxia} reported concentrations of MMP-1 in the range $1-100\,\mu$g ml$^{-1}$, so we take the intermediate concentration as reference MMP density, \textit{i.e.} $M=10\,\mu$g cm$^{-3}$.
We let the diffusion coefficient for the MMP be given by $D_m = 8\times 10^{-9}$ cm$^2$s$^{-1}$, which was experimentally determined by Saffarian and coworkers~\cite{saffarian2004interstitial}, although diffusion rates have been observed in the range $10^{-10}-10^{-8}$ cm$^2$s$^{-1}$~\cite{collier2011diffusion,kumar2018mmp}. 
While MMPs secretion rates by cells have been reported in a variety of works, these fail to provide parameter values in the appropriate unit of $\alpha_m$ in this model. For instance in~\cite{kumar2018mmp} secretion rates considered varying between $0.01-0.5$ s$^{-1}$ but the associated MMP concentration is unspecified, while in~\cite{ruggiero2017mathematical} the estimated rate is of $5.75\times10^{-10}$ and $4.44\times10^{-10}$ g cm$^{-3}$ secreted by stromal cells and macrophages respectively, without an indication of the considered time frame. We therefore here consider that molecular dynamics are generally faster than cellular ones and therefore the observed average MMP concentration satisfies \eqref{eq:m} at equilibrium, \textit{i.e.} $m = \alpha_m n / \lambda_m$. Under the chosen reference values $N$, $M$ and $\lambda_m$, this corresponds to an MMP production rate of $\alpha_m = 0.5 \times 10^{-12} \, \mu$g s$^{-1}$ per cell. Note that the resulting nondimensional parameter value is close to those chosen in previous mathematical models -- \textit{vid.} for instance~\cite{anderson2000mathematical,deakin2013mathematical,domschke2014mathematical,gerisch2008mathematical}. Nonetheless, we will also consider higher values of $\alpha_m$, since Deem and Cook-Mills~\cite{deem2004vascular} have reported up to a 4-fold increase in MMP production in the presence of reactive oxygen species, and in addition MMP production levels have been shown to be significantly upregulated in human cancers~\cite{shiomi2003mt1}.
Finally, Kim and Friedman~\cite{kim2010interaction} estimated -- from its half life -- the decay rate of MMP $\lambda_m=5\times 10^{-5}$ s$^{-1}$. 

\paragraph*{Chemoattractant (VEGF)}
We take the reference VEGF density to be $C=20$ ng cm$^{-3}$, in the range of values generally considered in \textit{in vitro} set ups -- \textit{vid.} for instance~\cite{hanjaya2009vascular,lee2007autocrine,serini2003modeling}.
Measures and estimates of the diffusion coefficient of the chemoattractant, usually identified as VEGF, are in the range $10^{-6}-10^{-9}$ cm$^2$s$^{-1}$ ~\cite{ambrosi2005review,gamba2003percolation,merks2008contact,miura2009vitro,serini2003modeling,singh2015role}, so we take $D_c=10^{-7}$ cm$^2$s$^{-1}$. 
Yen and coworkers~\cite{yen2011two} reported a VEGF secretion rate of 0.068 molecules cell$^{-1}$s$^{-1}$, which in combination with Avogadro's number ($6.022\times 10^{23}$ molecules per mole) and a molecular weight of 45kDa~\cite{yen2011two} results in a VEGF production rate of about $\alpha_c=0.5\times10^{-11}$ ng s$^{-1}$ per cell.
Serini \textit{et al.}~\cite{serini2003modeling} reported the half-life of VEGF-A to be approximately 64 minutes, corresponding to a decay rate of about $\lambda_c = 2.7 \times 10^{-4}$ s$^{-1}$, in line with the values chosen in~\cite{merks2008contact,singh2015role}.

\bibliographystyle{siam}
\bibliography{Vasculogenesis}
\end{document}


\title{A nonlocal partial differential equation model of endothelial progenitor cell cluster formation during the early stages of vasculogenesis\footnote{This paper was submitted to the \emph{Journal of Theoretical Biology}.}
\\
Supplementary Information}
\author{Chiara Villa\footnote{Corresponding author affiliation and email: Chiara Villa, School of Mathematics and Statistics, University of St Andrews, St Andrews KY16 9SS (UK), cv23@st-andrews.ac.uk} \and
Alf Gerisch  \and
Mark A. J. Chaplain              
}
\date{}
\maketitle

\renewcommand{\theequation}{S\arabic{equation}}
\renewcommand{\thesection}{S\arabic{section}} 
\renewcommand{\thefigure}{S\arabic{figure}}
In this document (supplementary information) we report supplementary material to the contents of the above named publication (main publication). In particular we include detailed linear stability analysis calculations in Section~\ref{sec:lsa}, provide supplementary figures in Section~\ref{sec:fig}, and explain a detail related to the numerical treatment of the zero-flux boundary condition of the nonlocal term, in Section~\ref{sec:num} of this document.

\section{Linear stability analysis}\label{sec:lsa}

We here conduct a linear stability analysis (LSA) on the model equations. To understand the role played by matrix degradation on the evolution of the solution of our model -- and therefore its role in the early stages of cluster based vasculogenesis -- particular emphasis will be put on the differences between the case in which $\gamma>0$ and $\gamma=0$.
First we look at the full nondimensional system (15) with saturation effects, that is the one making use of definition (16) for the function $f(n,\rho)$. We will also consider the simplified version of the model under definition (17) instead, corresponding to the absence of saturation effects, for a better understanding of the results of the LSA. The analysis is carried out for the 1D problem for simplicity, but in section \ref{sec:lsa:2d} we will briefly consider the corresponding 2D problem. 

\subsection{Spatially homogeneous steady states}\label{sec:lsa:ss}

The spatially homogeneous steady states of the system (15) ${\bf \bar{v}} = (\bar{n},\bar{\rho},\bar{m},\bar{c})^\intercal$ satisfy

\begin{equation}\label{eq:system:ss}
\begin{cases}
&p\bar{n}(1-\bar{n}-\bar{\rho})=0 \\[5pt]
&\mu (1-\bar{n}-\bar{\rho})_+ -\gamma\bar{\rho} \bar{m}=0 \\[5pt]
&\alpha_m \bar{n} -\lambda_m \bar{m} =0 \\[5pt]
&\alpha_c \bar{n} -\lambda_c \bar{c}=0
\end{cases} \;,
\end{equation}

\noindent giving either ${\bf \bar{v}} = (0,1,0,0)^\intercal$ or ${\bf \bar{v}} = (1,0,\alpha_m / \lambda_m,\alpha_c / \lambda_c)^\intercal$. Each spatially homogeneous steady state is therefore in the form

\begin{equation}\label{eq:ss}
{\bf \bar{v}} = \bigg(\bar{n},1-\bar{n},\frac{\alpha_m}{\lambda_m}\bar{n},\frac{\alpha_c}{\lambda_c}\bar{n}\bigg)^\intercal
\end{equation}
 
\noindent with either $\bar{n}=0$ or $\bar{n}=1$. The first one corresponds to the absence of cells, with the whole volume occupied by the ECM. On the other hand, the second one corresponds to the case in which the cells have completely degraded the ECM and are occupying the whole volume. Note that in absence of matrix degradation, \textit{i.e.} for $\gamma=0$, the spatially homogeneous steady states are still in the form \eqref{eq:ss}, this time with $0\leq\bar{n}\leq 1$. We therefore have, in addition to the two states already described, infinitely many steady states in which both cells and ECM are present, filling up the volume. These additional steady states might be referred throughout this section as ``intermediate'' steady states, as they correspond to intermediate values of $\bar{n}$, namely $0<\bar{n}<1$. \\

\noindent \textit{Remark \red{1}:} We are going to assume that $n+\rho\leq 1$ throughout the rest of the section. That is indeed satisfied by the steady states and we justify it through the biological argument that random perturbations to these states are likely to arise naturally if there is space available, while changes that would imply $n+\rho>1$ are much less likely as they would require high energy expenses to oppose a high pressure environment. 
Mathematically this allows us to avoid complications that arise when introducing perturbations in the typical ansatz (\textit{vid.} below) due to definitions (3) and (10). In fact, under this assumption the term $(1-n-\rho)_+ = (1-n-\rho)\geq0$ for all perturbations allowed. \\
\noindent \textit{Remark 2:} During the linear stability analysis reported in the following sections, we are going to use the fact that $\bar{n}+\bar{\rho}=1$ and $\gamma\bar{\rho}\bar{n} =\gamma\bar{\rho}\bar{m} = \gamma\bar{\rho}\bar{c} = 0$ for all spatially homogeneous steady states, as concluded above, to further simplify calculations.\\
\noindent \textit{Remark 3:} In the following we are going to consider biologically significant steady states, that is we are going to only investigate the stability of the steady states with $n>0$. The stability of the cell-free steady state $n=0$ could be investigated by considering the dynamics along a center manifold about the steady state, as linear stability analysis calculations reported below are inconclusive for this case. Such investigation, however, goes beyond the scope of this study and we leave the mathematical details to the interested reader, referring them to~\cite{carr2012applications,guckenheimer2013nonlinear}. However, simple numerical simulations indicate that this cell-free steady state is unstable.

\subsection{Stability under spatially homogeneous perturbations}\label{sec:lsa:pert:t}

\noindent Introducing a small spatially homogeneous perturbation ${\bf v} = {\bf \bar{v}} + {\bf \tilde{v}}(t)$, with $|{\bf \tilde{v}}|\ll 1$, in (15) and linearising leads to the following system for the perturbation ${\bf \tilde{v}}(t)$:

\begin{equation}\label{eq:pert:t}
\begin{cases}
&\dt \tilde{n} = -p\bar{n}(\tilde{n}+\tilde{\rho}) \\[5pt]
&\dt \tilde{\rho} = -\gamma(\bar{\rho} \tilde{m} +\bar{m} \tilde{\rho})  - \mu (\tilde{n}+\tilde{\rho})  \\[5pt]
& \dt \tilde{m} = \alpha_m \tilde{n} -\lambda_m \tilde{m} \\[5pt]
& \dt \tilde{c} = \alpha_c \tilde{n} -\lambda_c \tilde{c}
\end{cases} \;.
\end{equation}

\noindent Assuming small perturbations in the form $\tilde{n}, \,\tilde{\rho}, \,\tilde{m}, \,\tilde{c} \,\propto \exp{(\sigma t)}$,  the system \eqref{eq:pert:t} can be rewritten as

\begin{equation}\label{eq:pert:t2}
{\bf M}{\bf \tilde{v}}={\bf 0} \,, \quad \text{with} \quad {\bf M}= 
\begin{pmatrix}
\sigma +p\bar{n} &p\bar{n}&0&0\\
\mu&\sigma+\gamma\bar{m}+\mu&\gamma\bar{\rho} &0\\
-\alpha_m&0&\sigma+\lambda_m&0\\
-\alpha_c&0&0&\sigma+\lambda_c\\
\end{pmatrix} \;.
\end{equation}

\noindent For a non-trivial solution we require $\det{{\bf M}}=0$, leading to the dispersion relation

\begin{equation}\label{eq:disprel:t}
(\sigma +\lambda_m)(\sigma+\lambda_c)\Big[ \sigma^2 +\sigma\Big( p\bar{n}+\gamma\bar{m}+\mu\Big) +\gamma p\bar{n}\bar{m}\Big] = 0 \;,
\end{equation}

\noindent which can only be satisfied for Re$(\sigma)<0$, indicating the biologically significant spatially homogeneous steady state is stable under spatially homogeneous perturbations. 
On the other hand, in the absence of matrix degradation, \textit{i.e.} for $\gamma=0$, we obtain the corresponding version of the dispersion relation \eqref{eq:disprel:t}, which can also be satisfied for $\sigma=0$, thus linear stability analysis is inconclusive. In this case, under the assumption specified in Remark 1, the additional spatially homogeneous steady states are actually unstable to spatially homogeneous perturbations, as can be quickly verified numerically, and the system will converge to a nearby steady state depending on the initial conditions of $n$ and $\rho$, as well as the values of parameters $p$ and $\mu$.

\subsection{Stability under spatially inhomogeneous perturbations}\label{sec:lsa:pert:tx}

\noindent Let us now introduce spatially inhomogeneous perturbations ${\bf v} = {\bf \bar{v}} + {\bf \tilde{v}}(t,x)$, with $|{\bf \tilde{v}}|\ll 1$. After linearisation system (15) gives

\begin{equation}\label{eq:pert:tx}
\begin{cases}
&\dt \tilde{n} = D_n\Delta\tilde{n} -\bar{n}\,\nabla \cdot \mathcal{A}[ {\bf \bar{v}}+{\bf \tilde{v}}(t,\cdot)] -p\bar{n}(\tilde{n}+\tilde{\rho}) \\[5pt]
&\dt \tilde{\rho} = -\gamma(\bar{\rho} \tilde{m} +\bar{m} \tilde{\rho})  - \mu (\tilde{n}+\tilde{\rho}) \\[5pt]
& \dt \tilde{m} = D_m\Delta\tilde{m} +\alpha_m \tilde{n} -\lambda_m \tilde{m} \\[5pt]
& \dt \tilde{c} = D_c\Delta\tilde{c} +\alpha_c \tilde{n} -\lambda_c \tilde{c}
\end{cases} \;.
\end{equation}

\noindent We immediately notice that, as all steady states satisfy volume filling conditions, chemotaxis in presence of saturating effects does not play a role in the dynamics of small perturbations from any of these states. \\

\noindent Then, after linearisation, for the 1D problem $\mathcal{A}[ {\bf \bar{v}}+{\bf \tilde{v}}(t,\cdot)]$ in equation \eqref{eq:pert:tx}$_1$ becomes 

\begin{equation}\label{eq:a:pert:1d}
\mathcal{A}[ {\bf \bar{v}}+{\bf \tilde{v}}(t,\cdot)](x) =- \frac{1}{R} (S_{nn}\bar{n}+S_{n\rho}\bar{\rho}) \int_0^R \sum_{j=0}^1  \eta(j)\, \Gamma(r)\big(\tilde{n}(t,x+r\eta(j))+\tilde{\rho}(t,x+r\eta(j))\big)\, dr \;
\end{equation}

\noindent with $\Gamma(r)$ given by the corresponding definition in (8). \\

\noindent Assuming small perturbations in the form $\tilde{n}, \,\tilde{\rho}, \,\tilde{m}, \,\tilde{c} \,\propto \exp{(\sigma t +ikx)}$, note first of all that \eqref{eq:a:pert:1d} can be rewritten as

\begin{equation}\label{eq:a:pert2:1d}
\begin{split}
&\mathcal{A}[ {\bf \bar{v}}+{\bf \tilde{v}}(t,\cdot)](x) = -\frac{1}{R} (S_{nn}\bar{n}+S_{n\rho}\bar{\rho}) \big(\tilde{n}(x)+\tilde{\rho}(x) \big) \int_0^R\Gamma(r)\big(\exp{(ikr)}-\exp{(-ikr)}\big) \\
&\white{\mathcal{A}\{ {\bf \bar{v}}+{\bf \tilde{v}}(t,\cdot)\}(x)} = -\frac{1}{R} (S_{nn}\bar{n}+S_{n\rho}\bar{\rho}) \big(\tilde{n}(x)+\tilde{\rho}(x) \big) \frac{2i}{R}\int_0^R\Big( 1-\frac{r}{R}\Big)\sin(kr) \, dr  \\
&\white{\mathcal{A}\{ {\bf \bar{v}}+{\bf \tilde{v}}(t,\cdot)\}(x)} = -\frac{2i}{R^2k} (S_{nn}\bar{n}+S_{n\rho}\bar{\rho}) \big(\tilde{n}(x)+\tilde{\rho}(x) \big) \Big(1-\frac{1}{Rk}\sin(Rk)\Big) \,. 
\end{split}
\end{equation}

\noindent We will make use of the notation

\begin{equation}\label{eq:w1}
\text{w}_1(k) := \Big(1-\frac{1}{Rk}\sin(Rk)\Big) \,,
\end{equation}

\noindent for which we have that $\text{w}_1(k)\geq 0$ for all $k\in\mathbb{R}$. Then, for small perturbations in the form introduced above, the system \eqref{eq:pert:tx} can be rewritten as

\begin{equation}\label{eq:pert:tx2}
{\bf M}{\bf \tilde{v}}={\bf 0} \,, \; \text{with} \; {\bf M}= 
\begin{pmatrix}
\sigma + D_n k^2 +A(k)+p\bar{n} & A(k)+p\bar{n}&0&0\\
\mu&\sigma+\gamma\bar{m}+\mu&\gamma\bar{\rho} &0\\
-\alpha_m&0&\sigma+D_m k^2+\lambda_m&0\\
-\alpha_c&0&0&\sigma+D_c k^2+\lambda_c\\
\end{pmatrix} \,,
\end{equation}

\noindent where 

\begin{equation}\label{eq:ak:1d}
A(k) = \frac{2\bar{n}}{R^2} (S_{nn}\bar{n}+S_{n\rho}\bar{\rho}) \text{w}_1(k) \,,
\end{equation}

\noindent with $\text{w}_1(k)$ defined in \eqref{eq:w1}. This immediately implies $A(k)\geq 0 $ for all steady states, those with and without matrix degradation. The dispersion relation $\sigma(k^2)$ can then be obtained by imposing $\det{{\bf M}}=0$ for a non-trivial solution to \eqref{eq:pert:tx2}, and in order for the steady states to be unstable to spatially inhomogeneous perturbations -- and patterns to arise -- we require Re$\big(\sigma(k^2)\big)>0$ for some $k^2\in\mathbb{R}$. This, in its more general form, is given by

\begin{equation}\label{eq:disprel:tx}
(\sigma+D_ck^2+\lambda_c)\big(p\bar{n}+A(k) \big)\big[ \sigma+\gamma\bar{m} + (\sigma + D_n k^2) (\sigma+\gamma\bar{m}+\mu) \big] = 0 \;.
\end{equation}

\noindent It is clear from \eqref{eq:disprel:tx} that Re$\big(\sigma(k^2)\big)<0$ for all $k^2$. Therefore all steady states considered are also stable to spatially inhomogeneous perturbations. We claim this to be due to saturation effects introduced in (3), while we may still expect chemotaxis and cell-to-cell or cell-to-matrix adhesion to play an important role for cell aggregation when the initially perturbed states are far from being volume filling. Let us explore the matter further in the next section.

\subsection{Considerations in absence of saturation effects}\label{sec:lsa:nosat}

We here consider the pattern formation potential of the model in absence of saturation effects, that is we make use of definition (17) for the function $f(n,\rho)$, which influences both the chemotactic sensitivity of cells and their adhesion velocity. All results and considerations made in sections \ref{sec:lsa:ss} and \ref{sec:lsa:pert:t} on spatially homogeneous steady states and their stability under spatially homogeneous perturbations still hold. Introducing spatially inhomogeneous perturbations as in section \ref{sec:lsa:pert:tx} leads to the following linearised system alternative to \eqref{eq:pert:tx}:

\begin{equation}\label{eq:pert:tx:nosat}
\begin{cases}
&\dt \tilde{n} = D_n\Delta\tilde{n} -\chi\bar{n}\Delta\tilde{c} -\bar{n}\,\nabla \cdot \mathcal{A}[ {\bf \bar{v}}+{\bf \tilde{v}}(t,\cdot)] -p\bar{n}(\tilde{n}+\tilde{\rho}) \\[5pt]
&\dt \tilde{\rho} = -\gamma(\bar{\rho} \tilde{m} +\bar{m} \tilde{\rho})  - \mu (\tilde{n}+\tilde{\rho}) \\[5pt]
& \dt \tilde{m} = D_m\Delta\tilde{m} +\alpha_m \tilde{n} -\lambda_m \tilde{m} \\[5pt]
& \dt \tilde{c} = D_c\Delta\tilde{c} +\alpha_c \tilde{n} -\lambda_c \tilde{c}
\end{cases} \;,
\end{equation}

\noindent in which, for the 1D problem, $\mathcal{A}[ {\bf \bar{v}}+{\bf \tilde{v}}(t,\cdot)]$ is given by

\begin{equation}\label{eq:a:pert:1d:nosat}
\mathcal{A}[{\bf \bar{v}}+{\bf \tilde{v}}(t,\cdot)](x) = \frac{1}{R} \int_0^R \sum_{j=0}^1  \eta(j)\, \Gamma(r)\big(S_{nn}\tilde{n}(t,x+r\eta(j))+S_{n\rho}\tilde{\rho}(t,x+r\eta(j))\big)\, dr \;
\end{equation}

\noindent with $\Gamma(r)$ given by the corresponding definition in (8). Notice that under these modelling assumptions -- namely no saturation effects included in the adhesion velocity -- we do not need to assume $n+\rho\leq1$ for analytical tractability, and therefore have no crafty restrictions on the perturbations introduced. On the contrary, the contents of Remark 2 and Remark 3 still hold. Then, assuming small perturbations in the form $\tilde{n}, \,\tilde{\rho}, \,\tilde{m}, \,\tilde{c} \,\propto \exp{(\sigma t +ikx)}$, \eqref{eq:a:pert:1d:nosat} can be rewritten as

\begin{equation}\label{eq:a:pert2:1d:nosat}
\begin{split}
&\mathcal{A}[ {\bf \bar{v}}+{\bf \tilde{v}}(t,\cdot)](x) = \frac{1}{R} \big(S_{nn}\tilde{n}(x)+S_{n\rho}\tilde{\rho}(x)\big)  \int_0^R\Gamma(r)\big(\exp{(ikr)}-\exp{(-ikr)}\big) \\
&\white{\mathcal{A}\{ {\bf \bar{v}}+{\bf \tilde{v}}(t,\cdot)\}(x)} = \frac{1}{R} \big(S_{nn}\tilde{n}(x)+S_{n\rho}\tilde{\rho}(x)\big)  \frac{2i}{R}\int_0^R\Big( 1-\frac{r}{R}\Big)\sin(kr) \, dr  \\
&\white{\mathcal{A}\{ {\bf \bar{v}}+{\bf \tilde{v}}(t,\cdot)\}(x)} = \frac{2i}{R^2k}\big(S_{nn}\tilde{n}(x)+S_{n\rho}\tilde{\rho}(x)\big) \text{w}_1(k) \,, 
\end{split}
\end{equation}

\noindent with $\text{w}_1(k)\geq0$ defined in \eqref{eq:w1}. We can then rewrite \eqref{eq:pert:tx:nosat} as ${\bf M}{\bf \tilde{v}}={\bf 0}$, with ${\bf M}$ given by

\begin{equation}\label{eq:pert:tx2:nosat}
{\bf M}= 
\begin{pmatrix}
\sigma + D_n k^2 -A_{n}(k)+p\bar{n} & -A_{\rho}(k)+p\bar{n}&0&-\chi\bar{n}k^2\\
\mu&\sigma+\gamma\bar{m}+\mu&\gamma\bar{\rho} &0\\
-\alpha_m&0&\sigma+D_m k^2+\lambda_m&0\\
-\alpha_c&0&0&\sigma+D_c k^2+\lambda_c\\
\end{pmatrix} \;,
\end{equation}

\noindent where 

\begin{equation}\label{eq:ak:1d:nosat}
A_{n}(k) = \frac{2\bar{n}}{R^2} S_{nn}\text{w}_1(k) \qquad \text{and} \qquad  A_{\rho}(k) = \frac{2\bar{n}}{R^2} S_{n\rho}\text{w}_1(k) \,.
\end{equation}

\noindent We have both $A_{n}(k)\geq 0 $ and $A_{\rho}(k)\geq 0 $ for all $k\in\mathbb{R}$. Comparing these to \eqref{eq:pert:tx2} and \eqref{eq:ak:1d}, we see that in absence of saturation effects the contributions from cell-to-cell and cell-to-matrix adhesion are decoupled and both give a negative contribution to ${\bf M}$. Furthermore, the contribution from chemotaxis does not vanish for any of the biologically significant steady states (\textit{i.e.} $\bar{n}>0$). From this system the dispersion relation $\sigma(k^2)$ satisfies

\begin{equation}\label{eq:disprel:tx:nosat}
\begin{split}
(\sigma+D_mk^2+\lambda_m)\bigg\{ \, \bigg[ &(\sigma + D_n k^2)\Big(\sigma+\gamma\bar{m}+\mu\Big) + \big(p\bar{n}-A_{n}(k)\big) \Big(\sigma +\gamma\bar{m}\Big)  \\
&+ \mu\big(A_{\rho}(k)-A_{n}(k)\big) \bigg](\sigma+D_ck^2+\lambda_c) - \alpha_c\chi\bar{n}k^2(\sigma+\gamma\bar{m}+\mu) \, \bigg\}= 0 \;.
\end{split}
\end{equation}

\noindent 
We now have that any biologically significant steady state may be unstable to spatially inhomogeneous perturbations for strong enough cell-to-cell adhesion and/or chemotaxis, \textit{i.e.} for large enough $S_{nn}$ and $\alpha_c$ and/or $\chi$. In addition, the magnitude of cell-to-cell adhesion $A_n(k)$ in \eqref{eq:ak:1d:nosat} will be larger for smaller values of the sensing radius $R$ and larger values of $\bar{n}$, and the contribution from chemotactic movement will also increase for larger values of $\bar{n}$. Therefore in absence of saturation effects steady states corresponding to $\bar{n}>0$ may be unstable to spatially inhomogeneous perturbations and patterns may form. This suggests that perturbed initial conditions far from being volume filling, so that saturation effects do not play a big role, might result in cell aggregation thanks to cell-to-cell adhesion and chemotaxis, as long as the initial cell density -- interpreted as a small perturbation from a spatially homogeneous steady state -- is large enough for these dynamics to play a significant role.

\subsection{Extension of LSA results to the 2D problem}\label{sec:lsa:2d}

We here consider how the results obtained so far change in the 2D problem. Conclusions drawn in sections \ref{sec:lsa:ss} and \ref{sec:lsa:pert:t} remain unchanged. 

\paragraph*{Considerations in presence of saturation effects.}
When introducing spatially inhomogeneous perturbations ${\bf v} = {\bf \bar{v}} + {\bf \tilde{v}}(t,{\bf x})$, after linearisation $\mathcal{A}[ {\bf \bar{v}}+{\bf \tilde{v}}(t,\cdot)]({\bf x}) $ for the 2D problem is given by

\begin{equation}\label{eq:a:pert:2d}
\mathcal{A}[ {\bf \bar{v}}+{\bf \tilde{v}}(t,\cdot)]({\bf x}) = -\frac{1}{R} (S_{nn}\bar{n}+S_{n\rho}\bar{\rho}) \int_0^R r \int_0^{2\pi}  {\bm{\eta}}(\theta) \, \Gamma(r) \big(\tilde{n}(t,{\bf x}+r{\bm{\eta}}(\theta))+\tilde{\rho}(t,{\bf x}+r{\bm{\eta}}(\theta))\big)\, d\theta \,dr \;
\end{equation}

\noindent with $\Gamma(r)$ given by the corresponding definition in (8). Assuming small perturbations in the form $\tilde{n}, \,\tilde{\rho}, \,\tilde{m}, \,\tilde{c} \,\propto \exp{(\sigma t +i{\bf k}\cdot {\bf x})}$, with ${\bf k}=(k_1,k_2)^\intercal \in \mathbb{R}^2$, then \eqref{eq:a:pert:2d} can be rewritten as

\begin{equation}\label{eq:a:pert2:2d}
\begin{split}
&\mathcal{A}[{\bf \bar{v}}+{\bf \tilde{v}}(t,\cdot)]({\bf x}) = -\frac{1}{R} (S_{nn}\bar{n}+S_{n\rho}\bar{\rho}) \big(\tilde{n}({\bf x})+\tilde{\rho}({\bf x}) \big) \int_0^R r \int_0^{2\pi}  {\bm{\eta}}(\theta) \, \Gamma(r) \, \exp{(ir {\bf k}\cdot {\bm{\eta}}(\theta))}\, d\theta \,dr \\
&\white{\mathcal{A}\{ {\bf \bar{v}}+{\bf \tilde{v}}(t,\cdot)\}({\bf x})} = -\frac{1}{R} (S_{nn}\bar{n}+S_{n\rho}\bar{\rho}) \big(\tilde{n}({\bf x})+\tilde{\rho}({\bf x}) \big) \int_0^{2\pi}  {\bm{\eta}}(\theta) \int_0^R r\, \Gamma(r) \, \exp{(ir {\bf k}\cdot {\bm{\eta}}(\theta))}\, dr \, d\theta \\
&\white{\mathcal{A}\{ {\bf \bar{v}}+{\bf \tilde{v}}(t,\cdot)\}({\bf x})} = -\frac{1}{R} (S_{nn}\bar{n}+S_{n\rho}\bar{\rho}) \big(\tilde{n}({\bf x})+\tilde{\rho}({\bf x}) \big) \frac{3}{\pi R^2} \int_0^{2\pi}  {\bm{\eta}}(\theta) \int_0^R r\, \Big( 1-\frac{r}{R}\Big) \, \exp{(ir {\bf k}\cdot {\bm{\eta}}(\theta))}\, dr \, d\theta \\
&\white{\mathcal{A}\{ {\bf \bar{v}}+{\bf \tilde{v}}(t,\cdot)\}({\bf x})} = \frac{3}{\pi R^3} (S_{nn}\bar{n}+S_{n\rho}\bar{\rho}) \big(\tilde{n}({\bf x})+\tilde{\rho}({\bf x}) \big) \Big[ \Theta_I ({\bf k})+ i \, \Theta_R({\bf k})  \Big]
\end{split}
\end{equation}

\noindent where we have defined, with notation that will become more intuitive in a few steps, 

\begin{align}
& \Theta_R({\bf k}) = \int_0^{2\pi} \frac{{\bm{\eta}}(\theta)}{\big({\bf k}\cdot {\bm{\eta}}(\theta)\big)^2} \bigg[ \sin\big( R\, {\bf k}\cdot {\bm{\eta}}(\theta) \big) + \frac{2}{R\, {\bf k}\cdot {\bm{\eta}}(\theta)}\, \Big( \cos\big( R\, {\bf k}\cdot {\bm{\eta}}(\theta) \big) -1 \Big) \bigg] \, d\theta \;, \label{eq:theta1}\\
& \Theta_I({\bf k}) = \int_0^{2\pi} \frac{{\bm{\eta}}(\theta)}{\big({\bf k}\cdot {\bm{\eta}}(\theta)\big)^2} \bigg[ 1+ \cos\big( R\, {\bf k}\cdot {\bm{\eta}}(\theta) \big) + \frac{2}{R\, {\bf k}\cdot {\bm{\eta}}(\theta)} \, \sin\big( R\, {\bf k}\cdot {\bm{\eta}}(\theta) \big) \bigg] \, d\theta \label{eq:theta2}\;.
\end{align}

\begin{figure}
\centering
\includegraphics[width=0.7\linewidth]{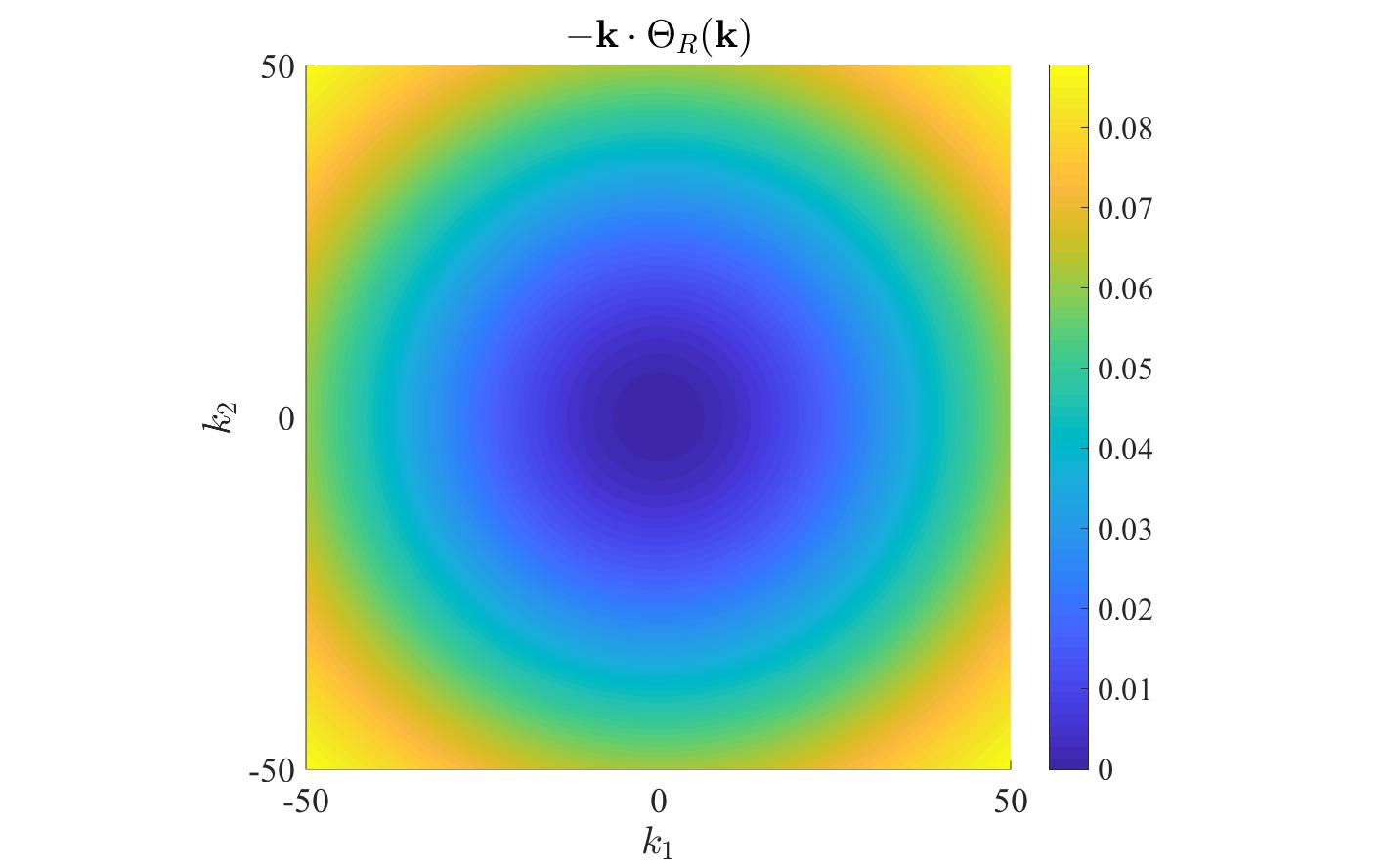}
\caption{\label{fig:LSA}Plot of the quantity $-{\bf k}\cdot\Theta_R({\bf k})$, where ${\bf k} = (k_1,k_2)^\intercal$ and $\Theta_R({\bf k})$ is defined in~\eqref{eq:theta1} for nondimensional $R=0.05$. We clearly have $-{\bf k}\cdot\Theta_R({\bf k})\geq 0$ for all ${\bf k}\in \mathbb{R}^2$, ensuring the real part of $A({\bf k})$ in~\eqref{eq:ak:2d}, as well as those of $A_n({\bf k})$ and $A_\rho({\bf k})$ in~\eqref{eq:ak:2d:nosat} are non-negative. The integral in~\eqref{eq:theta1} was evaluated numerically using the trapezoidal rule for numerical integration. }
\end{figure}

\noindent We eventually obtain the system \eqref{eq:pert:tx2}, in which $k^2=|{\bf k}|^2 = k_1^2 +k_2^2$ and, instead of $A(k)$ as in \eqref{eq:ak:1d}, we have $A({\bf k})$ given by

\begin{equation}\label{eq:ak:2d}
A({\bf k}) = - \frac{3\bar{n}}{\pi R^3} (S_{nn}\bar{n}+S_{n\rho}\bar{\rho}) \, {\bf k}\cdot  \Big[ \Theta_R ({\bf k})- i \, \Theta_I({\bf k})  \Big]  \,,
\end{equation}


\noindent with $\Theta_R({\bf k})$ and $\Theta_I({\bf k})$ defined in \eqref{eq:theta1} and \eqref{eq:theta2}. Despite the complexity of these integrals, we evaluated numerically the quantity $-{\bf k}\cdot \Theta_R ({\bf k})$ -- see Figure~\ref{fig:LSA} -- which allowed us to conclude that, just like in the 1D case, Re$\big(A({\bf k})\big)\geq0$ for all ${\bf k}\in \mathbb{R}^2$. For this problem we therefore obtain the corresponding version of equation \eqref{eq:disprel:tx} for the dispersion relation and reach the same conclusions as in section \ref{sec:lsa:pert:tx} for the full system including saturation effects. \\

\paragraph*{Considerations in absence of saturation effects.} In absence of saturation effects, $\mathcal{A}[ {\bf \bar{v}}+{\bf \tilde{v}}(t,\cdot)]({\bf x})$ is given by the 2D correspondent of equation \eqref{eq:a:pert:1d:nosat}, that is 

\begin{equation}\label{eq:a:pert:2d:nosat}
\mathcal{A}[ {\bf \bar{v}}+{\bf \tilde{v}}(t,\cdot)]({\bf x}) = \frac{1}{R} \int_0^R r \int_0^{2\pi} {\bm{\eta}}(\theta) \, \Gamma(r) \Big( S_{nn}\, \tilde{n}(t,{\bf x}+r{\bm{\eta}}(\theta))+ S_{n\rho}\,\tilde{\rho}(t,{\bf x}+r{\bm{\eta}}(\theta))\Big)\,  d\theta \,dr \,.
\end{equation}

\noindent Then assuming small perturbations in the form $\tilde{n}, \,\tilde{\rho}, \,\tilde{m}, \,\tilde{c} \,\propto \exp{(\sigma t +i{\bf k}\cdot {\bf x})}$, with ${\bf k}=(k_1,k_2)^\intercal \in \mathbb{R}^2$, \eqref{eq:a:pert:2d:nosat} can be rewritten as

\begin{equation}\label{eq:a:pert2:2d:nosat}
\begin{split}
&\mathcal{A}[ {\bf \bar{v}}+{\bf \tilde{v}}(t,\cdot)]({\bf x}) = \frac{1}{R} \big(S_{nn}\tilde{n}({\bf x})+S_{n\rho}\tilde{\rho}({\bf x})\big) \int_0^R r \int_0^{2\pi}  {\bm{\eta}}(\theta) \, \Gamma(r) \, \exp{(ir {\bf k}\cdot {\bm{\eta}}(\theta))}\, d\theta \,dr \\
&\white{\mathcal{A}\{ {\bf \bar{v}}+{\bf \tilde{v}}(t,\cdot)\}({\bf x})} = \frac{1}{R} (\big(S_{nn}\tilde{n}({\bf x})+S_{n\rho}\tilde{\rho}({\bf x})\big) \int_0^{2\pi}  {\bm{\eta}}(\theta) \int_0^R r\, \Gamma(r) \, \exp{(ir {\bf k}\cdot {\bm{\eta}}(\theta))}\, dr \, d\theta \\
&\white{\mathcal{A}\{ {\bf \bar{v}}+{\bf \tilde{v}}(t,\cdot)\}({\bf x})} = \frac{1}{R} \big(S_{nn}\tilde{n}({\bf x})+S_{n\rho}\tilde{\rho}({\bf x})\big) \frac{3}{\pi R^2} \int_0^{2\pi}  {\bm{\eta}}(\theta) \int_0^R r\, \Big( 1-\frac{r}{R}\Big) \, \exp{(ir {\bf k}\cdot {\bm{\eta}}(\theta))}\, dr \, d\theta \\
&\white{\mathcal{A}\{ {\bf \bar{v}}+{\bf \tilde{v}}(t,\cdot)\}({\bf x})} =- \frac{3}{\pi R^3} \big(S_{nn}\tilde{n}({\bf x})+S_{n\rho}\tilde{\rho}({\bf x})\big) \Big[ \Theta_I ({\bf k})+ i \, \Theta_R({\bf k})  \Big] \,,
\end{split}
\end{equation}

\noindent with $\Theta_R({\bf k})$ and $\Theta_I({\bf k})$ defined in \eqref{eq:theta1} and \eqref{eq:theta2}. We eventually end up with a system ${\bf M}{\bf \tilde{v}}={\bf 0}$, with ${\bf M}$ defined as in \eqref{eq:pert:tx2:nosat}, in which again we have $k^2=|{\bf k}|^2 = k_1^2 +k_2^2$  and, instead of $A_n(k)$ and $A_\rho(k)$ as in \eqref{eq:ak:1d:nosat}, $A_{n}({\bf k}) $ and $A_{\rho}({\bf k}) $ defined by

\begin{equation}\label{eq:ak:2d:nosat}
A_{n}({\bf k}) = - \frac{3\bar{n}}{\pi R^3} S_{nn}\,{\bf k}\cdot  \Big[ \Theta_R ({\bf k})- i \, \Theta_I({\bf k})  \Big]   \qquad \text{and} \qquad  A_{\rho}({\bf k}) = - \frac{3\bar{n}}{\pi R^3} S_{n\rho}\,{\bf k}\cdot  \Big[ \Theta_R ({\bf k})- i \, \Theta_I({\bf k})  \Big]  \,.
\end{equation}

\noindent Again, thanks to the numerically evaluated quantity plot in Figure~\ref{fig:LSA}, we have that Re$\big(A_n({\bf k})\big)\geq0$ and Re$\big(A_\rho({\bf k})\big)\geq0$ for all ${\bf k}\in \mathbb{R}^2$, and can therefore reach analogous conclusions to those drawn in section \ref{sec:lsa:nosat} for the corresponding 2D problem.

\section{Supplementary figures}\label{sec:fig}

\begin{figure}
\centering
\includegraphics[width=\textwidth]{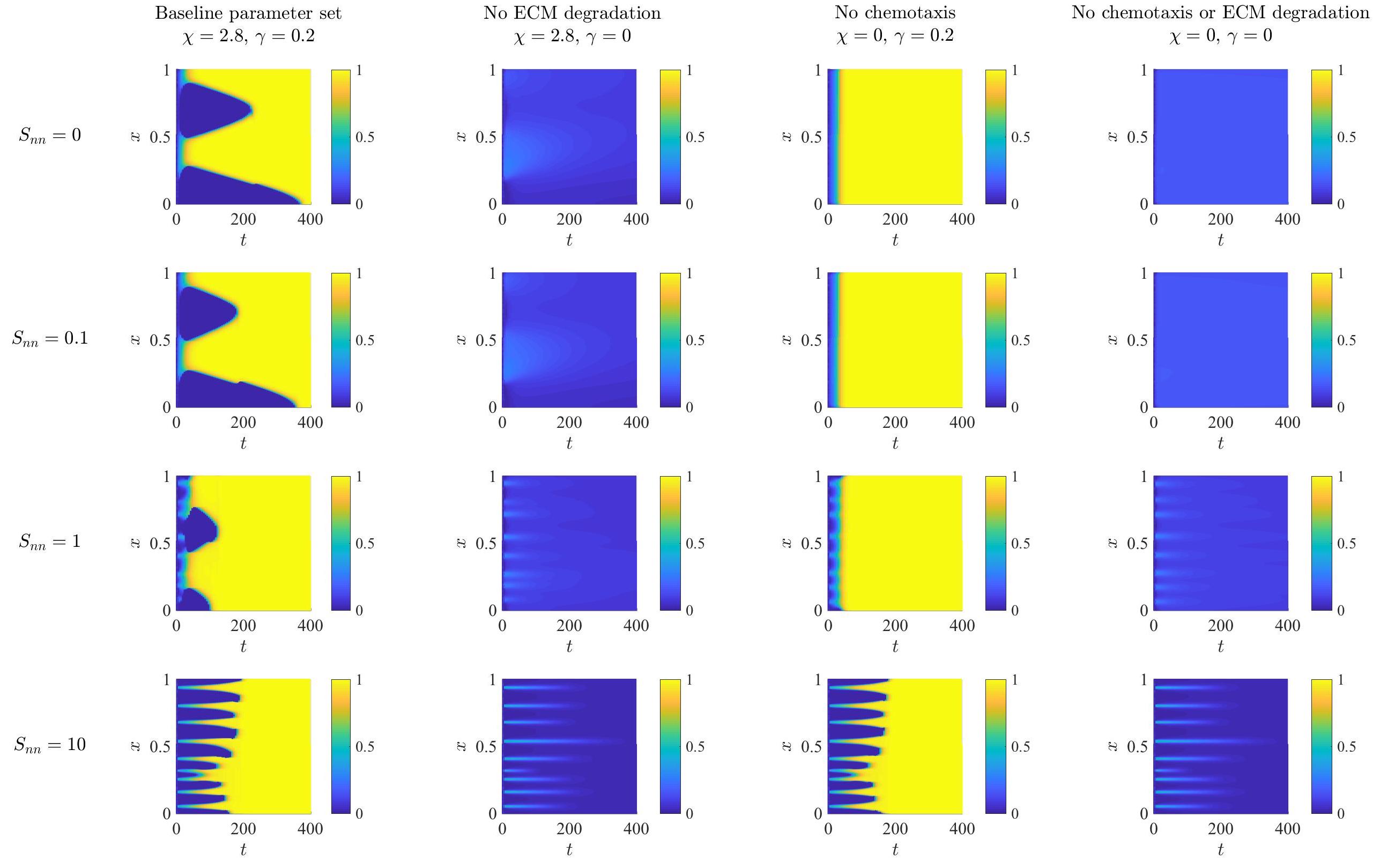}
\caption{\label{Fig:main:extra}\textbf{First row:} Plots of the cell density $n(t,x)$ up to $t=400$ obtained solving the system~(15), together with definitions~(6), (8) and~(16), initial conditions~(13) and~(14)$_1$, complemented with zero-flux boundary conditions, in absence of cell-to-cell adhesion, \textit{i.e} for $S_{nn}=0$: under the baseline parameter set (first column), in absence of matrix degradation, \textit{i.e.} for $\gamma=0$ (second column), in absence of chemotaxis, \textit{i.e.} $\chi=0$ (third column), and in absence of both chemotaxis and matrix degradation, \textit{i.e.} $\chi=\gamma=0$ (fourth column). \textbf{Second, third and fourth rows:} Same as first row but in the presence of cell-to-cell adhesion, with $S_{nn}=0.1$ (second row), $S_{nn}=1$ (third row) and $S_{nn}=10$ (fourth row) respectively. }
\end{figure}

Figure~\ref{Fig:main:extra} contains the plots up to $t=400$ of the same solutions plot in Figure~6 in the main publication, in order to show long-time asymptotics of such solutions.

\begin{figure}
\begin{minipage}[c][9cm][t]{.5\textwidth}
  \vspace*{\fill}
  \centering
   \includegraphics[width=\linewidth]{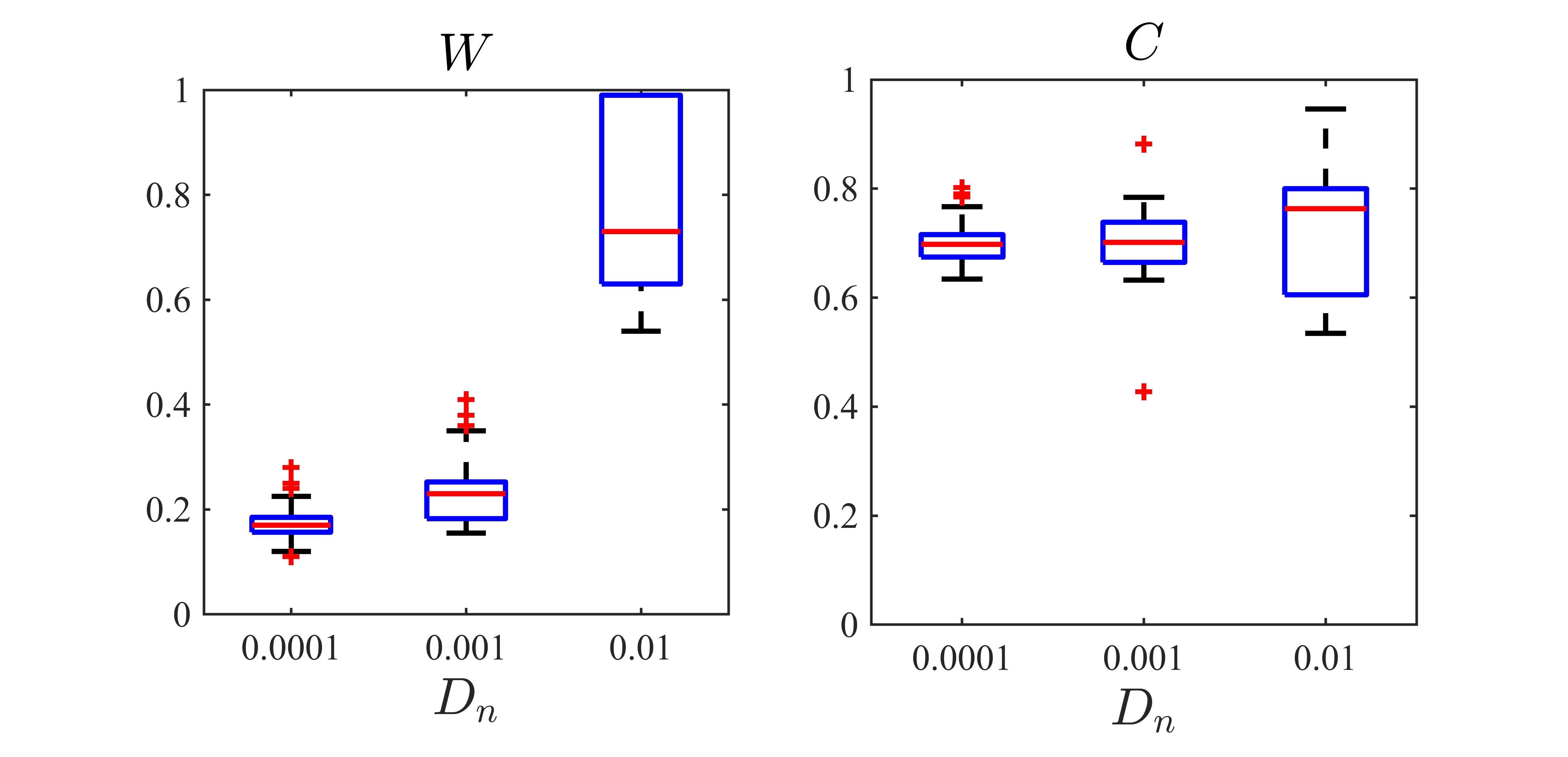}
  \subcaption{Varying cell diffusion $D_n$}
  \label{fig:size:Dn:extra}
  \par\vfill
    \vspace{1\baselineskip}
     \includegraphics[width=\linewidth]{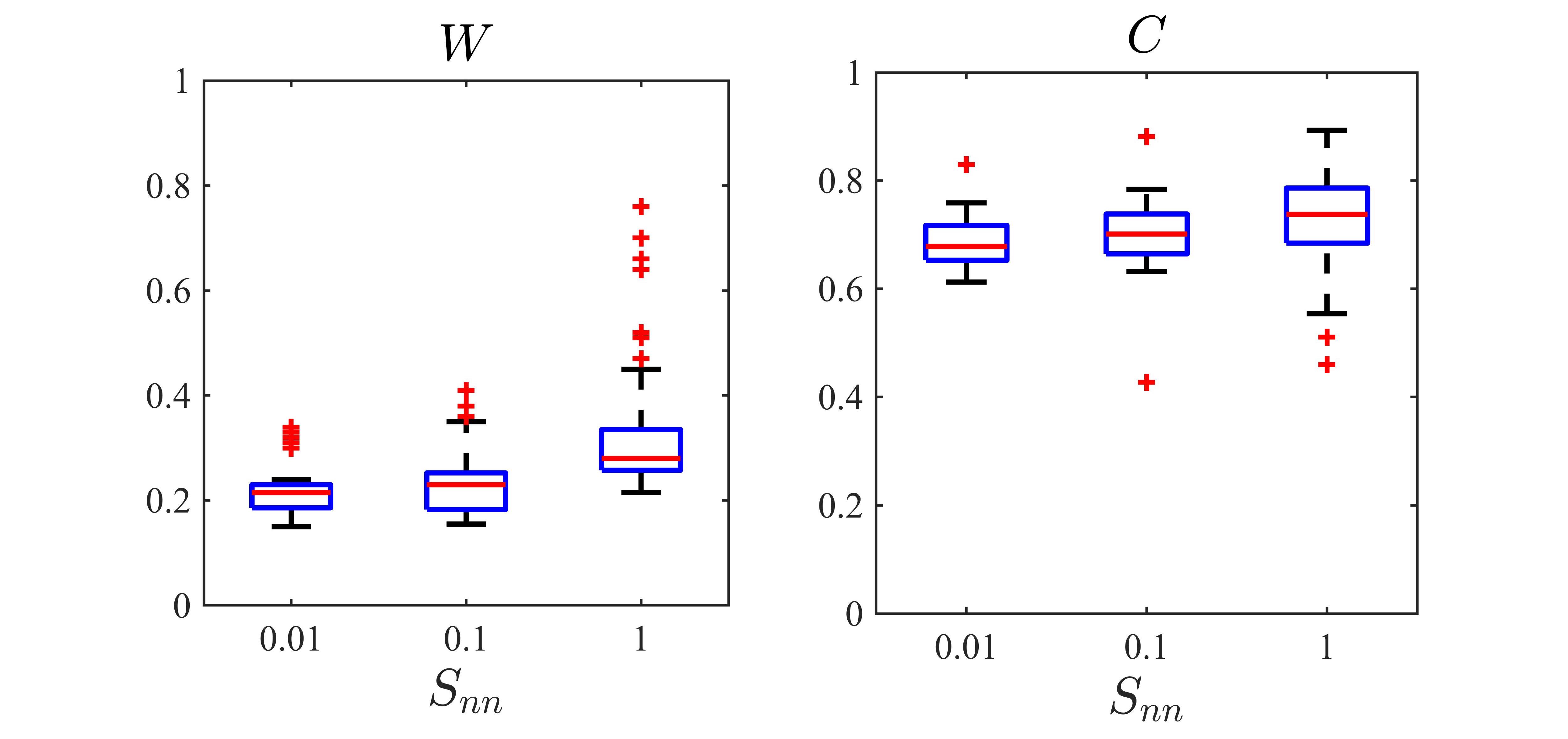}
  \subcaption{Varying cell-to-cell adhesion coefficient $S_{nn}$}
  \label{fig:size:Snn:extra}
\end{minipage}%
\begin{minipage}[c][9cm][t]{.5\textwidth}
  \vspace*{\fill}
  \centering
  \includegraphics[width=\linewidth]{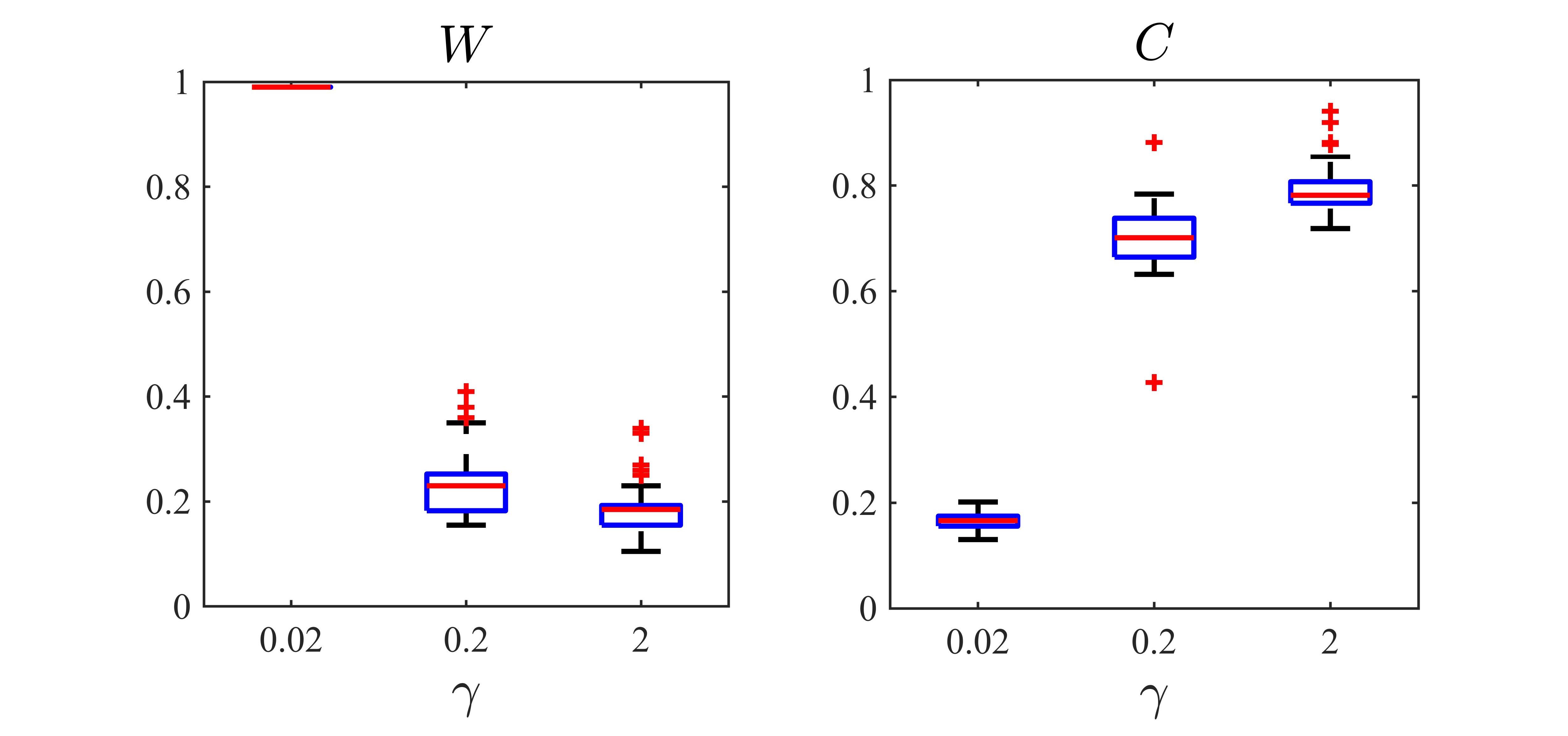}
  \subcaption{Varying ECM degradation rate $\gamma$}
  \label{fig:size:gamma:extra}\par\vfill
    \vspace{1\baselineskip}
  \includegraphics[width=\linewidth]{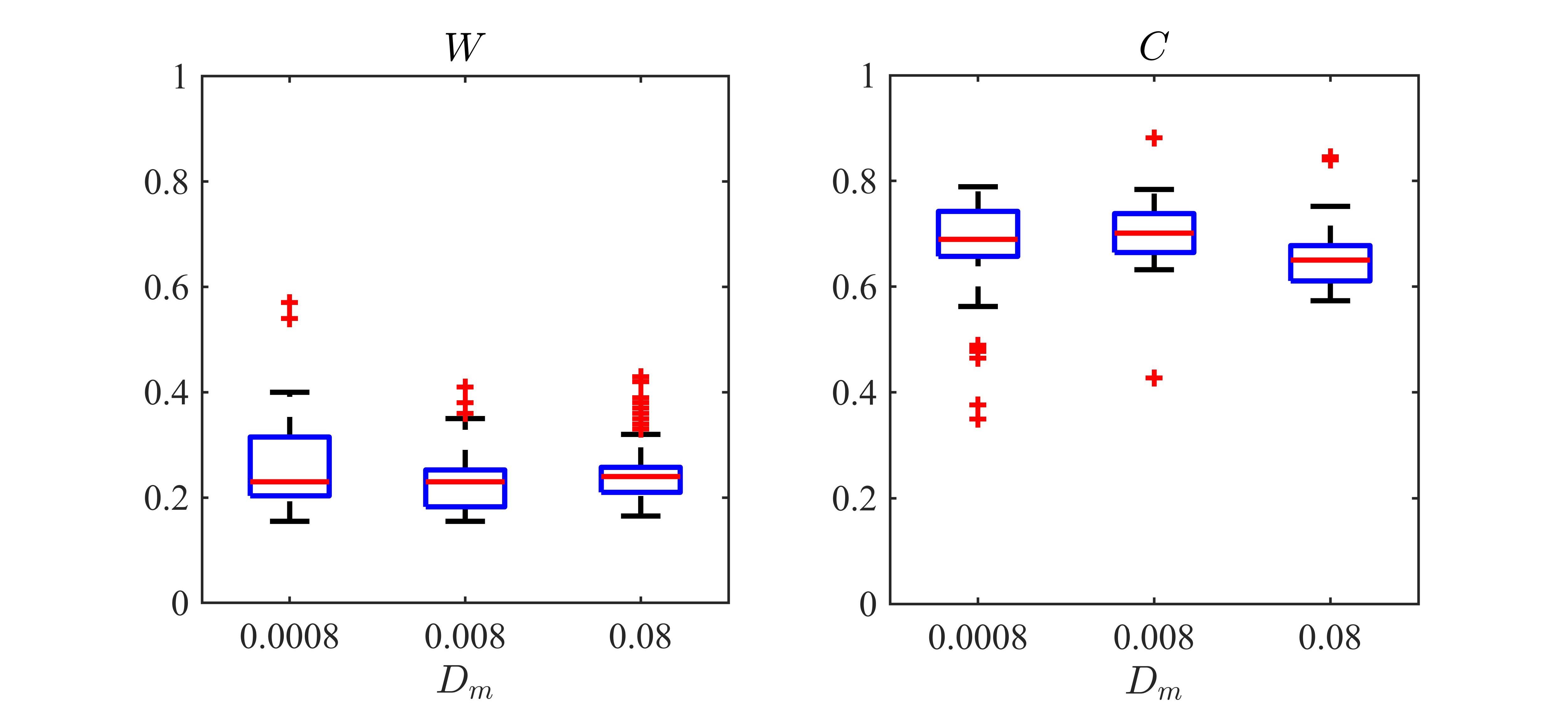}
  \subcaption{Varying MMP diffusion coefficient $D_m$}
  \label{fig:size:Dm:extra}
\end{minipage}
\centerline{\begin{minipage}[c][4.5cm][t]{.5\textwidth}
  \vspace*{\fill}
  \centering
  \vspace{2\baselineskip}
  \includegraphics[width=\linewidth]{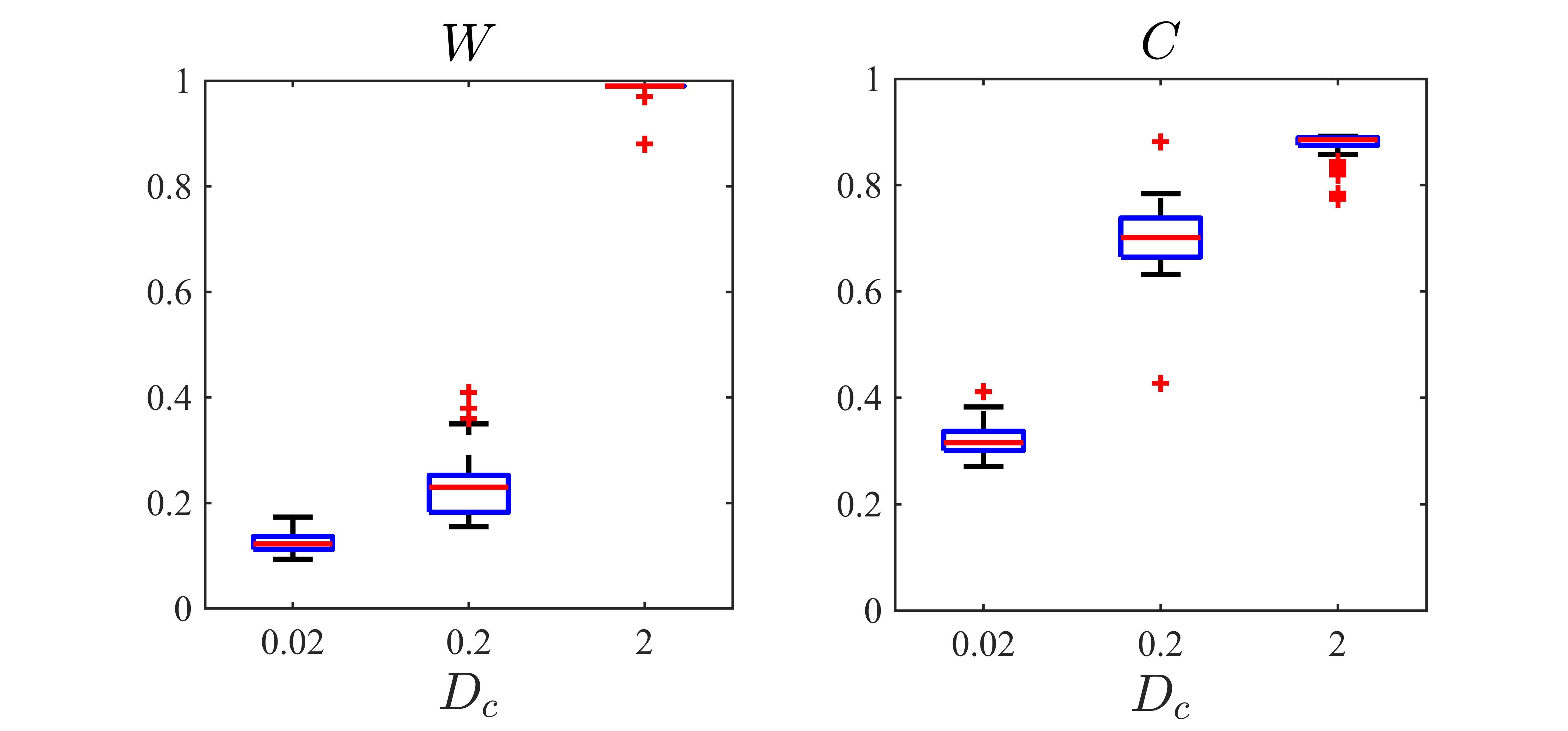}
  \subcaption{Varying VEGF diffusion coefficient $D_c$}
  \label{fig:size:Dc:extra}
\end{minipage}}
\vspace{2\baselineskip}
\caption{\label{Fig:Size:extra} Cluster width $W$ and compactness $C$, defined in (20) and (21), under variations of certain parameters from the baseline parameter set (BPS), in Table~1 of the main publication. 
In (a) we have boxplots of $W$ and $C$ measured on the numerical solution of the system~(15) at $t=50$, for $D_n$ taking its value in the BPS (center), one order of magnitude higher (left) and one order of magnitude lower (right) than the one in the BPS. Each boxplot collects data from 100 simulations under randomised initial conditions~(13) and~(14)$_1$.
In (b)-(e) we have the same as in (a) but varying parameters $S_{nn}$ (b), $\gamma$ (c), $D_m$ (d) and $D_c$ (e).}
\end{figure}

Figure~\ref{Fig:Size:extra} contains plot analogous to those in Figures~7 and~8 in the main publication, with additional information complementing the results presented in Section~3.4 of the main publication. In particular the plots display how choosing parameter values of different orders of magnitude compared to those chosen for the baseline parameter set (see Table 1 in the main publication) will influence average cluster width W and compactness C, for the parameter  $D_n$ (Figure~\ref{fig:size:Dn:extra}), $S_{nn}$ (Figure~\ref{fig:size:Snn:extra}), $\gamma$ (Figure~\ref{fig:size:gamma:extra}), $D_m$ (Figure~\ref{fig:size:Dm:extra}) and $D_c$ (Figure~\ref{fig:size:Dc:extra}).

\begin{figure}
\centering
\includegraphics[width=\textwidth]{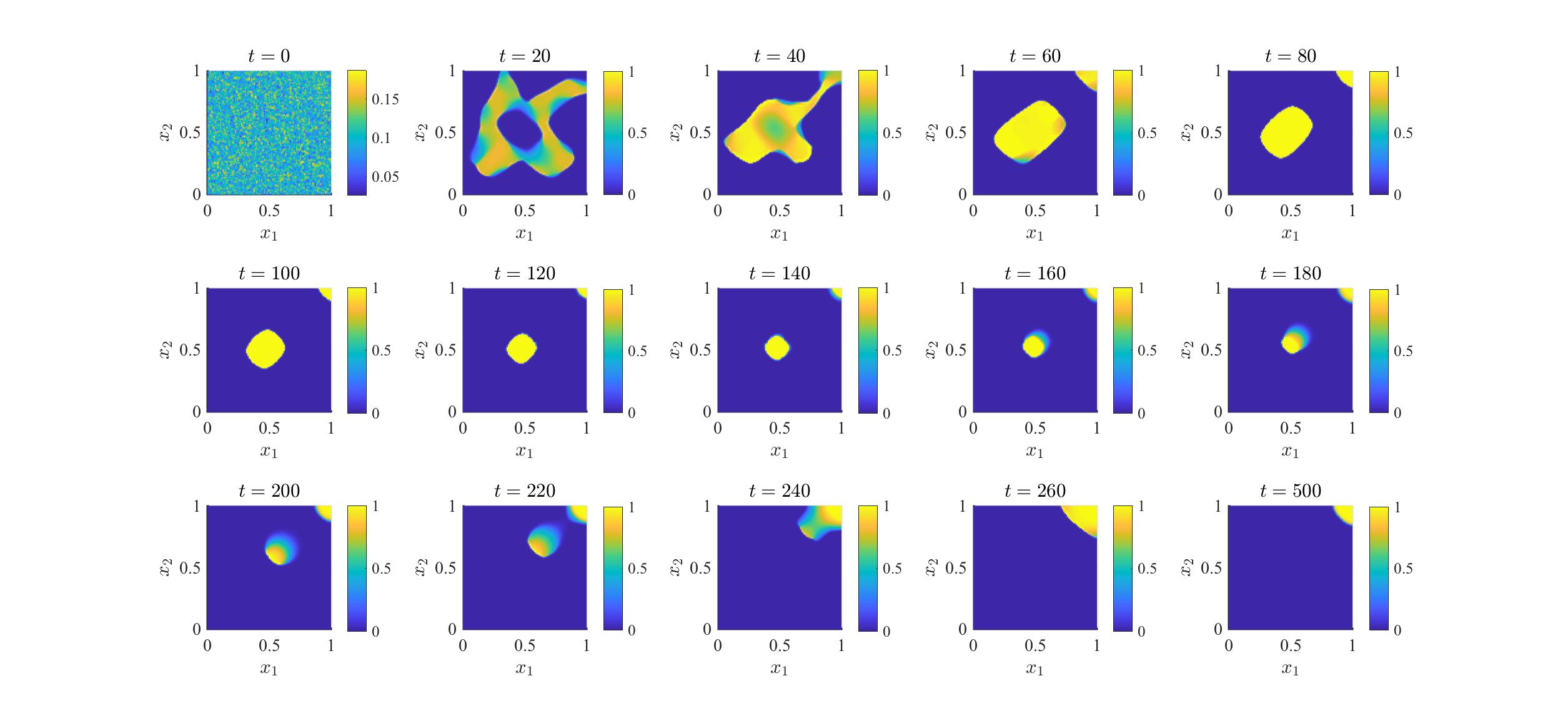}
\caption{\label{Fig:2D:main:extra} Plots of the cell density $n(t,{\bf x})$ obtained solving the system~(15), together with definitions~(6), (8) and~(16), initial conditions~(13) and~(14)$_2$, complemented with zero-flux boundary conditions, under the parameter choices reported in Table~1. The solution is plotted at time $t=0, 20, 40, 60, 80$ (first row, left to right), $t=100, 120, 140, 160, 180$ (second row, left to right) and $t=200, 220, 240, 260, 500$ (third row, left to right). }
\end{figure}

Figure~\ref{Fig:2D:main:extra} contains plots of the two-dimensional simulations under the baseline parameter set shown in the second row of Figures 9, 10 and 11 of the main publication, displaying the solution at additional times, showing the long-term behaviour of the solution.

\begin{figure}
\centering
\includegraphics[width=\textwidth]{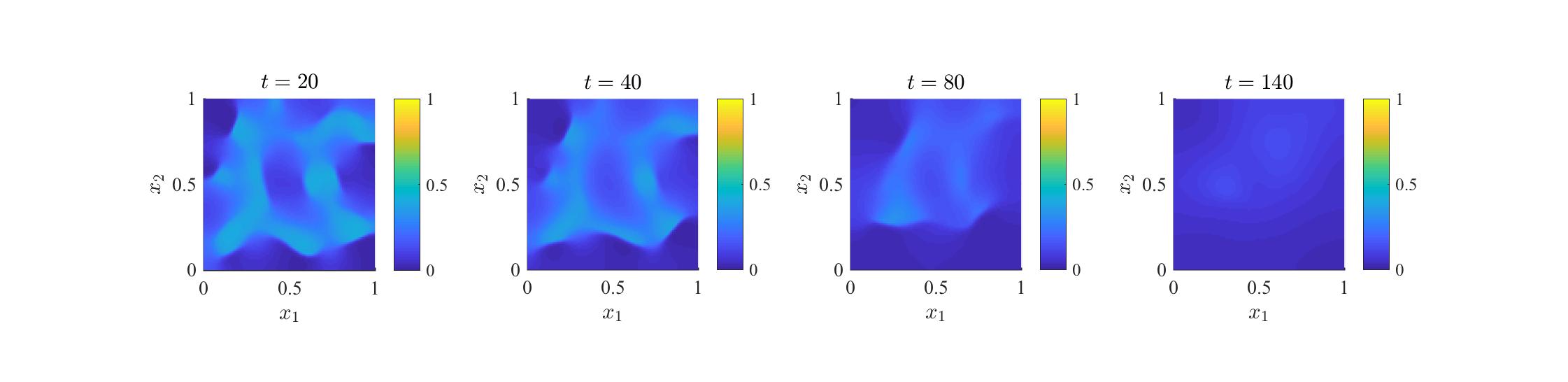}
\caption{\label{Fig:2D:gamma0:extra} Plots of the cell density $n(t,{\bf x})$ obtained solving the system~(15), together with definitions~(6), (8) and~(16), initial conditions~(13) and~(14)$_2$, complemented with zero-flux boundary conditions, under the parameter choices reported in Table~1, except for $\gamma=0$. The solution is plotted at time $t=20$ (first panel), $t=40$ (second panel), $t=80$ (third panel) and $t=140$ (fourth panel). }
\end{figure}

Figure~\ref{Fig:2D:gamma0:extra} contains plots of the two-dimensional simulations under the baseline parameter set in absence of matrix degradation (\textit{i.e.} $\gamma=0$), starting from the same initial conditions (\textit{cf.} first panel in first row of Figure~\ref{Fig:2D:main:extra}).

\begin{figure}
\centering
\includegraphics[width=\textwidth]{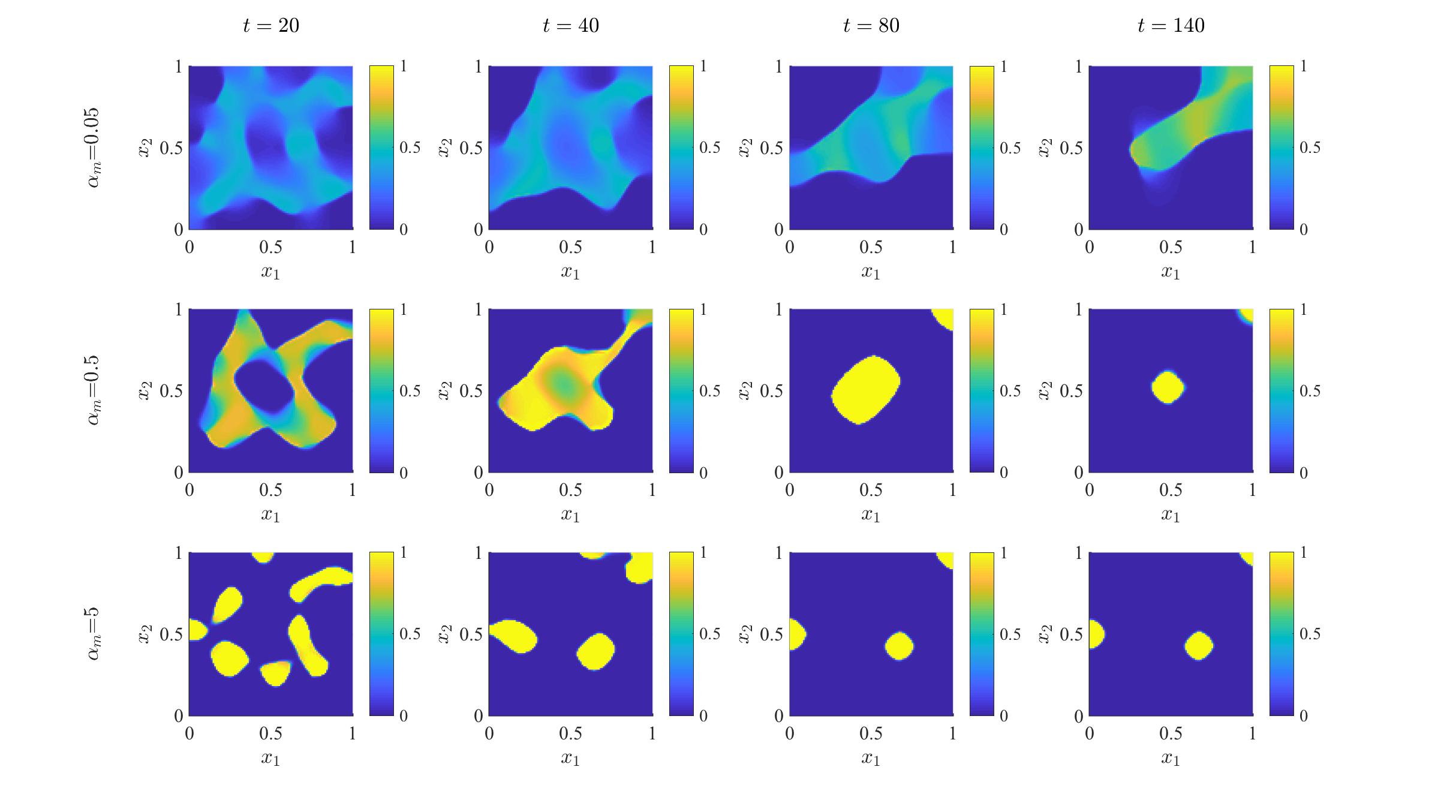}
\caption{\label{Fig:2D:alpham:extra} \textbf{First row:} Plots of the cell density $n(t,{\bf x})$ obtained solving the system~(15), together with definitions~(6), (8) and~(16), initial conditions~(13) and~(14)$_2$, complemented with zero-flux boundary conditions, under the parameter choices reported in Table~1, except for $\alpha_m=0.05$. The solution is plotted at time $t=20$ (first panel), $t=40$ (second panel), $t=80$ (third panel) and $t=140$ (fourth panel). \textbf{Second and third row:} Same as first row, except for $\alpha_m=0.5$ (second row) and $\alpha_m=5$ (third row).}
\end{figure}

Figure~\ref{Fig:2D:alpham:extra} contains plots of the two-dimensional simulations under the baseline parameter set (second row) and under different values of the MMP secretion rate $\alpha_m$, starting from the same initial conditions (\textit{cf.} first panel in first row of Figure~\ref{Fig:2D:main:extra}).

\begin{figure}[ht!]
\centering
\includegraphics[width=\textwidth]{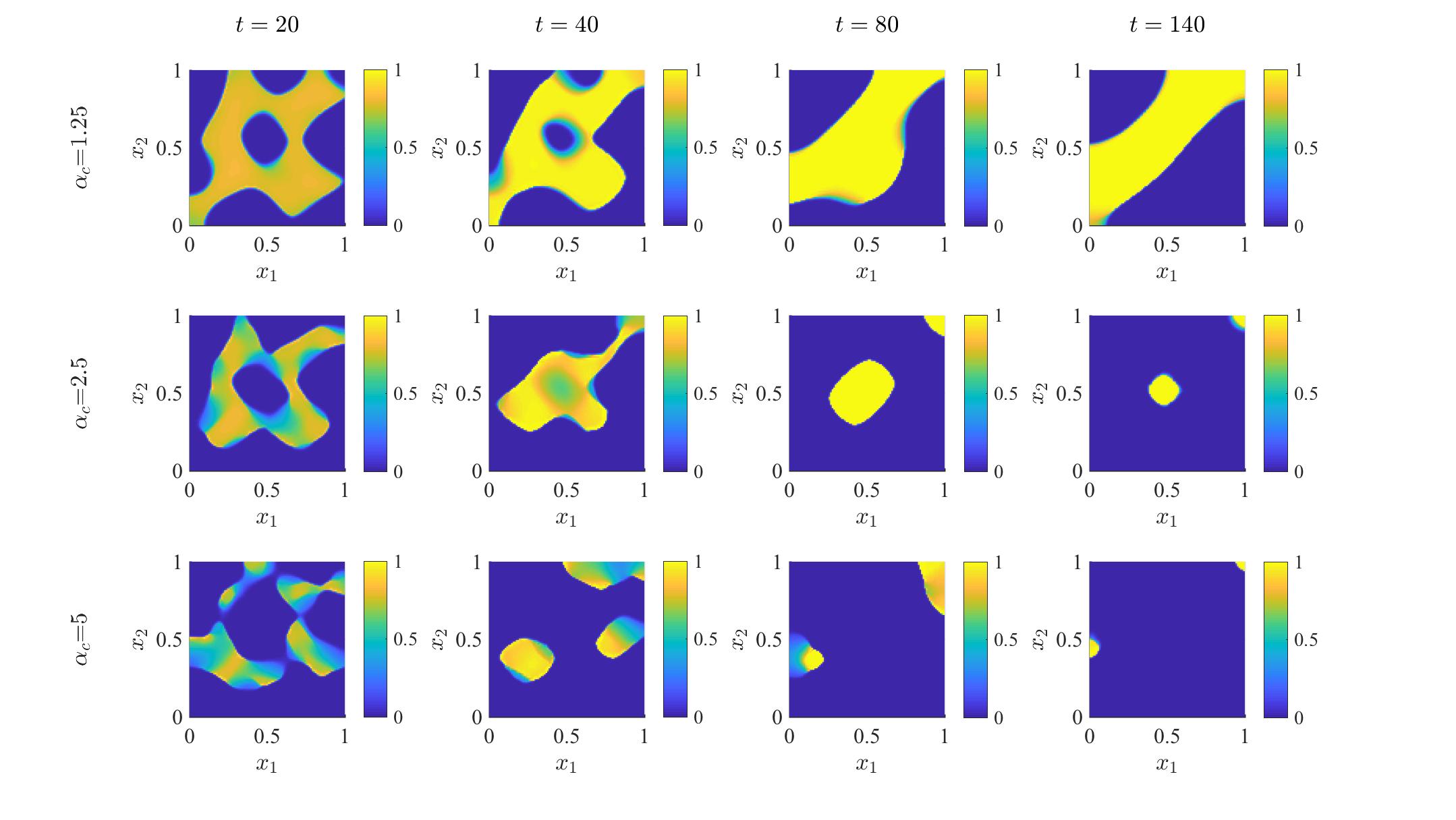}
\caption{\label{Fig:2D:alphac:extra} \textbf{First row:} Plots of the cell density $n(t,{\bf x})$ obtained solving the system~(15), together with definitions~(6), (8) and~(16), initial conditions~(13) and~(14)$_2$, complemented with zero-flux boundary conditions, under the parameter choices reported in Table~1, except for $\alpha_c=1.25$. The solution is plotted at time $t=20$ (first panel), $t=40$ (second panel), $t=80$ (third panel) and $t=140$ (fourth panel). \textbf{Second and third row:} Same as first row, except for $\alpha_c=2.5$ (second row) and $\alpha_c=5$ (third row).}
\end{figure}

Figure~\ref{Fig:2D:alphac:extra} contains plots of the two-dimensional simulations under the baseline parameter set (second row) and under different values of the VEGF secretion rate $\alpha_c$, starting from the same initial conditions (\textit{cf.} first panel in first row of Figure~\ref{Fig:2D:main:extra}).

\pagebreak

\section{Toeplitz-to-circulant embedding in the evaluation of the nonlocal term with zero-flux boundary conditions }\label{sec:num}
Consider the nonlocal term as given in Equation~(5) in Section~2.1.1 of the main document, that is
\begin{equation}\label{eq:a:def:1d}
\mathcal{A}[{\bf v}(t,\cdot)](x) := \frac{1}{R}\int_0^R \sum_{j=0}^1  \eta(j)\, \Gamma(r)g\big({\bf v}(t,x+r\eta(j))\big)\, dr \;,
\end{equation}
together with our realisation of zero-flux boundary conditions, that is $g\big({\bf v}(t,x)\big) = 0$ if $x$ is outside of the spatial domain $\Omega$. \\
A detailed account of the accurate and efficient numerical treatment of~\eqref{eq:a:def:1d} with \emph{periodic} boundary conditions, also for the corresponding spatially two-dimensional case, is given in \cite{gerisch2010approximation}.
In the case of periodic boundary conditions, an efficient numerical scheme is possible if the spatial domain is covered by a uniform grid with $N$ grid cells because then, on each grid cell interface $x$, the same discretization of the nonlocal term in $x$ can be applied. This gives rise to an evaluation of the nonlocal term on all $N$ grid cell interfaces, recall the periodic boundary condition, simultaneously.
We denote the vector of these $N$ values by $A\in\mathbb{R}^N$, which can be computed in the form of a matrix-vector product
\[
 A= C G\,.
\]
In the above $G\in\mathbb{R}^N$ is the evaluation of $g\big({\bf v}\big)$ where ${\bf v}$ takes the average values of the PDE approximation in the $N$ grid cells as provided by the finite volume discretization and $C\in\mathbb{R}^{N,N}$ is a square matrix.
The entries of $C$, the integration weights, are obtained by inserting in~\eqref{eq:a:def:1d} the piecewise-constant reconstruction of function $g\big({\bf v}(t,x)\big)$ from the values in $G$  in place of the function  $g\big({\bf v}(t,x)\big)$ itself.
The uniform grid together with the periodic boundary conditions amount to $C$ being a circulant matrix.
Matrix-vector products involving a circulant matrix can be evaluated efficiently using Fast Fourier Transform (FFT) techniques with complexity $\mathcal{O}(N\log N)$ as opposed to $\mathcal{O}(N^2)$ for the naive evaluation of the product.
Note that matrix $C$ has many zeros if the sensing radius $R$ is small compared to the size of the domain $\Omega$, as is typically the case, but the number of non-zero diagonals grows with increasing the number of grid cells covering $\Omega$ such that the total number of non-zero entries of $C$ is fixed at $\mathcal{O}(N^2)$ and $C$ cannot be considered a sparse matrix. Similar considerations are possible for the spatially two-dimensional version of~\eqref{eq:a:def:1d} leading to an even more significant reduction of computational complexity when the matrix-vector product if evaluated using FFT techniques as opposed to a naive evaluation. We refer to \cite{gerisch2010approximation} for full details. \\
The process outlined above can be carried over to the evaluation of~\eqref{eq:a:def:1d} with zero-flux boundary conditions.
The vector $G\in\mathbb{R}^N$ is defined as above.
Since we consider zero-flux boundary conditions, we do not need to evaluate~\eqref{eq:a:def:1d} on the two grid cell boundaries coinciding with the two boundary points of $\Omega$, as the flux is defined as zero there anyway. This leaves us with the computation of~\eqref{eq:a:def:1d} in the $N-1$ interior grid cell interfaces, that is $A\in\mathbb{R}^{N-1}$.
The integration weights are computed as above, in fact, for grid cell interfaces sufficiently in the interior of $\Omega$ they coincide with the weights for periodic boundary conditions above. Closer to the boundary of $\Omega$, the weights which would apply outside of $\Omega$ are multiplied with zero-values for $g$ and can thus be ignored. Collecting all the necessary weights in a matrix $T\in\mathbb{R}^{N-1,N}$ gives rise to the matrix-vector product
\[
 A = T G\,.
\]
Here now $T$ is a banded (and rectangular) Toeplitz matrix. Efficient FFT techniques are not directly applicable but a naive evaluation is expensive for the same reason as outlined in the periodic case above. However, we can embed $T$ into a slightly larger circulant matrix $\tilde C$ and then apply FFT techniques to this extended matrix-vector product. This is most easily seen in a specific example.
Let $N=7$ and consider the following banded Toeplitz matrix
\[
 T=\begin{pmatrix}
 0 & 1 & 2 & 3 & 0 & 0 & 0\\
 -1& 0 & 1 & 2 & 3 & 0 & 0\\
 -2& -1& 0 & 1 & 2 & 3 & 0\\
 0 & -2& -1& 0 & 1 & 2 & 3\\
 0 & 0 & -2& -1& 0 & 1 & 2\\
 0 & 0 & 0 & -2& -1& 0 & 1
 \end{pmatrix}\,,
\]
where the lower band width is $2$ and the upper band width is $3$ and the weights take symbolic values (indicating the sub/super diagonal they are on) and not specific weight values. For real matrices $T$ the value of $N$ is much larger and the lower and upper band width depends on $R$ and $N$. 
The vectors $A$ and $G$ for our example have the form
\[
 A = \begin{pmatrix}A_1\\A_2\\A_3\\A_4\\A_5\\A_6\end{pmatrix}
 \quad\text{ and }\quad
 G = \begin{pmatrix}G_1\\G_2\\G_3\\G_4\\G_5\\G_6\\G_7\end{pmatrix}\,.
\]
We can now embed $T$ into a circulant matrix $\tilde C\in\mathbb{R}^{N+2,N+2}$ of minimal size in the following way
\[
 \tilde C = \begin{pmatrix}
 0 & 1 & 2 & 3 & 0 & 0 & 0  &{\color{red}-2} & {\color{red}-1}\\
 -1& 0 & 1 & 2 & 3 & 0 & 0  &{\color{red}0} & {\color{red}-2}\\
 -2& -1& 0 & 1 & 2 & 3 & 0  &{\color{red}0} & {\color{red}0}\\
 0 & -2& -1& 0 & 1 & 2 & 3  &{\color{red}0} & {\color{red}0}\\
 0 & 0 & -2& -1& 0 & 1 & 2  &{\color{red}3} & {\color{red}0}\\
 0 & 0 & 0 & -2& -1& 0 & 1  &{\color{red}2} & {\color{red}3}\\
 \color{red}3 & \color{red}0 & \color{red}0 & \color{red}0 & \color{red}-2& \color{red}-1& \color{red}0  &\color{red}1 & \color{red}2\\
 \color{red}2 & \color{red}3 & \color{red}0 & \color{red}0 & \color{red}0 & \color{red}-2& \color{red}-1& \color{red}0 & \color{red}1\\
 \color{red}1 & \color{red}2 & \color{red}3 & \color{red}0 & \color{red}0 & \color{red}0 & \color{red}-2& \color{red}-1& \color{red}0 
     \end{pmatrix}\,.
\]
Note that the actual dimension of $\tilde C$ depends on the upper and lower bandwidth of matrix $T$.
Finally, defining
\[
 \tilde A =\begin{pmatrix}A\\\ast\\\ast\\\ast\end{pmatrix}\in\mathbb{R}^{N+2}
 \quad\text{ and }\quad
 \tilde G =\begin{pmatrix}G\\0\\0\end{pmatrix}\in\mathbb{R}^{N+2}\,,
\]
we can compute $\tilde A = \tilde C \tilde G$ using FFT techniques as in the periodic boundary condition case and recover the desired vector $A$ as the first $N-1$ entries of $\tilde A$. \\
Thanks to the Toeplitz-to-circulant embedding described here, also nonlocal models with zero-flux boundary conditions can be simulated efficiently. The techniques outlined here carry over to the spatially two-dimensional setting, where a block-Toeplitz matrix with Toeplitz blocks is embedded in a block-circulant matrix with circulant blocks and 2D-FFT is applied, see \cite{gerisch2010approximation}.

\bibliographystyle{siam}
\bibliography{Vasculogenesis_Supp}